\documentclass[aps,prd,twocolumn,groupedaddress,superscriptaddress,letterpaper,10pt]{revtex4-1}
\usepackage{booktabs}
\usepackage{graphicx}  
\usepackage{color}     
\usepackage{amsmath}   
\usepackage{amsfonts}  
\usepackage{setspace,subcaption}  
\usepackage{dcolumn}   
\usepackage{enumitem}  
\usepackage{bm}        
\usepackage[usenames,dvipsnames]{xcolor}
\usepackage{geometry}

\begin{document}

\title{Generalized Density Functional Equation of State for Astrophysical Simulations with 3-body forces and Quark Gluon Plasma}

\author{J. Pocahontas Olson}
\affiliation{Center for Astrophysics, Department of Physics, University of Notre Dame, Notre Dame, IN 46556}

\author{MacKenzie Warren}
\email[]{mwarren@msu.edu}
\affiliation{Center for Astrophysics, Department of Physics, University of Notre Dame, Notre Dame, IN 46556}
\affiliation{Joint Institute for Nuclear Astrophysics, Department of Physics, University of Notre Dame, Notre Dame, IN 46556}
\affiliation{Department of Physics and Astronomy, Michigan State University, East Lansing, Michigan 48824}

\author{Matthew Meixner}
\email[]{matthew.meixner@jhuapl.edu}
\affiliation{ Space Exploration Sector, Johns Hopkins University Applied Physics Laboratory,  \\
Laurel, Maryland 20723 USA}

\author{Grant J. Mathews}
\email[]{gmathews@nd.edu}
\affiliation{Center for Astrophysics, Department of Physics, University of Notre Dame, Notre Dame, IN 46556}
\affiliation{Joint Institute for Nuclear Astrophysics, Department of Physics, University of Notre Dame, Notre Dame, IN 46556}

\author{N. Q. Lan}
\email[]{nquynhlan@hnue.edu.vn}
\affiliation{Center for Astrophysics, Department of Physics, University of Notre Dame, Notre Dame, IN 46556}
\affiliation{Joint Institute for Nuclear Astrophysics, Department of Physics, University of Notre Dame, Notre Dame, IN 46556}
\affiliation{Hanoi National University of Education, 136 Xuan Thuy, Hanoi, Vietnam}
    
\author{H. E. Dalhed}
\affiliation{Lawrence Livermore National Laboratory,
    Livermore, CA, 94550}

\date{\today} 

\begin{abstract}
We present an updated general purpose nuclear equation of state (EoS) for use in simulations of core-collapse supernovae, neutron star mergers and black hole collapse. This EoS is formulated in the context of Density Functional Theory (DFT) and is generalized to include all DFT EoSs consistent with known nuclear and astrophysical constraints.
This EoS also allows for the possibility of the formation of material with a net proton excess ($Y_p > 0.5$)
and has an improved treatment of the nuclear statistical equilibrium and the transition to heavy nuclei as the density approaches nuclear matter density.
We  include the effects of pions in the regime above nuclear matter density and incorporate all of the known mesonic and baryonic states at high temperature.
 We analyze how   a 3-body nuclear force term in the DFT at high densities  stiffens the EoS to satisfy the maximum neutron star constraint, however the density dependence of the symmetry anergy and the formation of pions at high temperatures allows for a softening of the central core in supernova collapse calculations leading to a robust explosion.  We also add 
 the possibility of a transition to a QCD chiral-symmetry-restoration and deconfinement phase at densities above nuclear matter density.  
This paper details the physics, and constraints on, this new EoS and presents an illustration of  its implementation in both neutron stars and core-collapse supernova  simulations.
We present the first results from core-collapse supernova simulations with this EoS.
\end{abstract}

\maketitle
\section{Introduction}
To describe the hydrodynamics of compact matter, be it in heavy-ion nuclear collisions, supernovae or neutron stars, an equation of state (EoS) is needed to relate the physics of the various state variables.  In supernovae the EoS determines the dynamics of the collapse and the outgoing shock, and in part determines whether the remnant ends up as a neutron star or a black hole.  In a neutron star, it determines the maximum mass, mass-radius relationship, internal composition, cooling timescales, and dynamics of neutron star mergers.

The two most commonly used equations of state in astrophysical simulations are the EoS of Lattimer and Swesty~(LS91)~\cite{LS91} and that of H.~Shen~et~al.~(Shen98)~\cite{Shen98a, Shen98b}.  The former utilizes a non-relativistic parameterization of nuclear interactions in which nuclei are treated as a compressible liquid drop including surface effects. The latter is based upon a Relativistic Mean Field~(RMF) theory using the TM1 parameter set in which nuclei are calculated in a Thomas-Fermi approximation. Subsequently, H.~Shen~et~al.~\cite{Shen2011} released updates of the Shen98 EoS table. The first update \cite{Shen2011}, EoS2, increased the number of temperature points as well as switching to a linear grid spacing in the proton fraction. In the second update \cite{Shen2011}, EoS3, the effects of $\Lambda$ hyperons were taken into account. It should also be noted that several extensions to the Shen98 table have also been developed, either by the implementation of hyperons~\cite{Ishizuka2008} or, of particular relevance to the present work, including a mixed phase transition to a quark gluon plasma~\cite{Fischer2011,Sagert}. 

Over the last several years much progress has been made on other popular formulations of the nuclear EoS for astrophysical simulations, which we briefly summarize here.
The EoS of Hempel~et~al.~\cite{Hempel2010} is described by a RMF in nuclear statistical equilibrium (NSE) for an ensemble of nuclei and interacting nucleons.  Steiner~et~al.~\cite{Steiner2012} also constructed several EoSs to match recent neutron star observations. In these models the nucleonic matter was parameterized with a RMF model that treats nuclei and non-uniform matter with the statistical model of Hempel~et~al.~\cite{Hempel2010}.


The EoS described here, the Notre Dame-Livermore (NDL) EoS, complements the two most popular EoSs, in that it is formulated in the context of Density Functional Theory~(DFT) rather than the liquid drop or RMF formalism.  
All three approaches are an approximation to the exact many-body problem with the true strong interaction.  Since DFT connects transparently with the many-body Hamiltonian and can be constrained by nuclear structure \cite{brown2013,brown2014} it may be  closer to the true many body problem.  Nevertheless, all three EoSs represent different approaches.  Thus, one measure of the uncertainty in the supernova EoS is to compare these three approaches.  The purpose of the present paper is to summarize the  formulation and first application of this general DFT EoS.

Moreover, the present EoS is of particular interest in the context of modern supernova  simulations.  Even after decades of research the mechanism of core collapse supernova explosions is not yet understood in detail.  Indeed, most supernova simulations that impose spherical symmetry do not explode except for low-mass progenitor stars \cite{janka2012, kitaura2006, burrows2007}. Thus,  it is currently thought that a successful explosion requires some other subtle effects such as  neutrino heated convection \cite{burrows2012} and the standing accretion shock instability (SASI)  \cite{blondin2003}, or micro-turbulent heating behind the shock \cite{Couch15}.  It  is also worthy of note that it still possible to obtain an explosion in spherical symmetry either by invoking  a low mass progenitor \cite{kitaura2006,burrows2007}, by enhancing the flux of neutrinos emanating from the core via convection below the neutrinosphere \cite{wilson1988,wilson1993,WilsonMathews}, via a magnetic-rotation instability \cite{wilson2005},  a second shock produced by a transition to quark-gluon plasma within the nascent neutron star \cite{Sagert2009,FischerSagert2011}, or a resonant oscillation between a $\sim$ keV sterile neutrino and an electron neutrino \cite{Warren14,Warren16}.

It has also been argued  \cite{wilson2005} that at least part of the reason for a successful spherical explosion could be attributed the utilization of an EoS that was sufficiently soft (high compressibility) near nuclear matter density to produce a heated high-density proto-neutron star core (with associated high-temperature neutrinos). 
However, constraints from the observed masses \cite{Demorest,Antoniadis} and radii \cite{steiner2010} of neutron stars require the existence of a stiff (low compressibility) EoS.  In view of the importance of clarifying the contributions of the EoS to the explosion mechanism it is important to update the physics and also to include the possibility of a transition to quark gluon plasma~(QGP) at high density.

In this work we describe  the new NDL EoS, that  is publicly available at {\url{www.crc.nd.edu/~astro/NDLEOS/}}. 
This EoS evolves from the original formulation of Bowers and Wilson~\cite{Bowers82} and somewhat updated in Wilson and Mathews~\cite{WilsonMathews}. The NDL~EoS is updated to be consistent all available experimental nuclear matter constraints and recent mass \cite{Demorest,Antoniadis} and radius \cite{Steiner2012} constraints from neutron stars.  
In  particular, we re-formulate this EoS in the context of a generalized DFT with a Skyrme force near the nuclear saturation density.  We also add a transition to QGP in the regime above the nuclear saturation density. We also add a more realistic transition through the  sub-nuclear pasta phases.

With regards to the DFT  formulation we note that there are already hundreds of DFT Skyrme parameterizations available (see for example \cite{stone2003,Xu09,Dutra2012}).  Of these, only a small fraction can satisfy both the nuclear structure constraints and  properties of neutron stars \cite{Dutra2012}. Here, we develop a new generalized nuclear EoS capable of incorporating any or all of the Skyrme parameterizations in realistic astrophysical simulation.s  Moreover, this EoS can be easily updated as new data and/or Skyrme DFT parameterizations become available.
We present an illustration of the first results of core collapse supernova simulations based upon the set of DFT formulations that satisfy all of the nuclear and neutron-star mass-radius constraints.  We show that the these new EoSs lead to an enhancement in the supernova kinetic energy at early times ($\sim 250$ ms) compared to the earlier version of the EoS.

\section{The NDL Equation of State}

Depending upon the density and temperature there are a variety of matter components that may or may not contribute significantly to the equation of state  during various epochs of supernova collapse, the interiors of neutron stars, and black hole collapse. These include photons, electrons, positrons, neutrinos, mesons, excited mesonic and baryonic states~\cite{Beringer2012}, free neutrons, protons, and atomic nuclei, and even the possibility of a transition to quark gluon plasma. 

At low density and high temperatures we assume a meson gas; consisting of thermally created, pair-produced mesons with zero chemical potential.  In the high density, but low temperature limit, pions are constrained by chemical equilibrium among the neutrons, protons and the other baryonic states.  Baryons are  assigned a non-zero chemical potential  that guarantees baryon number conservation.  The inclusion of the additional mesonic and baryonic states is yet another improvement appearing in this updated EoS.

Below nuclear matter density, the conditions for nuclear statistical equilibrium (NSE) are imposed in the NDL~EoS above a temperature of $T \approx 0.5$ MeV.  Below this temperature in dynamical astrophysical simulations the nuclear matter is solved using a nine element reaction network which must be evolved during collapse. Above this temperature, the nuclear constituents are represented by free nucleons, alpha particles, and a single ``representative'' heavy nucleus.   As nuclear matter density is approached an approximation to the transitions among pasta phases is adopted that is consistent with the nuclear matter Skyrme density functional adopted at high density.

The high density hadronic phase of the EoS is treated with  parameterized Skyrme energy density functionals.  The effects of pions and other mesonic and baryonic resonances on the state variables at high densities are also included as well as  a phase transition to a QGP.

Since at the relevant densities the material is optically thick to photons, one can include photons along with matter particles in the equation of state. 
The electrons and positrons are approximated as a uniform background and are treated as a non-interacting ideal Fermi-Dirac gas.  Photons are approximated as a black-body and obey the usual Stefan-Boltzmann law. Neutrinos, however, are not necessarily confined and must be transported dynamically. In astrophysical simulations most matter except neutrinos can be assumed to be in local thermodynamic equilibrium (one temperature in a zone) but not necessarily in chemical equilibrium (i.e. in supernovae the weak reactions have not necessarily equilibrated).  The independent variables generally chosen for the equation of state are then the temperature $T$, the matter rest-mass density $\rho$ (or alternatively, the number density $n$), and the net charge per baryon $Y_e = n_e / n_B$.  The previous formulation \cite{Bowers82} required $Y_e < 0.5$, but we have removed that restriction in this new version based upon the possibility \cite{fischer2012} for proton rich ejecta above the proto-neutron star in core-collapse supernovae.

The baryonic contribution to the NDL~EoS is divided into five regimes: 
\begin{enumerate}
	\item Baryons below nuclear matter density and not in NSE; 
	\item Baryons below nuclear matter density and in NSE, including the effects of pasta phases of nuclear matter;
	\item Hadronic matter above saturation density including pions;
	\item A phase transition to quark gluon plasma; and
    \item A pure quark-gluon plasma.
\end{enumerate}

The description of matter is completely determined by three input state variables:  the density ($n$ in [fm$^{-3}$]), temperature ($T$ in [MeV]) and the ``electron fraction'' ($Y_e$). 
We define the electron fraction and constituent number  fractions as 
\begin{gather}
	Y_e = \frac{n_{e}}{n_{B}}  	\\
    Y_i = \frac{n_i}{n_{B}}		~.
\end{gather}
where $n_{B}$ is the baryon number density and $n_i$  denotes the the number density of species $i$.


\subsection{Baryons Below Saturation \\and not in NSE}
Below nuclear saturation density and above T $\approx 0.5$~MeV  nuclei are in NSE. However, below this temperature the isotopic abundances should be evolved dynamically. To achieve this, the nuclear constituents are approximated by a 9 element nuclear burn network consisting of $n$, $p$, $^4$He, $^{12}$C, $^{16}$O, $^{20}$Ne, $^{24}$Mg, $^{28}$Si, and $^{56}$Ni~\cite{Bowers82}.

The free energy per baryon is taken to be the sum of contributions from an ideal gas $F_g$~[Eq.~(\ref{eqn:Freegas})] and a coulomb correction $F_c$~[Eq.~(\ref{eqn:Coulomb})]. 
The ideal gas contribution is simply,
\begin{equation}  
F_g = \frac{T}{m_B} \sum_i X_i \left[\frac{1}{A_i} \ln{\biggl(\frac{X_i n_B m_B \mathfrak{\cal A}}{g_i T^{3/2} A_i^{5/2}}\biggr)}\right]	~,
\label{eqn:Freegas}
\end{equation}
where $X_i$ is  the nuclear mass fraction,  $T$ is  the temperature, $A_i$ is the atomic mass number of each species, and $g_i$ is the spin degeneracy. 
The index $i$ runs over the entire reaction network, and ${\cal A}$ is  given by
\begin{equation}  
\mathfrak{\cal A} = \frac{2\pi^3/2}{e\left(m_B\right)^{3/2}}	~.
\label{eqn:thermalwave}
\end{equation}
[Note, that natural units ($\hbar = c = k = 1$) have been adopted here and throughout this manuscript.
We also maintain capital letters for total energy per baryon [MeV/baryon] and lowercase for energy densities [MeV/fm$^3$].]

In this regime the baryonic Coulomb contribution to the free energy is approximated by
\begin{equation}  
	F_C = -\frac{1}{3}n_B^{1/3}e^2\langle A\rangle^{2/3}Y_e^2	~,
   	\label{eqn:Coulomb}
\end{equation}
where $\langle  A \rangle$ is the  average atomic mass of the dynamic composition. From these relations the baryonic pressure and energy per unit mass can be calculated from the ideal gas thermodynamic relations~\cite{WilsonMathews}.

\begin{gather}	  
P_M = n_BT\left( \sum_i \frac{X_i}{A_i} \right) -\frac{1}{9m_B}n_B^{4/3}e^2\langle A \rangle^{2/3}Y_e^2	~, \\
\epsilon_{M} = \frac{3}{2}\frac{T}{m_B} \left( \sum_i \frac{X_i}{A_i} \right) - \frac{1}{3m_B}n_B^{1/3}e^2\langle A \rangle^{2/3}Y_e^2	~.
\end{gather}

\subsection{Baryons Below Saturation and in NSE}

When nuclear statistical equilibrium~(NSE) is valid, the baryonic nuclear material is approximated as consisting of a four component fluid of free protons, neutrons, alpha particles and an average representative heavy nucleus. This formulation is reasonably accurate and convenient in that it leads to fast analytic solutions for the NSE. One should exercise caution, however,~\cite{Bowers82} when considering detailed thermonuclear burning or a precise value of $Y_e$ in NSE is desired. In such cases an extended NSE network should be employed.  

In the absence of weak interactions the neutron and proton mass fractions are constrained by charge conservation (i.e. constant electron fraction $Y_e$),
\begin{equation}
 \label{chargec}
    \sum_i \left(Z_{i}/A_{i}\right)X_i  = Y_e 	~,
\end{equation}
and baryon conservation, i.e.
\begin{equation}
\label{massc}
    \sum_i X_i    = 1 	~.
\end{equation}

The thermodynamic quantities are determined from the free energy per baryon, which is given as a sum of the various constituents,
\begin{equation}
F = F_n + F_p + F_\alpha + F_{\langle A \rangle}	~,
\label{free-energy}
\end{equation}
where $F_{n}$ and $F_{p}$ are contributions from free neutrons and protons respectively, $F_\alpha$ is the free energy of the alpha particles, and $F_{\langle A \rangle}$ is the free energy of heavy nuclei.  These can each be expanded  \cite{WilsonMathews} in  terms of their various contributions,

\begin{widetext}
\begin{align}
    \label{Eq:fn}
    F_n &= X_B Y_n \left\{ \epsilon_{n 0} W + \epsilon_N(1 - W)                    
           + \frac{3}{2} T \left[ \sqrt{1 + \zeta_n^2} - \ln \left(
            \frac{1 + \sqrt{1 + \zeta_n^2}}{\beta \zeta_n} \right) \right] \right\}  , \\
    \label{Eq:fp}
    F_p &= X_B Y_p \left\{ \epsilon_{p 0} W + \epsilon_N(1 - W)                  
        + \frac{3}{2} T \left[ \sqrt{1 + \zeta_p^2} - \ln \left(
            \frac{1 + \sqrt{1 + \zeta_p^2}}{\beta \zeta_p} \right) \right] \right\}  , \\
    \label{Eq:falpha}
    F_\alpha &= X_\alpha  \left\{ \epsilon_{\alpha 0} W + \epsilon_N(1 - W)
             +  \frac{T}{4} \text{ln}\left(\frac{ X_\alpha n m_B \alpha }{ T^{3/2} 4^{5/2}}\right) \right\}  ,  \\
    \label{Eq:fA}
    \begin{split}
    F_{\langle A \rangle}
        &= F_\text{bulk} + F_{S} + F_{C} + F_\text{thermal} \\
	%
      \end{split}
\end{align}
\end{widetext}
where the various terms in \mbox{Eqs.~(\ref{Eq:fn}) - (\ref{Eq:fA})} are defined as the following.  

The mass number of the representative heavy nucleus $\langle A \rangle$ is taken to be $A=100$ if $\langle A \rangle \geq 100$,  while for $\langle A \rangle < 100$ we approximate
the density dependent mass of the average heavy nucleus as:
\begin{equation}
   \langle A \rangle = 194.0 (1 - Y_e)^2(1 + X + 2X^2 + 3 X^3)    \label{eq:avgA}~~.
   \end{equation}
  This expression arises~\cite{bethe1979} from enforcing that the nuclear surface energy be twice the Coulomb energy.

  The density parameter $X$ in Eq.~(\ref{eq:avgA}) is defined by
\begin{equation}
    X \equiv \biggl( \frac{ \rho }{ 7.6 \times 10^{13} ~{\rm g/cm^{3}}} \biggr)^{1/3} .
\end{equation}
In Eqs.~(\ref{Eq:fn}) and (\ref{Eq:fp}) $X_B$ is the free baryon mass fraction,  while in Eqs.~(\ref{Eq:falpha}) and (\ref{Eq:fA}) $X_\alpha$ and $X_A$ are the mass fractions of $^4$He and the average heavy nucleus respectively.  The quantities $Y_p$ and $Y_n$ are the relative number fractions of free baryons in protons or neutrons, respectively.

The quantity $Y_A$ is the average $Z/A$ for heavy nuclei determined by the minimization of the free energy as described below.  The quantity $W$ in Eqs.~(\ref{Eq:fn})-(\ref{Eq:fA}) is a weighting factor that interpolates between the low density and high density regimes. It is defined by $W = (1 - \rho/\rho_N )^2$. The transition from subnuclear to supra-nuclear density is expected to be continuous. The reason for this is that, as the density increases, the equilibrium continuously shifts to progressively heavier nuclei.


The quantity $\rho_N $ is the density at which nuclear matter becomes a uniform sea of nucleons. In the formulation of Bowers and Wilson~\cite{WilsonMathews}, this was found by fitting the saturation density of nuclear matter [i.e. $P_M(n_{0}, T=0, Y_e) = 0$] as a function of $\rho$ and $Y_e$. The zero-temperature result was chosen to simplify the problem of making a smooth transition between the three equation of state regimes. The result is
\begin{equation}
    \rho_N = 2.66 \times 10^{14} \left[1 - (1 - 2 Y_e)^{5/2} \right]  	~~~ \rm g~cm^{-3}~.
    \end{equation}
We caution, however, that the true crust-core transition density is temperature dependent and the transition density is correlated with some of the EOS properties such as the symmetry energy slope \cite{Xu09}.  However, in supernova collapse the the passage through this transition is rapid and thus has little observable affect on the explosion.  Also,  the maximum neutron star mass constraint utilized here  is not sensitive to the crust-core transition as shown in Xu et al. \cite{Xu09}.

The normal $^{56}$Fe ground state is taken as the zero of binding energy.  This is unlike most other equations of state for which the zero point is chosen relative to dispersed free nucleons. The reason for   these choices is that it avoids the numerical complication of negative internal energies in the hydrodynamic state variables at low temperature and density due to the binding energy of nuclei. The energy per nucleon required to dissociate $^{56}$Fe into free nucleons is  fixed at $\epsilon_{p 0} = 8.37$~MeV for protons, while for neutrons it is $\epsilon_{n 0} = 9.15$~MeV \cite{WilsonMathews}.  Thus we take
\begin{equation}
F_{\text{bulk}} = X_{\langle A \rangle}\epsilon_N(1 - W)~.
\end{equation}

The last component of the Helmholtz free energy for heavy nuclei is the thermal contribution,
\begin{equation}
F_{\text{thermal}} =X_{\langle A \rangle}  \frac{T}{A} \ln 
            	\left(\frac{ X_A n m_B \alpha}{g_A T^{3/2} A^{5/2}} \right)~.
\end{equation}
where $A$ is the mass number of the average heavy nucleus, as given in Eq.~\ref{eq:avgA}.

The quantities $\zeta_n$ and $\zeta_p$  in Eqs ({\ref{Eq:fp}) and (\ref{Eq:fn}) are a measure of the degeneracy
of the free baryons.  They are defined \cite{Bowers82,WilsonMathews} by
\begin{equation}
    \zeta_n = \frac{{\cal B}(\rho Y_n X_B)^{2/3}}{T} ;
    \quad
    \zeta_p = \frac{{\cal B}(\rho Y_p X_B)^{2/3}}{T}  ,
\end{equation}
where the quantity ${\cal B}(\rho Y_i X_B)^{2/3}$ is the energy per baryon of a zero-temperature, non-relativistic ideal fermion gas and the constant ${\cal B}$ is
\begin{equation}
    {\cal B} = \frac{3}{10} \biggl(\frac{3}{8 \pi} \biggr)^{2/3} \frac{h^2}{m_B^{5/3}} ~.
\end{equation}

The dimensionless constant $\beta$ appearing in Eqs. \eqref{Eq:fp} and \eqref{Eq:fn} is determined such that the translational part of $f_p$ and $f_n$ reduces to the correct non-degenerate limit ($T\rightarrow \infty$, $\zeta_i \rightarrow 0$).
That is,
\begin{align}
\begin{split}
    \frac{3}{2} &T \biggl[ \sqrt{1 + \zeta_n^2} - \ln{\biggl(\frac{1 + \sqrt{1 + \zeta_n^2} }{\beta \zeta_n} \biggr)} \biggr] \\
        &\rightarrow T \ln{\biggl(\frac{X_B \rho Y_i {\cal A}}{T^{3/2}} \biggr)} .
\end{split}
\end{align}
This requirement implies
\begin{equation}
    \beta = \biggl(\frac{{\cal A}}{2} \biggr)^{2/3}\biggl(\frac{3}{e {\cal B}} \biggr) = 0.781   ~, 
\end{equation}
where ${\cal A}$ is given in Eq.~(\ref{eqn:thermalwave}).

The function $b(Y_e)$ in Eq.~(\ref{Eq:fA}) is determined by the condition that the Coulomb contribution
to the pressure at $\rho = \rho_N$ be canceled by the term proportional to $b(Y_e)$.  This requires,
\begin{equation}
    b(Y_e) = \frac{ e^2}{18} \biggl(\frac{\langle A \rangle^2}{m_B} \biggr)^{\frac{1}{3}} \biggl[ \frac{1}{\rho_N} + 2 \biggl( \frac{\partial \ln{\langle A \rangle}}{\partial \rho} \biggr)_{\rho_N} \biggr]     .
\end{equation}

The expression for the statistical weight of the heavy nucleus $g_A$ appearing in Eq.~(\ref{Eq:fA}) is taken to be
\begin{gather}
    \begin{split}
        \frac{1}{A} \ln{g_A} &= \frac{3}{2} \Biggl\{ \biggl[ 1 - \sqrt{1+ \biggl( \frac{T}{T_S} \biggl)^2} \biggr] \frac{T}{T_S}  \\
                             &\quad + \ln{ \biggl[ \frac{T}{T_S} + \sqrt{1 + \biggl( \frac{T}{T_S} \biggl) }  \biggr] } \Biggr\} ,
    \end{split}  \\
 \intertext{where}
    T_S = (8 \text{ MeV}) \biggl( 1 + 2 \frac{\rho}{\rho_N} \biggr)   .
\end{gather}

\subsection{Nuclear Pasta Phases}

There is a great deal of interesting nuclear physics in this regime at low temperatures in which the interplay between the Coulomb and surface energies lead to various forms of ``pasta'' nuclei, with growing mass number and geometries varying from spherical to sheet-like to cylinder-like geometries~\cite{RavenhallPethickWilson1983}. 
Moreover, although this regime is not important during the collapse itself it does matter for the nascent proto-neutron star.  This is because  convection near the surface and in this density regime of the star can have  a significant impact on the early ($\sim 0.1-0.5$ sec) transport of neutrino flux and its associated heating of material behind the shock. Additionally, pasta phases will have a significant impact upon crust cooling timescale for neutron stars \cite{horowitz2015,schneider2016}.

  Therefore,  in the interest of providing a deeper physical underpinning of the current EoS we  include the transition among the pasta phases.
   A great deal of effort \cite{PastaRef1} has gone into describing this interesting regime, however, 
in the spirit of the current phenomenological Skyrme-force approach of the current work, we can follow the Wigner-Seitz cell derivation of \cite{LS91,RavenhallPethickWilson1983} updated to self-consistently transition  the current Skyrme parameters of the EoS employed here.  This approach was based upon an adoption of the Skyrme interaction, but is applicable to a broad class of density functionals such as the ones of interest here, and hence is a natural means to extend the model developed here.
   
   Within the Wigner-Seitz cell one begins as above by dividing the nuclear free energy into contributions from the formation of very large bulk heavy nuclei that occupy a fraction of the volume in addition to an exterior fluid composed of neutrons, protons, and alpha particles. Heavy nuclei (Eq.~(14)) are characterized by a bulk energy plus surface and coulomb energies.
   Hence,  for the free energy in Eq.~(\ref{free-energy}), we replace the volume factor $W = (1 - \rho/\rho_N)^2 $ with  
   \begin{equation}
   W =(1 - u/X_A)^2~~, 
   \end{equation}
   where
  \begin{equation}
   u = \frac{V_A}{V_c} = X_A \frac{\rho}{\rho_N} ~~.
  \end{equation}
  Here, $V_A$ is the volume of heavy nuclei and $V_c$ is the cell volume.
  The nuclear volume is expressed $V_A = (4/3) \pi r_A^3$ with $r_A$ the effective nuclear radius  corrected for various shapes as described  below. 

  In this case, the surface and Coulomb terms in the free energy of the heavy nucleus $F_{\langle A\rangle} $ are described with modified terms due to the exotic shapes.  
For the formation of nuclear pasta phases in bulk nuclear matter in the Wigner-Seitz cell approximation one can express~\cite{LS91} the sum of $F_S + F_C$ during the passage through this transition as a simple analytic function of the volume parameter $u$, charge to mass ratio $Y_A$  for the average nucleus, and the temperature $T$ as:
\begin{equation} 
F_S + F_C = \beta[c(u) s(u)^2]^{1/3}/n_B = \beta {\cal D}(u)/n_B~~,
\end{equation}
where ${\cal D}(u)$ was deduced in Ref.~\cite{LS91}, based upon a fit to the Thomas-Fermi Skyrme-force calculations of Ref.~\cite{RavenhallPethickWilson1983}:
\begin{widetext}
\begin{equation}  
	{\cal D}(u) = u \left(u-1\right)\frac{\left(\left(1-u\right) D\left(u\right)^{1/3} +u D\left(1-u\right)^{1/3}\right)}{u^{2} + \left(1-u\right)^2 + \alpha u^2 (1-u)^{2}}
\end{equation}
\end{widetext}
where,    
$D(u) \equiv 1 - (3/2)u^{1/3} + (1/2)u$ 
is a Coulomb correction for spherical bubbles in the Wigner-Seitz approximation, and  $\alpha = 0.6$ is a parameter adjusted to optimize the fit to the Thomas-Fermi calculations of \cite{RavenhallPethickWilson1983}.

The normalization factor $\beta$ then contains the dependence of the Coulomb correction upon the charge-to-mass ratio $Y_A$ and temperature T.  This can also be written analytically 
\begin{equation}  
	\beta = 9 \left[\frac{\pi \sigma(Y_A,T)^2 e^2 Y_A^2 n^2}{15}\right]^{1/3} ~.
\end{equation}
Here,   $n$ is the nuclear number density, while $\sigma(Y_A,T)$ is the temperature dependent surface energy per unit area \cite{LS91}.  For a broad range of density functionals can be written \cite{LS91}:
\begin{align}  
	\sigma(Y_A,T) &= \sigma(0.5,0) h(T) \nonumber  \\
				  &\times \frac{ 16 + q}{Y_A^{-3} + q + (1 - Y_A)^{-3}}
        	      	~,
	\label{eq:sigma}
\end{align}
where $\sigma(0.5,0) \approx 1.15$ MeV fm$^{-2}$  
is the surface tension of cold symmetric nuclear matter deduced~\cite{LS91} from fits to individual nuclei. The temperature dependence of the surface tension is taken to diminish quartically up to a critical temperature according to:
\begin{equation}  
	h(T) =
		\begin{cases}
			[1 - (T/T_c(Y_A))^2]^2		&	T \le T_c(Y_A) \\
            0 							&	T  >  T_c(Y_A),
		\end{cases}
\end{equation}
where $T_{c}$ is the critical temperature above  which nuclear pasta phases do not exist and is related the frequency of the giant monopole resonance~\cite{LS91}. 
Here, we express this in terms of the nuclear compressibility parameter $K$ described in Section~\ref{skyrme_sec} and density $n$,
\begin{equation}  
	T_c(Y_A) = 2.4344 K^{1/2} n_B^{-1/3} Y_A (1 - Y_A)	~\text{MeV.}
\end{equation}

The  dimensionless quantity  $q$ in Eq.~(\ref{eq:sigma}) relates to surface symmetry energy $S_0$,
\begin{equation}  
	q = 384 \pi r_0^2 \sigma(0.5,0) /S_0 - 16	~,
\end{equation}
with $r_0 =(3/4 \pi n_0)^{1/3}$ is the nuclear radius parameter here written in terms of the nuclear saturation density $n_0$.

This specifies the transition to pasta nuclei in terms of $W,~Y_A$,~and $T$.  What remains is to specify the dependent variables $W$ and $Y_A$ in terms of the EoS variables $\rho$ and $Y_e$.  This is obtained from the conditions of mass and charge balance in the solution of the  chemical potentials  as described in the next subsection.

  The quantities $n_p = Y_p n_B$,   $n_n = Y_n n_B$, and $n_\alpha = Y_\alpha n_B$  are the fractions of unbound protons, neutrons,  and $\alpha$ particles, respectively.  These quantities are determined from the minimization of the free energy as described below. This then provides a treatment of pasta phases consistent with the Skyrme parametrization above the saturation density which we describe in Section~\ref{skyrme_sec}. 

\subsection{Chemical Potentials}
Having specified the free energies above, the chemical potentials are found from the minimization of the Helmholtz free energy per baryon ($F$),
\begin{gather}
  \label{Eq::ChemPotn}
    \mu_n = \biggl( \frac{\partial F}{\partial X_B} - \frac{Y_p}{X_B} \frac{\partial F}{\partial Y_p} \biggr)  ,\\
 \label{Eq::ChemPotp}
    \mu_p = \biggl( \frac{\partial F}{\partial X_B} + \frac{Y_n}{X_B} \frac{\partial F}{\partial Y_p} \biggr)  ,\\
 \label{Eq::ChemPotalpha}
    \mu_\alpha =  4 \biggl(\frac{\partial F}{\partial X_\alpha}\biggr) ,\\
  \label{Eq::ChemPotnA}
   \mu_{nA} = \biggl( \frac{\partial F}{\partial X_A} - \frac{Y_A}{X_A} \frac{\partial F}{\partial Y_A} \biggr)  ,\\
 \label{Eq::ChemPotpA}
    \mu_{pA} = \biggl( \frac{\partial F}{\partial X_A} + \frac{(1 - Y_A)}{X_A} \frac{\partial F}{\partial Y_A} \biggr)  ,
\end{gather}
where $\mu_p$, $\mu_n$ and $\mu_\alpha$ are the chemical potentials of free protons, neutrons, and alpha particles.  The quantities $\mu_{nA}$ and $\mu_{pA}$ are the chemical potentials of neutrons and protons within heavy nuclei.  These quantities are related by the Saha equation:
\begin{gather}
   \label{cp1}
    2 \mu_n + 2 \mu_p = \mu_\alpha   \\
\label{cp2}
    2 \mu_{nA} + 2 \mu_{pA} = \mu_\alpha  \\
\label{cp3}    \mu_{nA} - \mu_{pA} =  \mu_n - \mu_p =  \hat{\mu}.
\end{gather}

This set of conditions is sufficient to specify the relative mass fractions of the constituent species.
In the current implementation, the three chemical potential constraints [Eqs.~(\ref{cp1})-(\ref{cp3})] combined with charge and baryon number conservation [Eqs.~(\ref{chargec})and (\ref{massc})] are solved self consistently to determine the matter composition. This leads to a 20\% increase in the mass fraction of heavy nuclei when compared to the original approximation scheme of Ref.~\cite{Bowers82}.

\subsection{Baryonic Matter Above Saturation Density \label{skyrme_sec}}
Above nuclear matter density, the baryons are treated as a continuous fluid. In this regime, the free energy per nucleon is given in the form 
\begin{equation}
	F = F_{bulk}(n_B,Y_p) + F_{therm}(n_B, T) + 8.79 \text{ MeV,}
	\label{eqn:free_energy}
\end{equation}
where the addition of 8.79 MeV sets the zero for the free energy to be the ground state of~$^{56}$Fe as discussed above.

Above the saturation density we include both 2-body ($v_{ij}^{(2)}$) and 3-body ($v_{ijk}^{(3)}$) interactions in the many-nucleon system.  The Hamiltonian of this system is thus given by
\begin{equation}
	\hat{H} =  \sum_i\hat{t}_i + \sum_{i<j}v_{ij}^{(2)} + \sum_{i<j<k}v_{ijk}^{(3)}	~,
	\label{eqn:Hamiltonian}
\end{equation}
where $\hat{t}_i$ is the one body contribution while $v_{ij}$ and $v_{ijk}$ are the 2- and 3-body interactions, respectively.
In the density functional approach one can parametrize these interactions to describe the ground-state properties of finite nuclei and nuclear matter~\cite{Brink, Moszkowski, Skyrme1956}.  The microscopic interactions, such as meson exchange, are embedded in the parameters of the density dependent forces.  

Among the most widely used interactions are those of the Skyrme type forces.  In this approach, the two-body potential is given in the form introduced by  Vautherin and Brink~\cite{Vautherin72}.

That is, for most of the  Skyrme potentials considered here  the high density behavior can be dominated by a 3-body repulsive interaction. This term is taken to be a zero range force of the form $v_{123} = t_3\delta \left({\bf r_1 - r_2}\right)\delta \left({\bf r_2 - r_3}\right)$.  If the assumption is made that the medium is spin-saturated, which is valid for neutron star matter and nuclei~\cite{Ring}, the three-body term is then equivalent to a density dependent two-body interaction ~\cite{Vautherin72} given by
\begin{equation}
v_{12}^{(3)'} = \frac{1}{6}t_3\left(1+x_3 \hat{P_{s}}\right)\delta \left({\bf r_1 - r_2}\right) n_B^\sigma \left(\frac{{\bf r_1+r_2}}{2}\right).
\label{eqn:3body}
\end{equation}
This modified Skyrme potential can then be written as in~\cite{Mansour}. This modification has been introduced~\cite{Brink} to enhance the incompressibility of nuclear matter at high densities. A value of $\sigma$ = 1/3 is a common choice~\cite{Kohler, Krivine}, although this parameter varies in the range of $\sigma\sim 0.14-1.0$ as seen in Table~\ref{tab:skyrmemodels}.

The main advantage of the Skyrme density functional is that the variables that characterize nuclear matter can be expressed as analytic functions. 
We use $T_F$ to denote the kinetic energy of a particle at the Fermi surface
\begin{equation}
	T_F = \frac{\hbar^2}{2m}\left(\frac{3\pi^2}{2}\right)^{2/3}n_B^{2/3}  ~.
\end{equation} 
Then, calculating the expectation value of the Hamiltonian [Eq.~(\ref{eqn:Hamiltonian})] in a Slater determinant, the energy per nucleon for symmetric nuclear matter can be derived~\cite{Vautherin72, Dutra2012},  
\begin{widetext}
	\begin{equation}  
		\begin{split}	\label{eqn:skyrme}
			\frac{E}{A} &= \frac{3}{5}T_F H_{5/3} + \frac{t_0}{8} n_B [2(x_0+2)-(2x_0+1)H_2]  + \frac{1}{48} \sum_{i=1}^3 t_{3i} n_B^{\sigma_i+1} [2(x_{3i}+2) - (2x_{3i}+1)H_2]  \\
		 			&+ \frac{3}{40} \left(\frac{3 \pi^2}{2}\right)^{2/3}n_B^{5/3}\left(a H_{5/3}+b H_{8/3}\right)  + \frac{3}{40} \left(\frac{3\pi^{2}}{2}\right)^{2/3} n_B^{5/3 + \delta} \left[t_{4}(x_{4}+2) H_{5/3} - t_{4}(x_{4} +\frac{1}{2} )H_{8/3} \right] \\
					&+\frac{3}{40}\left(\frac{3 \pi^{2}}{2}\right)^{2/3} n_B^{5/3 + \gamma} \left[t_{5}(x_{5}+2)H_{5/3} + t_{5}(x_{5}+\frac{1}{2})H_{8/3} \right]~,
		\end{split}
	\end{equation}
    \end{widetext}
  where
	\begin{align}
		a &= t_1 (x_1+2) + t_2(x_2+2)	, \\
		b &= \frac{1}{2}	[t_2(2x_2+1)-t_1(2x_1+1)] 	, \\
		H_m(Y_p) &= 2^{m-1}[Y_p^m+(1-Y_p)^m]	.
	\end{align}
Notice that for symmetric matter $H_m (Y_P = 1/2) = 1$ for all $m$.  We have included here more non-standard terms, such as those involving $t_4, x_4, t_5$ and $x_5$, which do appear in some parameterizations \cite{Dutra2012}.

All quantities and coefficients for symmetric nuclear matter can be  obtained from Eq.~(\ref{eqn:skyrme}).
The pressure, is deduced from 
	\mbox{$P = n_B^2 \frac{\partial}{\partial n_B} \left( \frac{E}{A} \right)$} 
and is given in Eq.~(\ref{eqn:skyrme_press}).
Eq.~(\ref{eqn:skyrme_comp}) gives the volume compressibility of symmetric nuclear matter. This is calculated from the derivative of the pressure with respect to number density: 
	\mbox{$K = 9 \left(\frac{\partial P}{\partial n_B}\right) $} 
    \mbox{$= 18\frac{P}{n_B}+9n_B^2 \frac{\partial^2}{\partial n_B^2} \left( \frac{E}{A}\right)$}.  
Finally, the skewness coefficient, 
	\mbox{$Q = 27n_B^3\frac{\partial^3}{\partial n_B^3} \left(\frac{E}{A} \right) $},
is deduced from the third derivative of the free energy per nucleon~\cite{Dutra2012} and is given in Eq.~(\ref{eqn:skyrme_skew}).	
	
\begin{widetext}
	\begin{align}
		\label{eqn:skyrme_press} 	\begin{split}
		P &= \frac{2}{5}T_F n_B H_{5/3} + \frac{t_0}{8} n_B^2 [2(x_0+2)-(2x_0+1)H_2]  + \frac{1}{48} \sum_{i=1}^3 t_{3i} (\sigma_i+1) n_B^{\sigma_i+2} [2(x_{3i}+2) - (2x_{3i}+1)H_2]  \\
		 			&+ \frac{1}{8} \left(\frac{3 \pi^2}{2}\right)^{2/3}n_B^{8/3}\left(a H_{5/3}+b H_{8/3}\right) +\frac{1}{40} \left(\frac{3 \pi^{2}}{2} \right)^{2/3} (5+ 3\delta) n_B^{8/3 + \delta} \left[t_{4} (x_{4} + 2) H_{5/3} - t_{4} (x_{4} + \frac{1}{2} )H_{8/3} \right] \\
					&+\frac{1}{40} \left(\frac{3\pi^{2}}{2} \right)^{2/3} (5+3\gamma) n_B^{8/3 + \gamma} \left[t_{5} (x_{5}+1) H_{5/3} + t_{5}(x_{5}+\frac{1}{2}) H_{8/3} \right] 
					\end{split}\\
		\label{eqn:skyrme_comp} 	\begin{split}
		K &= 6 T_F H_{5/3} + \frac{9 t_0}{4} n_B [2(x_0+2)-(2x_0+1)H_2]  \\
		&+ \frac{3}{16} \sum_{i=1}^3 t_{3i} (\sigma_i+1)(\sigma_i+2)n_B^{\sigma_i+1} [2(x_{3i}+2) - (2x_{3i}+1)H_2]  + 3 \left(\frac{3 \pi^2}{2}\right)^{2/3} n_B^{5/3}\left(a H_{5/3}+b H_{8/3}\right)   \\
		& +\frac{3}{40} \left(\frac{3\pi^{2}}{2} \right)^{2/3} (5+3\delta) (8+3 \delta) n_B^{5/3+\delta} \left[t_{4} (x_{4}+2) H_{5/3} - t_{4}(x_{4}+\frac{1}{2} )H_{8/3} \right] \\
		& + \frac{3}{40} \left(\frac{3 \pi^{2}}{2} \right)^{2/3} (5+3 \gamma)(8+3\gamma) n_B^{5/3+\gamma} \left[t_{5} (x_{5}+2) H_{5/3} + t_{5} (x_{5} +\frac{1}{2}) H_{8/3} \right]
			\end{split} \\ 
		\label{eqn:skyrme_skew} \begin{split}
		Q &= \frac{24}{5}T_F H_{5/3} + \frac{9}{16} \sum_{i=1}^3 t_{3i} \sigma_i (\sigma_i+1) (\sigma_i-1) n_B^{\sigma_i+1} 
						[2(x_{3i}+2) - (2x_{3i}+1)H_2] 	\\
						&- \frac{3}{4} \left(\frac{3 \pi^2}{2}\right)^{2/3} n_B^{5/3}\left(a H_{5/3}+b H_{8/3}\right) \\
						& +\frac{3}{40} \left(\frac{3\pi^{2}}{2} \right)^{2/3} (2+ 3\delta) (5 +3 \delta)(3 \delta -1) n_B^{5/3 + \delta} \left[t_{4} (x_{4} +2) H_{5/3} - t_{4} (x_{4} - \frac{1}{2} ) H_{8/3} \right] \\
				& + \frac{3}{40} \left(\frac{3\pi^{2}}{2} \right)^{2/3} (2+ 3\gamma)(5+3\gamma)(3\gamma -1) n_B^{5/3 + \gamma} \left[t_{5} (x_{5} + 2) H_{5/3} + t_{5} (x_{5} + \frac{1}{2} ) H_{8/3} \right]
		\end{split}
	\end{align}
\end{widetext}

Equations (\ref{eqn:skyrme}) and (\ref{eqn:skyrme_press})-(\ref{eqn:skyrme_skew}) completely describe the properties of symmetric nuclear matter. Values for the coefficients can be constrained by fixing the density of nuclear saturation, as well as imposing the observational constraint that the maximum mass of a neutron star must exceed $2.01 \pm 0.04 ~M_\odot$~\cite{Demorest,Antoniadis}.

Table~\ref{tab:skyrmemodels} summarizes the Skyrme parameterizations considered in this work.   These versions were identified  in Dutra \textit{et al}~\cite{Dutra2012} as the EoSs that satisfy  all available constraints from properties of nuclear matter and nuclei (e.g. see review in Ref.~\cite{Newton2013}).  The EoS's that satisfy the neutron star maximum mass constraint M$\ge 2.01 \pm 0.04$ \cite{Antoniadis} are also identified.  In what follows we will run core collapse simulations with the EoS's compatible with those EoS's that are compatible with the maximum neutron star mass constraint.

\begin{table*}
   \caption{Skyrme parameterizations considered in this work, taken from Ref.~\cite{Dutra2012}. All models considered here have $t_4$, $x_4$, $t_5$, and $x_5$ equal to zero.     
    \label{tab:skyrmemodels}}
    \begin{tabular}{ c c cccccccccccccc} %
    \hline
Model & $t_{0}$ & $t_{1}$ & $t_{2}$ & $t_{31}$ & $t_{32}$ & $t_{33}$ & $x_{0}$ & $x_{1}$ & $x_{2}$ & $x_{31}$ & $x_{32}$ & $x_{33}$& $\sigma_{1}$ & $\sigma_{2}$ & $\sigma_{3}$ \\ 
\hline
    GSkI\footnote{Maximum neutron star mass $> 2$ M$_\odot$ \cite{Antoniadis}} & -1855.5 & 397.2 & 264.6 & 13858.0 & -2694.1 & -319.9 & 0.12 & -1.76 & -1.81 & 0.31 & -1.19 & -0.46 & 0.33 & 0.67 & 1.00\\
    GSkII & -1856.0 & 393.1 & 266.1 & 13842.9 & -2689.7 & -- & 0.09 & -0.72 & -1.84 & -0.10 & -0.35 & -- & 0.33 & 0.67 & -- \\
    KDE0v1$^a$ & -2553.1 & 411.7 & -419.9 & 14603.6 & -- & -- & 0.65 & -0.35 & -0.93 & 0.95 & -- & -- & 0.17 & -- & -- \\
    LNS & -2485.0 & 266.7 & -337.1 & 14588.2 & -- & -- & 0.06 & 0.66 & -0.95 & -0.03 & -- & -- & 0.17 & -- & -- \\
    MSL0$^a$ & -2118.1 & 395.2 & -64.0 & 12875.7 & -- & -- & -0.07 & -0.33 & 1.36 & -0.23 & -- & -- & 0.24 & -- & --\\
    NRAPR$^a$ & -2719.7 & 417.6 & -66.7 & 15042.0 & -- & -- & 0.16 & -0.05 & 0.03 & 0.14 & -- & -- & 0.14 & -- & -- \\
    Ska25s20$^a$ & -2180.5 & 281.5 & -160.4 & 14577.8 & -- & -- & 0.14 & -0.8 & -- & 0.06 & --& --& 0.25 & -- & -- \\
	Ska35s20$^a$ & -1768.8 & 263.9 & -158.3 & 12904.8 & -- & -- & 0.13 & -0.80 & 0.00 & 0.01 & -- & -- & 0.35 & -- & --\\
	SKRA & -2895.4 & 405.5 & -891. & 16660.0 & -- & -- & 0.08 & -- & 0.20 & -- & -- & -- & 0.14 & -- & -- \\
    SkT1$^a$ & -1794.0 & 298.0 & -298.0 & 12812.0 & -- & -- & 0.15 & -0.50 & -0.50 & 0.09 & -- & -- & 0.33 & -- & -- \\
    SkT2$^a$ & -1791.6 & 300.0 & -300.0 & 12792.0 & -- & -- & 0.15 & -0.50 & -0.50 & 0.09 & -- & -- & 0.33 & -- & -- \\
    SkT3$^a$ & -1791.8 & 298.5 & -99.5 & 12794.0 & -- & -- & 0.14 & -1.00 & 1.00 & 0.08 & -- & -- & 0.33 & -- & -- \\
    Skxs20 & -2885.2 & 302.7 & -323.4 & 18237.5 & -- & -- & 0.14 & -0.26 & -0.61 & 0.05 & -- & --& 0.17 & -- & --\\
    SQMC650 & -2462.7 & 436.1 & -151.9 & 14154.5 & -- & --& 0.13 & -- & -- & -- & -- & -- & 0.17 & -- & -- \\
    SQMC700 & -2439.1 & 371.0 & -96.7 & 13773.6 & -- & -- & 0.10 & -- & -- & -- & -- & -- & 0.16 & -- & -- \\
    SV-sym32 & -1883.3 & 319.2 & 197.3 & 12559.5 & -- & -- & 0.01 & -0.59 & -2.17 & -0.31 & -- & -- & 0.30 & -- & -- \\
    \hline
   \end{tabular}

\end{table*}

\subsection{Thermal Correction~\label{sec:therm}}
For energetic environments such as core collapse supernovae or heavy ion collisions it is necessary to consider nuclear matter at finite temperature.  This is addressed via a thermal correction.  
For the thermal contribution to the free energy per particle in bulk nuclear matter we follow the approach described in Refs.~\cite{WilsonMathews,Mayle}. We assume a degenerate gas 
of the mesonic and baryonic states~\cite{Beringer2012}, 
as a function of temperature $T$ and baryon density $n$.  Since the zero-temperature contribution to the free energy is already properly taken into account by the Skyrme contribution, only the thermal portion needs to be added. We also assume that the 
baryonic states
are in chemical equilibrium. The expression for the thermal contribution is written: 
\begin{equation}
	F_{\text{thermal}}(n_B,T) = \alpha-\alpha_0 + \frac{1}{n_B}\left(\omega - \omega_0\right) ~,
\end{equation}
where $\alpha$ and $\omega$ are  a finite temperature ``chemical potential'' and grand} potential density, respectively. The quantities, $\alpha_0$ and $\omega_0$ are the zero-temperature limits of the ``chemical potential'' and grand potential density and $n_B$ is the local baryon number density.

$\alpha_0$ is constrained from the number density of baryons and is determined by the relation
\begin{equation}
n_B = \sum_i\frac{g_i}{2\pi^2}\left(\alpha_0^2 - \widetilde{m}^2_i\right)^{3/2} ~,
\label{eqn:numdensity}
\end{equation}
while the zero-temperature limit of the grand potential density is \cite{WilsonMathews}
\begin{widetext}
\begin{equation}
\omega_{0} = -\sum_i\frac{g_i}{2\pi^2}\left[\alpha_0\sqrt{\alpha_0^2-\widetilde{m}^2_i}\left(\alpha_0^2-\frac{5}{2}\widetilde{m}_i^2\right)+\frac{3}{2}\widetilde{m}_i^4\text{ln}\left(\frac{\alpha_0+\sqrt{\alpha_0^2-\widetilde{m}_i^2}}{\widetilde{m}_i}\right)\right]   ~.
\label{eqn:omega0}
\end{equation}
\end{widetext}
In Eqs (\ref{eqn:numdensity}) and (\ref{eqn:omega0}) $\widetilde{m}_i$ is an effective particle mass deduced from fits to results from relativistic Bruckner Hartree-Fock theory~\cite{Migdal}
\begin{equation}
%
%
\label{eqn:eff_mass}
\widetilde{m}_i = \frac{m_i}{1+\xi\eta} ~,
\end{equation}
where $\xi = 0.027$ and the sum over $i$ includes both nucleons and delta particles.

The finite temperature $\alpha$ is then found by enforcing baryon number conservation and assuming an ideal Fermi gas, 
\begin{equation}
	n_B = \sum_i\frac{g_i}{2\pi^2}\int_0^{\infty}\left(h(p_i,\alpha_i)-h(p_i,-\alpha_i)\right)p^2 dp	,
\label{eqn:numberdensity}
\end{equation}
where, $h(p_i,\alpha_i)$ is the usual Fermi distribution function 
\begin{equation}
	h(p_i,\alpha_i) = \frac{1}{\exp[(\epsilon_i - \alpha_i)/T] + 1}  	~,
\end{equation}
and $g_i$ is the spin-isospin degeneracy factor.  

The finite temperature grand potential density is given as
\begin{widetext}
\begin{equation}
\omega_{\text{baryon}} = - \sum_{i} \frac{g_{i} T}{2 \pi^2} \int_{0}^{\infty} p^{2} dp \left(\ln{(1+\exp{[-(\epsilon-\alpha)/T]})}+\ln{(1+\exp{[-(\epsilon+\alpha)/T]})}\right)
\label{eqn:omega}
\end{equation}
\begin{equation}
\omega_{\text{meson}} = 2 \sum_{i} \frac{g_{i} T}{2 \pi^2} \int_{2 \pi \hbar n^{1/3}}^{\infty} p^{2} dp \ln{\left(1-\exp{\left(-\epsilon/T\right)}\right)}
\end{equation}
\end{widetext}
where the index $i$ runs over baryonic and mesonic resonances and the effective energy $\epsilon_i$ is given by
\begin{equation}
\epsilon_i = \sqrt{p^2+\widetilde{m}_i^2}.
\end{equation}

It should be noted that $\alpha$ and $\alpha_0$ are only used to construct $F_{thermal}$ and do not correspond to a real chemical potential, since they assume an ideal gas behavior. The actual chemical potentials are found from derivatives of the total free energy~[Eq.~\ref{eqn:free_energy}] with respect to density as in Eqs.~(\ref{Eq::ChemPotn})-(\ref{Eq::ChemPotpA}). For nuclear matter they simplify to:
\begin{align}
\label{eqn:mun}
\mu_n &= F + n_B\left(\frac{\partial{F}}{\partial{n_B}}\right)_{T,Y_p} - Y_p\left(\frac{\partial{F}}{\partial{Y_p}}\right)_{T,n_B} \\
\mu_p &= F + n_B\left(\frac{\partial{F}}{\partial{n_B}}\right)_{T,Y_p} + Y_n\left(\frac{\partial{F}}{\partial{Y_p}}\right)_{T,n_B}
\label{eqn:mup}
\end{align}
where $Y_n + Y_p = 1$. 

\subsection{Thermodynamic State Variables}
Once the thermal contribution to the free energy is constructed, the thermodynamic quantities can be calculated. Of particular interest are the total internal energy ($E$), the total pressure ($P$), the entropy per baryon ($S$) and the adiabatic index $\Gamma$. The total internal energy is calculated from the free energy:
\begin{equation}
E = F - T\left(\frac{\partial F}{\partial T}\right)_{n_B,Y_e} + E_{e} + E_{\gamma}    ~,
\end{equation}
where $E_e$ is the electron energy determined by numerically integrating over Fermi-Dirac distributions and $E_{\gamma}$ is the photon energy contribution determined from the usual Stefan-Boltzmann law.

The pressure is calculated from \mbox{$P = n_B^2\left(\partial F/ \partial n_B\right)$}. It is important to note that the thermal contribution to the pressure is not the simple form of $\omega_0 - \omega$ as one would expect from the usual application of the thermodynamic potential, but is given by a slightly more complicated form
\begin{equation}
\label{eqn:therm_press}
P_{therm} = \omega_0 - \omega +n_B\frac{\partial}{\partial n_B}\left[\omega - \omega_0\right].
\end{equation}
This is due to the fact that the effective mass [cf.~Eq.~(\ref{eqn:eff_mass})] is density dependent. If this dependence were removed we would recover the usual form of the pressure from the thermodynamic potential. 

We thus determine the total pressure from the free energy 
\begin{equation}
\label{eqn:total_press}
P = n_B^2\left(\frac{\partial F_{bulk}}{\partial n_B}\right) + P_{therm} + P_e + P_{\gamma} ~,
\end{equation}
where again the electron and photon contributions are determined from the energies discussed previously. The entropy per baryon in units of Boltzmann's constant is given by a simple derivative \mbox{$S = -\left(\partial F/ \partial T\right)$}, and the adiabatic index is given by the usual form \mbox{$\Gamma = \left[\partial \ \text{ln} P/ \partial \ln{(\rho)}\right]$}.  We note, however, that the neutrino pressure and energy contributions must be independently solved and accounted for in a simulation.

\begin{figure}
\centering
\includegraphics[width = 0.5\textwidth]{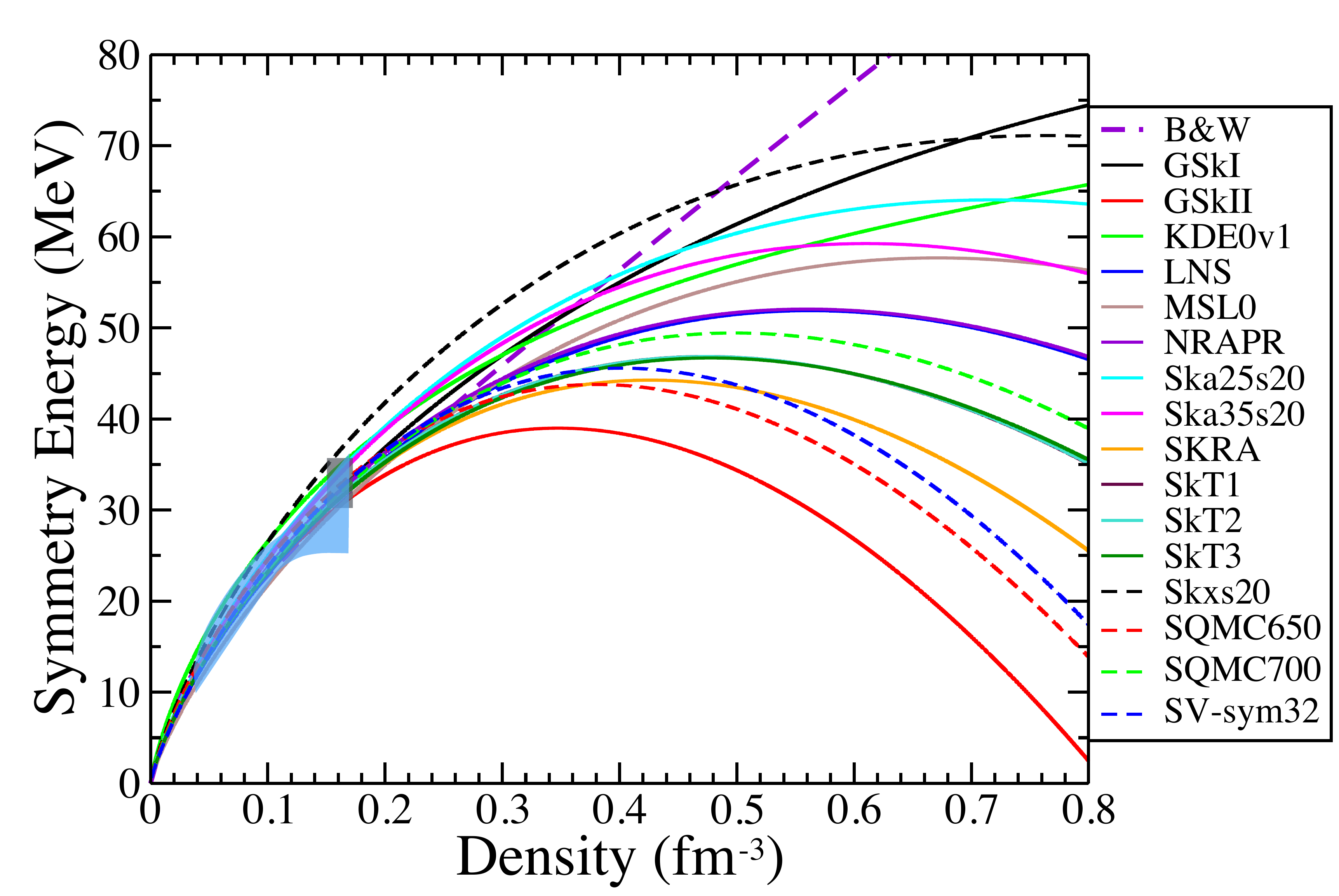}
\caption{Symmetry energy versus density for various Skyrme parameter sets listed in Table~\ref{tab:skyrmemodels}.  Existing constraints  at nuclear saturation density ($n_{0} = 0.16\pm 0.01 \text{fm}^{-3}$) \cite{stone2007} are indicated by the gray shaded region ($S_{0} = 30-35$ MeV) and constraints below saturation density \cite{danielewicz2013} are indicated by the blue shaded region.    The behavior of the symmetry energy above saturation density is unconstrained.  Many parameterizations become negative at supranuclear densities, which leads to erroneous behavior in neutron stars.  (Color available online.) }
\label{fig:sym_energy}
\end{figure}

The density dependence of the symmetry energy beyond saturation is highly uncertain. Fig.~\ref{fig:sym_energy} shows the symmetry energy versus density for various parameter sets found in Table~\ref{tab:skyrmemodels} and Ref.~\cite{Dutra2012}.  Note that, for these models, the non-standard terms $t_{4},~t_{5},~x_{4},$ and $x_{5}$ are all zero.   The symmetry energy at saturation is known~\cite{Lattimer2012,Dutra2012} to lie within the range \mbox{$S_0 = 30 - 35$ MeV} \cite{stone2007}. 
For all parameter sets considered here, this constraint is met.  However, for many Skyrme models the symmetry energy either saturates at high densities, or in the worst case becomes negative. This results in a negative pressure deep inside the neutron star core. For this work, we will only consider parameter sets with a fairly stiff symmetry energy. 

\section{Pions in the nuclear environment}

Pions are a crucial ingredient in simulations of core collapse supernovae.  The interior temperatures can reach to a significant fraction ($\sim 50 $ MeV) of the pion rest mass so that some thermally produced pions exist in the tail of the Boltzmann distrobution.  The existence of these thermally produced pions has the effect of softening the EoS.  This increases the central densities neutrino luminosities and thereby enhances the explosion.

The contribution to  the hadronic EoS from the lightest mesons (i.e. the pions) has been constrained~\cite{McAbee} from a comparison between relativistic heavy-ion collisions and one-fluid nuclear collisions.  In that work the formation and evolution of the pions was computed in the context of Landau-Migdal theory~\cite{Migdal} to determine the pion effective energy and momentum. In this approach the pion energy is given by a dispersion relation~\cite{Migdal}
\begin{equation}
\epsilon_\pi^2 = p_\pi^2 + \widetilde{m}_\pi^{2}    ~,
\label{eqn:dispersion}
\end{equation}
where $\widetilde{m}_\pi$ is the pion ``effective mass'' defined to be
\begin{equation}
\widetilde{m}_\pi = m_\pi\sqrt{1+\Pi\left(\epsilon_\pi,p_\pi,n_B\right)}   ~.
\label{eqn:pionmass}
\end{equation}

Following~\cite{Mayle} and~\cite{Friedman} the polarization parameter $\Pi$ can be written, 
\begin{equation}
\Pi\left(\epsilon_\pi,p_\pi,n_B\right) = \frac{p_\pi^2 \Lambda^2\left(p_\pi\right)\chi \left(\epsilon_\pi,p_\pi,n_B\right)}{m_\pi^2-g'm_\pi^2\Lambda^2\left(p_\pi\right)\chi\left(\epsilon_\pi,p_\pi,n_B\right)}   ~.
\label{eqn:propagator}
\end{equation}
where the denominator is the Ericson-Ericson-Lorentz-Lorenz correction~\cite{Ericson}. The quantity \mbox{$\Lambda \equiv \text{exp}(-p_\pi^2/b^2)$} with \mbox{$b = 7m_\pi$}, is a cutoff that ensures that the dispersion relation~[Eq.~(\ref{eqn:dispersion})] asymptotically approaches the high momentum limit, 
\begin{equation}
\epsilon_{\infty}\equiv\epsilon\left(p_\pi\rightarrow\infty,n_B\right) = \sqrt{m_\Delta^2+p_\pi^2}-m_N ~.
\end{equation}

Following~\cite{Ericson} we take the polarizability to be
\begin{equation}
\chi\left(\epsilon_\pi,p_\pi,n_B\right) = -\frac{4a\epsilon_{\infty}n_B}{\epsilon_\infty^2-\epsilon_\pi^2} ~,
\end{equation}
where \mbox{$a = 1.13/m_\pi^2$}.  This form for the polarizability ensures that the effective pion mass is always less than or equal to the vacuum rest mass $m_\pi$.

A key quantity in the above expressions is the Landau parameter $g'$. This is an effective nucleon-nucleon coupling strength.  To ensure consistency with observed Gamow-Teller transition energies a constant value of \mbox{$g' = 0.6$} was deduced in~\cite{Friedman}. However, in~\cite{McAbee}, Monte-Carlo techniques were used to statistically average a momentum dependent $g'$ with particle distribution functions. It was found that $g'$ varies linearly with density and is approximately given by
\begin{equation}
	g' = g_1 + g_2\eta   ~,
\end{equation}
 where $\eta \equiv n_B/n_0$.   A value of \mbox{$g_1 = 0.5$} was chosen to be consistent with known Gamow-Teller transitions. A value for $g_2$ was then obtained~\cite{McAbee} by optimizing fits to a range of pion multiplicity measurements obtained at the Bevlac~\cite{Harris}. These data were best fit for a value of \mbox{$g_2 = 0.06$}. 
 
 The pions are assumed to be in chemical equilibrium with the surrounding nuclear matter. We consider the pion-nucleon reactions:
\begin{equation}
p\leftrightarrow n + \pi^+  ~, \quad   n\leftrightarrow p + \pi^-     ~.
\end{equation}
This leads to the following relations among the chemical potentials for neutrons, protons, and pions
\begin{equation}
\mu_p = \mu_n + \mu_{\pi^+}  ~, \quad   \mu_n = \mu_p + \mu_{\pi^-}  ~.
\end{equation}
These equilibrium conditions let us express the pion chemical potentials in terms of the neutron and proton chemical potentials: \mbox{$\hat{\mu} \equiv \mu_n - \mu_p$} $= \mu_{\pi^-} = -\mu_{\pi^+}$.
Using the definitions of  $\mu_n$ and $\mu_p$ from \mbox{Eqs.~(\ref{eqn:mun}) - (\ref{eqn:mup})},  
the expressions for the pion chemical potentials are found to be
\begin{equation}
\label{eqn:mupi}
\mu_{\pi^-} = -\mu_{\pi^+} = -\frac{1}{n_B} \frac{\partial F}{\partial Y_p} ~.
\end{equation}

For a given temperature ($T$) and number density ($n_B$) the pion number densities are given by the standard Bose-Einstein integrals
\begin{equation}
\label{eqn:pin}
n_i = \int_0^\infty\frac{p^2}{2\pi^2}\frac{dp}{e^{(\epsilon_\pi - \mu_i)/T}-1}   ~,
\end{equation}
where $i$ is for $\{ \pi^+, \pi^-, \pi^0 \}$, and $\epsilon_\pi$ is given by Eq.~(\ref{eqn:dispersion}). Note that the $\pi^0$ chemical potential is taken to be zero, since these particles can be created or destroyed without charge constraint.  

The charge fraction per baryon for the charged pions is defined as \mbox{$Y_{\pi^-} = n_{\pi^-}/n_B$}.  From Eq.~(\ref{eqn:mupi})  we can calculate the pion number densities from the pion chemical potentials. Then, electric charge conservation gives,
\begin{equation}
\label{eqn:Yp}
Y_e = Y_p  - Y_{\pi^-} + Y_{\pi^+}   ~.
\end{equation}
Thus, we can solve Eq.~(\ref{eqn:Yp}) for the unknown quantity $Y_p$.

Once $Y_p$ is determined, the pionic energy densities and partial pressures can be calculated from
\begin{equation}
E_i = \int^\infty_0\frac{p^2 dp}{2\pi^2}\frac{\epsilon_\pi}{\text{exp}\left[\left(\epsilon_\pi-\mu_i\right)/T\right]-1}  ~,
\end{equation}
and
\begin{equation}
P_i = \int^\infty_0\frac{p^2 dp}{2\pi^2}\frac{(1/3)p(\partial{\epsilon_\pi}/\partial{p})}{\text{exp}\left[\left(\epsilon_\pi-\mu_i\right)/T\right]-1}   ~.
\end{equation}  
In the high temperature, low density limit the pionic mass approaches the bare pion mass and other pionic excitations are become relevant. Hence, to properly treat other density and temperature regimes, we also include all mesonic and baryonic states using bare masses as discussed in Sec.~\ref{sec:therm}.

We note that this treatment of the pions is preferable to including a condensate, as was done in Refs.~\cite{mueller1997} and \cite{cavagnoli2011}.  In Refs.~\cite{mueller1997} and \cite{cavagnoli2011}, pions were treated as a non-interacting Bose condensate.  In our treatment, pion-nucleon couplings are accounted for.

\section{QCD Phase Transition}
It is generally expected~\cite{McLerran} that for sufficiently high densities and/or temperature, a transition from hadronic matter to quark-gluon plasma (QGP) can occur.  Recent progress~\cite{Kronfeld} in lattice gauge theory (LGT) has shed  light on the transition to a  QGP in the low baryon-chemical-potential, high-temperature limit. It is now believed that at high temperature and low density a deconfinement and chiral symmetry restoration occur simultaneously at the crossover boundary. In particular, at low density and high temperature, it has been found~\cite{Kronfeld} that the order parameters for deconfinement and chiral symmetry restoration changes abruptly for temperatures of \mbox{$T = 145 - 170$ MeV}~\cite{Borsanyi, Bazavov1}. However, neither order parameter exhibits the characteristic change expected from a first order phase transition. An analysis of many~\cite{Aoki,Bazavov2} thermodynamic observables confirms that the transition from a hadron phase to a high temperature QGP is a smooth crossover. 

However, at low density the hadron phase can be approximated as a pion-nucleon gas, while the QGP phase can be approximated as a non-interacting relativistic gas of quarks and gluons~\cite{Fuller}. Equating the pressures in the hadronic and QGP phases, the critical temperature $T_c$ for the low density transition can be approximated~\cite{Fuller} as:
\begin{equation}
T_c \approx \left(g_q - g_h\right)^{-1/4}\left(\frac{90}{\pi^2}\right)^{1/4}B^{1/4}  ~,
\label{eqn:Tc}
\end{equation}
where the statistical weight $g_q$ for a low-density high-temperature QGP gas with three relativistic quarks is \mbox{$g_q \approx 51.25$}, while \mbox{$g_h \approx 17.25$} was found for the hadronic phase by summing over all known meson data. 

Adopting the lattice gauge theory results~\cite{Kronfeld} that  \mbox{$145 \lesssim T_c \lesssim 170$ MeV}, then implies~\cite{Fuller} that a reasonable range for the QCD vacuum energy is \mbox{$165 \lesssim B^{1/4} \lesssim 240$ MeV}. 
This provides an initial range for the QCD vacuum energy to be adopted in this work.  In Section~\ref{sec:bag}, we will further constrain this parameter by requiring that the maximum mass of a neutron star exceed \mbox{$2 ~M_\odot$}~\cite{Demorest,Antoniadis}.

Another parameter that impacts the thermodynamic properties of the system is the strong coupling constant $\alpha_s$.  For this manuscript we adopt a value of $\alpha_s = 0.33$ as this is a representative value for the energy regime under consideration~\cite{Beringer2012}.

A transition to a QGP phase during the collapse can have a significant impact on the dynamics and evolution of the nascent proto-neutron star. In Ref.~\cite{Gentile} it was shown that a first order phase transition to a deconfined QGP phase resulted in the formation of two distinct but quickly coalescing shock waves.  More recently, it has been shown~\cite{Fischer2010} that if the transition is first order, and global conservation laws are imposed, then the two shock waves can be time separated by as much \mbox{as $\sim 150$ ms}. Neutrino light curves showing such temporally separated spikes might even be resolvable in modern terrestrial neutrino detectors~\cite{Fischer2010}. Moreover, the arrival of the second shock can significantly enhance the explosion.

The observation of \mbox{$ M > 2 M_\odot$} neutron stars  \cite{Demorest,Antoniadis}, however, highly restricts the possibility of a first order phase transition to a quark gluon plasma taking place inside the interiors of stable cold neutron stars. Nevertheless, a transition to quark-gluon plasma would always occur for initial stellar masses \mbox{ $\gtrsim 20~M_\odot$} that result in failed supernova events, because every phase of matter must be traversed during the formation of stellar mass black holes.  Even if neutron stars do not have pure- or mixed- quark-gluon plasma interiors, this transition to QGP may have an impact~\cite{Nakazato} on the neutrino signals during black hole formation in addition to  its possible impact on core-collapse supernovae.

\subsection{The Quark Model}
For the description of quark-gluon plasma we use
a bag model with 2-loop corrections \cite{McLerran}, and construct the EoS from a phase-space integral representation over scattering amplitudes.  We allow for the possibility of a coexistence mixed phase in a first order transition, or a simple direct crossover transition. In the hadronic phase the thermodynamic state variables, are calculated from the Helmholtz free energy $F(T,V, n_B)$ as described in the previous sections.  However, it is convenient to compute the QGP phase in terms of the grand potential, $\Omega(T,V,\mu)$.  Both descriptions are equivalent and are related  by a Legendre transform: 
	$\Omega = F - \sum_i \mu_i N_i$.  Adopting the convention of  Landau and Lifshitz \cite{Landau69},  the thermodynamic potential can be written   in terms of the partition function $Z$ as 
\begin{equation}
\Omega =  -\frac{1}{\beta} \ln{Z} ~~,
\end{equation}
where $\beta = 1/T$.  In the Feynman path integral formulation the partition function is represented as a functional integral of the exponential of an effective action integrated over all fields \cite{Kapusta}.  

Once the grand potential is specified, the state variables can be easily deduced, e.g.
\begin{equation}
P =  -\biggl(\frac{\partial \Omega}{\partial V}\biggr)_{T, \mu} ~~,
\end{equation}
\begin{equation}
n =  -\frac{1}{V} \biggl(\frac{\partial \Omega}{\partial \mu}\biggr)_{V,T} ~~.
\end{equation}

The grand potential for  the quark-gluon plasma summed over  each flavor $i$ takes the form:
\begin{equation}
	\Omega = \sum_i(\Omega_{q0}^i + \Omega_{q2}^i) +  \Omega_{g0} + \Omega_{g2} + B V
\label{eq:GrandPotential}
\end{equation}
where $q_0$ and $g_0$ denote the $0^\text{th}$-order bag model grand potentials for quarks and gluons, respectively, while $q_2$ and $g_2$ denote the 2-loop corrections. The fifth term $BV$ is the usual QCD vacuum energy.  
In most calculations, sufficient accuracy is obtained by using fixed current algebra masses (e.g. \mbox{$m_u \sim m_d \sim 0$}, \mbox{$m_s \sim 95 \pm 5$ MeV}). For this work we choose  a bag constant $B^{1/4} = 165 - 240$ MeV as noted above.  

The quark contributions to the grand potential are  given~\cite{McLerran} from phase-space integrals over Feynman amplitudes~\cite{Kapusta}:
\begin{widetext}
	\begin{align}
		\label{eqn:idealgas} 
		\Omega_{q0}^{i} =& -2N_cT V\int_0^{\infty} \frac{d^3p}{(2 \pi)^3} 
			\left[\text{ln}\left(1+e^{-\beta\left(E_i-\mu_i\right)}\right) 
				+ \text{ln}\left(1+e^{-\beta\left(E_i+\mu_i\right)}\right)\right]    \\
		\label{eqn:twoloop}
		\Omega_{q2}^i =& \alpha_s \pi N_g V\Bigg[
			\frac{1}{3} T^2 \int_0^\infty \frac{d^3p}{(2 \pi)^3}\frac{N_i(p)}{E_i(p)}
				+ \int_0^\infty \frac{d^3p}{(2 \pi)^3} \frac{d^3p'}{(2 \pi)^3}
					\frac{1}{E_i(p)E_i(p')}\left[N_i(p)N_i(p')+2\right] 	\nonumber \\
			  &\qquad \qquad \quad \times \left[
			  \frac{N_i^+(p)N_i^+(p')+N_i^-(p)N_i^-(p')}{\left(E_i(p)-E_i(p')\right)^2-\left({\bf p-p'}\right)^2} 
			  + \frac{N_i^+(p)N_i^-(p')+N_i^-(p)N_i^+(p')}{\left(E_i(p)+E_i(p')\right)^2-\left({\bf p-p'}
			  \right)^2}\right]\Bigg] ,
\end{align}
\end{widetext}
where $N_c$ is the number of colors, and $N_g$ is the number of gluons ($N_g = 8$).   The $N_i^{\pm}$ denote the quark Fermi-Dirac distributions:
\begin{equation}
N_i^{\pm}(p) = \frac{1}{e^{\beta\left(E_i(p)\mp\mu_i\right)} + 1}   ~.
\end{equation}

The one- and two-loop gluon and ghost contributions to the thermodynamic potentials can be evaluated in a similar fashion to that of the quarks.
\begin{align}
\Omega_{g0} =& 2 N_g T V\int_0^\infty\frac{d^3p}{(2 \pi)^3}\text{ln}\left(1-e^{-\beta\lvert p \rvert}\right) \nonumber \\
                         =& -\frac{\pi^2}{45}N_gT^4   ~. 
\end{align}
\begin{equation}
\Omega_{g2} = \frac{\pi}{36}\alpha_sN_c N_gT^4.
\end{equation}

For massless quarks, \mbox{Eqs.~(\ref{eqn:idealgas}-\ref{eqn:twoloop})} are easily evaluated \cite{McLerran} to give
\begin{align}
\label{q20}
\Omega_{q0}^i =& -\frac{N_c V}{6} \left(\frac{7\pi^2}{30}T^4 + \mu_i^2T^2 +\frac{\mu_i^4}{2\pi^2}\right) \\
\Omega_{q2}^i =& \frac{N_g \alpha_s V}{8\pi}  \left(\frac{5\pi^2}{18}T^4 + \mu_i^2T^2 +\frac{\mu_i^4}{2\pi^2}\right)  .
\end{align}
For the massive strange quark Eq.~(\ref{eqn:idealgas}) can be easily integrated. However, Eq.~(\ref{eqn:twoloop}) cannot be integrated numerically, due to the divergences inherent in the integral. We therefore, approximate~\cite{McLerran} the two-loop strange quark contribution with the zero mass limit. This may over estimate the contribution due to a finite strong coupling constant.  However,  sincethe quark mass is relatively small compared to its chemical potential, this is a reasonable approximation.

\subsection{Conservation Constraints in the Mixed Phase}

The matter deep inside the core of a collapsing star consists of a multi-component system constrained by the conditions of both charge and baryon number conservation.   
All thermodynamic quantities in a first order quark-hadron phase transition vary in proportion to the volume fraction~[\mbox{$\chi \equiv V^Q/(V^Q+V^H)$}] throughout the mixed phase regime, where $V^{Q}$ is the volume of material composed of quark-gluon plasma and $V^{H}$ is the volume composed of hadronic matter.

For the description of a first order phase transition we utilize a Gibbs construction. In this case the two phases are in equilibrium when the chemical potentials, temperatures, and pressures are equal. For the description of the phase transition from hadrons to quarks, this construction can be written
\begin{eqnarray}
\mu_p &=& 2\mu_u + \mu_d \\
\mu_n &=& 2\mu_d + \mu_u \\
\mu_d &=& \mu_s \\
T_H &=& T_Q   \\
P^H\left(T,Y_e, \{\mu_i^H\}\right) &=& P^Q\left(T,Y_e,\{\mu_i^Q\}\right)   ,
\end{eqnarray}
where \{$\mu_i^H$\} $\equiv$ \{$\mu_n, \mu_p,\mu_{\pi}, \mu_e,\mu_\nu$\} and \{$\mu_i^Q$\} $\equiv$ \{$\mu_u,\mu_d,\mu_s,\mu_e,\mu_\nu$\}.  

The Gibbs construction ensures that a uniform background of photons and leptons exists within the differing phases. Therefore, the contribution from the photon, neutrino, electron, and other lepton pressures cancel out in phase equilibrium. Also, with this choicethe two conserved quantities, charge and baryon number, vary linearly in proportion to the degree of completion of the phase transition (or volume fraction $\chi$), i.e.,
\begin{align}
n_B Y_e =&  \left(1-\chi\right)n_B^HY_c^H + \chi n_B^Q Y_c^Q \\
n_B =& \left(1-\chi\right)n_B^H + \chi n_B^Q ,
\end{align}
where we have defined \mbox{$Y_c^H = Y_p + Y_{\pi^+} - Y_{\pi^-}$} and \mbox{$n_B^Q Y_c^Q = 1/3~(2n_u - n_d - n_s)$}.  The internal energy and entropy densities likewise vary in proportion to the degree of phase transition completion,
\begin{align}
\epsilon =& \left(1-\chi\right)\epsilon^H + \chi \epsilon^Q \\
             s =&  \left(1-\chi\right)s^H + \chi s^Q    .
\end{align}

\section{Results and Comparisons}
In this section we compare properties of the new NDL~EoS with the two most commonly employed equations of state used in astrophysical collapse simulations as well as the original EoS of Bowers and Wilson.
\begin{figure}[h]
  \centering
  \includegraphics[width=0.5\textwidth]{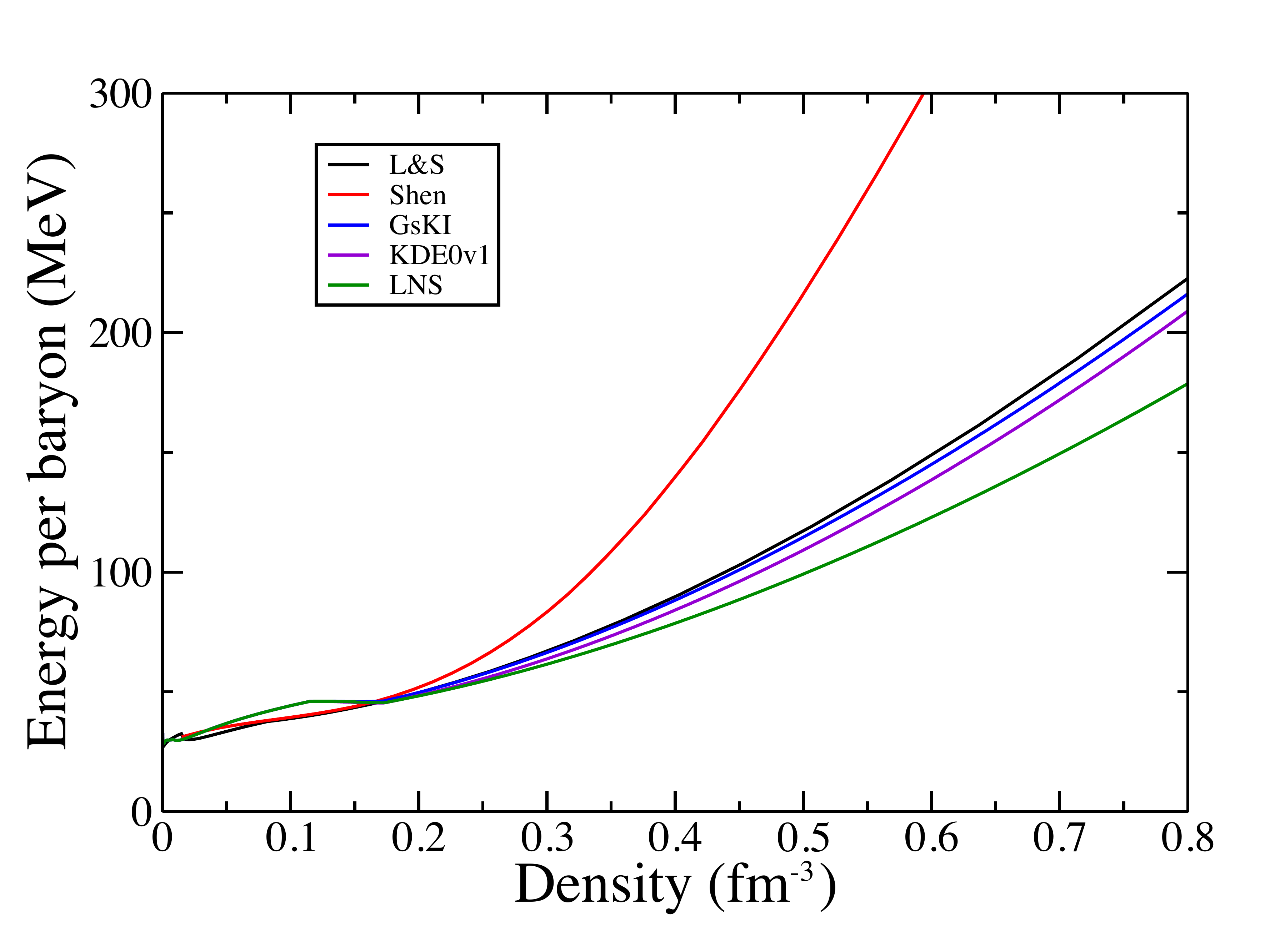}
  \caption{The energy per particle as a function of density comparing the Lattimer~and~Swesty EoS (with compressibility of $K_0$ = 220 MeV), the Shen EoS, and the NDL~EoS with several Skyrme parametrizations from Table~\ref{tab:skyrmemodels}.  (Color available online)}
 \label{fig:had_energy}
\end{figure}

Fig.~\ref{fig:had_energy} shows the total internal energy as a function of local proper baryon density. We compare the Lattimer and Swesty EoS~\cite{LS91} with \mbox{$K_0 = 220$ MeV}, the Shen EoS~\cite{Shen98a,Shen98b}, and the NDL~EoS with several choices of Skyrme parameterization as labeled from Ref.~\cite{Dutra2012}. 
These plots correspond to a fixed electron fraction and temperature of $Y_e = 0.3$ and $T = 10$~MeV.  A steep rise in the energy per baryon at high densities occurs for larger values of the compressibility as expected.

\begin{figure}[h]
  \centering
  \includegraphics[width=0.5\textwidth]{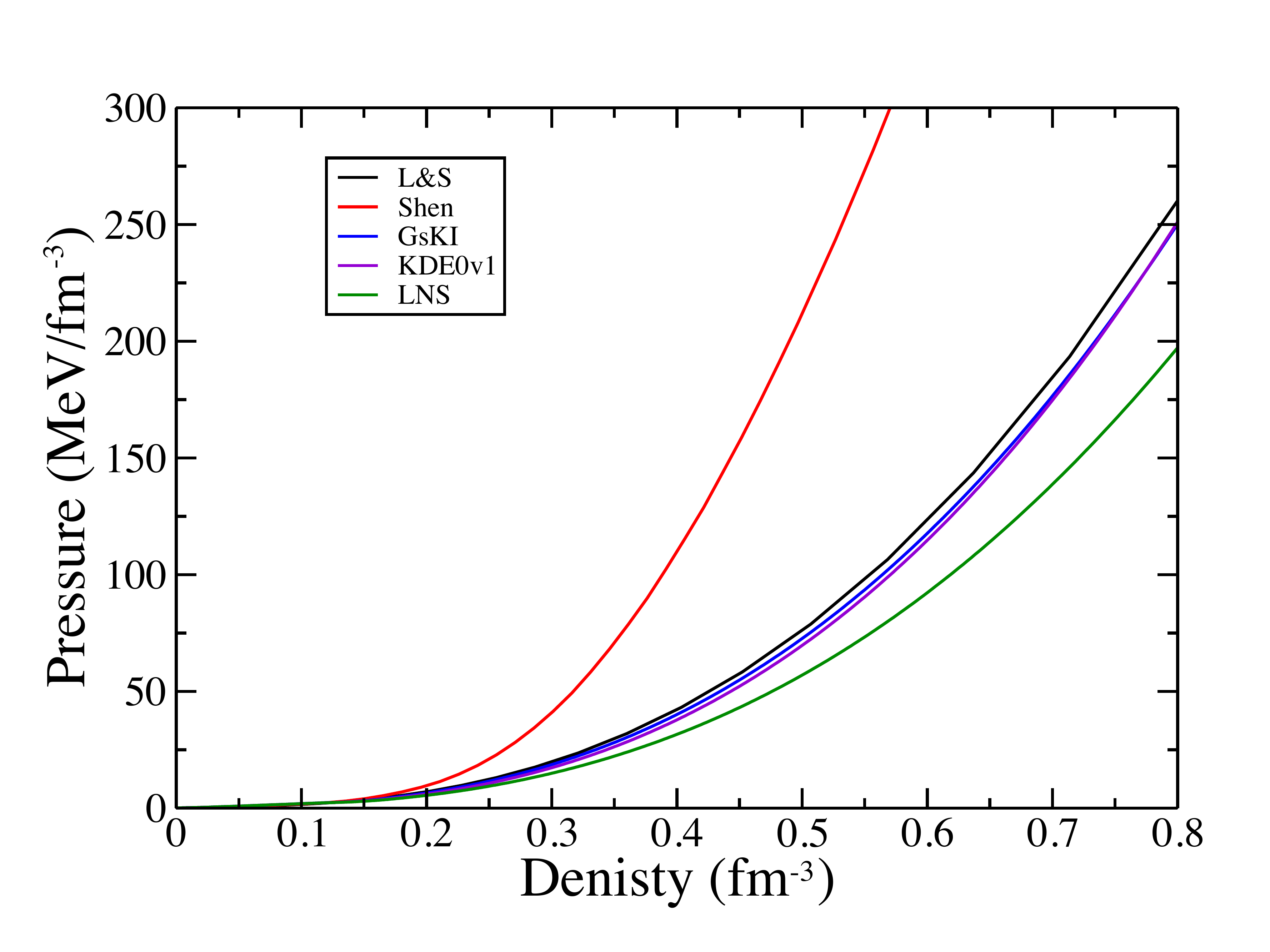}
  \caption{Pressure as a function of density comparing the Lattimer~and~Swesty EoS (with compressibility of $K_0$ = 220 MeV), the Shen EoS, and the NDL~EoS with several Skyrme parameterizations as labeled from Table~\ref{tab:skyrmemodels}. (Color available online) }
    \label{fig:had_pressure}
\end{figure}

Similarly, Fig.~\ref{fig:had_pressure} depicts the pressure vs. density for the Shen EoS~\cite{Shen2011}, the Lattimer~and~Swesty EoS~\cite{LS91}, and several Skyrme parameterizations of the NDL EoS. The Shen EoS consistently leads to higher pressure.  This results from the use of the TM1 parameter set \cite{Shen98a,Shen98b} that contains relatively high values for both the symmetry energy at saturation and the nuclear compressibility, typical of RMF approaches.

\subsection{Pion Effects on the EoS}
\begin{figure}[h]
  \centering
  \includegraphics[width=0.5\textwidth]{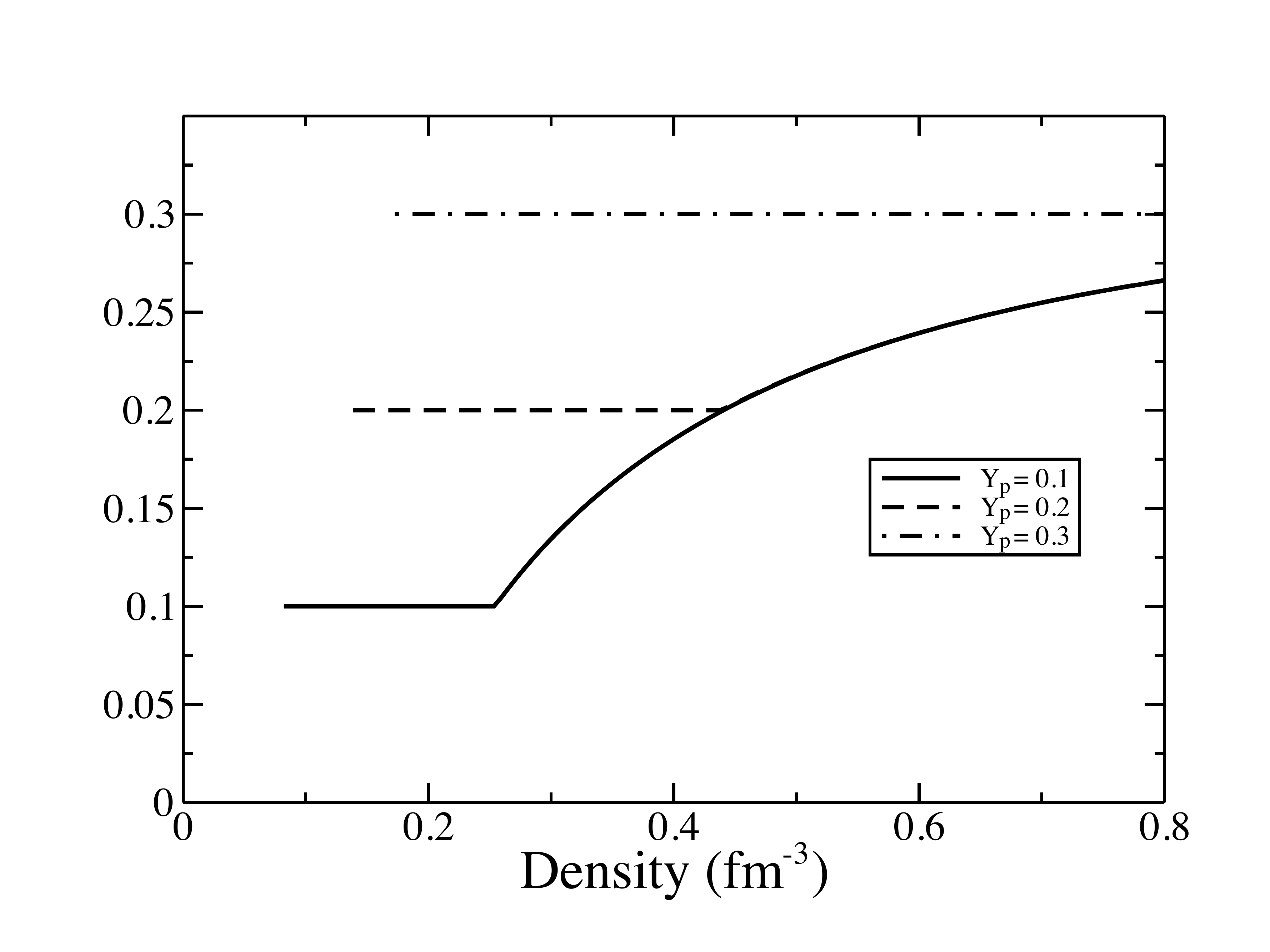}
  \caption{The proton fraction above nuclear saturation showing the effects of pions in a hot dense astrophysical environment with $T = 10$~MeV, such as occurs in core collapse supernovae. For small electron fractions more pions are created due to the dependence of the chemical potential on the isospin asymmetry parameter~$I$.}
 \label{fig:pion_prot_frac}
\end{figure}

The solution to the pion dispersion relation, Eq.~(\ref{eqn:dispersion}), does not lead to  conditions within the proto-neutron star necessary to generate a pion condensate. The pions considered here are thermal pionic excitations calculated from the pion propagator in Eq.~(\ref{eqn:propagator}). In very hot and dense nuclear matter the number density of pionic excitations is greatly enhanced by the $\pi N \Delta$ coupling~\cite{Friedman}. Hence, it becomes energetically favorable to form pions in the nuclear fluid when the chemical potential balance shifts from electrons to negative pions. This allows the charge states to equilibrate with these newly formed pions.  Since the pions are assumed to be in chemical equilibrium with the surrounding nuclear fluid, this has a profound effect on the proton fraction within the medium, particularly for low electron fractions (see Fig.~\ref{fig:pion_prot_frac}). 

\begin{figure}[h]
  \centering 
  \includegraphics[width=0.5\textwidth]{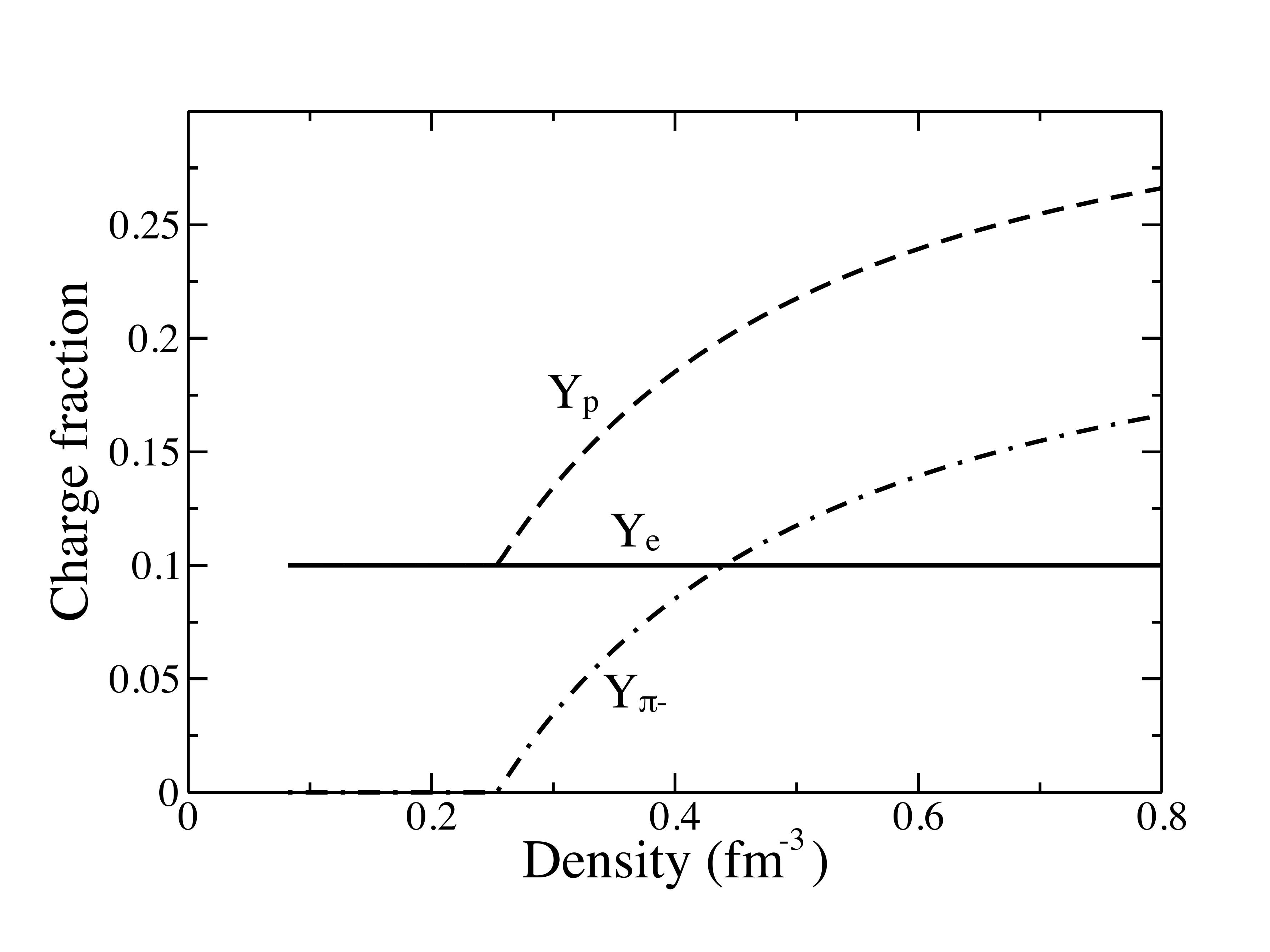}
  \caption{Pion charge fraction versus density at a temperature $T = 10$~MeV using the GsKI Skyrme parameter set. At a density of about $n_B \approx 0.45 \text{ fm}^{-3}$, the pion charge fraction exceeds the electron fraction of the medium. }
 \label{fig:pion_charge_frac}
\end{figure}

The pion charge fraction as a function of baryon number density is shown in Fig.~\ref{fig:pion_charge_frac}. For a low fixed $Y_e$, the charge fraction of negative pions can actually become greater than the electron fraction, and the negative pions essentially replace the electrons in equilibrating the charge.

From the solution to the pion chemical potentials [Eq.~(\ref{eqn:mupi})], one can see that as the density increases, negative pions are created.  This is  due to the chemical potential constraints and the dispersion relation [Eq.~(\ref{eqn:dispersion})]. At the same time the number density of positively charged pions remains negligible due to the fact that they have a negative chemical potential.  The pion chemical potential [Eq.~(\ref{eqn:mupi})] increases linearly with respect to density but decreases  linearly with respect to $Y_p$, due to its dependence on the slope of the free energy with respect to $Y_{p}$. 
Therefore, for high electron fractions the pion chemical potential will remain small and charge equilibrium can be maintained solely among the electrons and protons.  

It should be noted, however, that as treated here, pions would not exist in the ground state configuration of a cold neutron star. As the temperature approaches zero the pionic effects diminish, until only the nucleon EoS contributes to the neutron star structure.

\begin{figure}[h]
  \centering
  \includegraphics[width=0.5\textwidth]{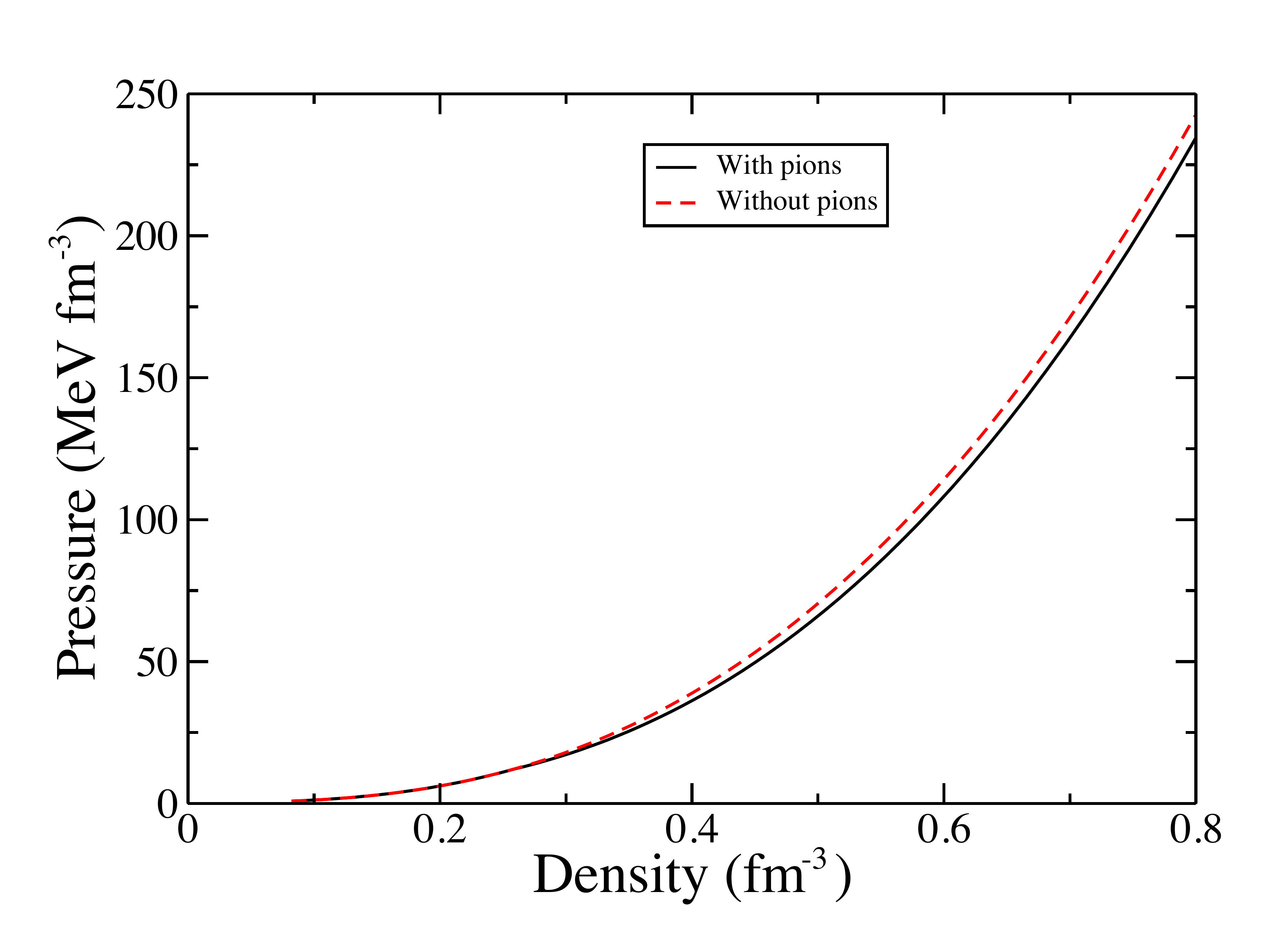}
  \caption{Pressure versus baryon number density at $T = 10$~MeV showing a few percent reduction in the pressure at high densities due to the presence of pions. This figure was made with the GsKI parameter set. (Color available online)}
 \label{fig:pion_pressure}
\end{figure}

In the hot dense medium of supernovae, however, these pions tend to soften the hadronic EoS since they relieve some of the degeneracy pressure due to the electrons.  We have found that the reduction in pressure is relatively insensitive to the temperature of the medium and is lowered by a few percent for all representative temperatures found in the supernova environment, as shown in Fig.~\ref{fig:pion_pressure}.  

This will affect SN core collapse models since it allows collapse to higher densities and temperatures without violating the requirement that the maximum neutron star mass exceed \mbox{$2.01 \pm 0.04 ~M_\odot$} for cold neutron stars \cite{Antoniadis}.

\subsection{Hadron-QGP Mixed Phase}
The constraint of global charge neutrality exploits the isospin restoring force experienced by the confined hadronic matter phase.  The hadronic portion of the mixed phase becomes more isospin symmetric than the pure hadronic phase because charge is transferred from the quark phase to the hadron phase in equilibrium. 

\begin{figure}[t!]
  \centering
  \includegraphics[width=0.5\textwidth]{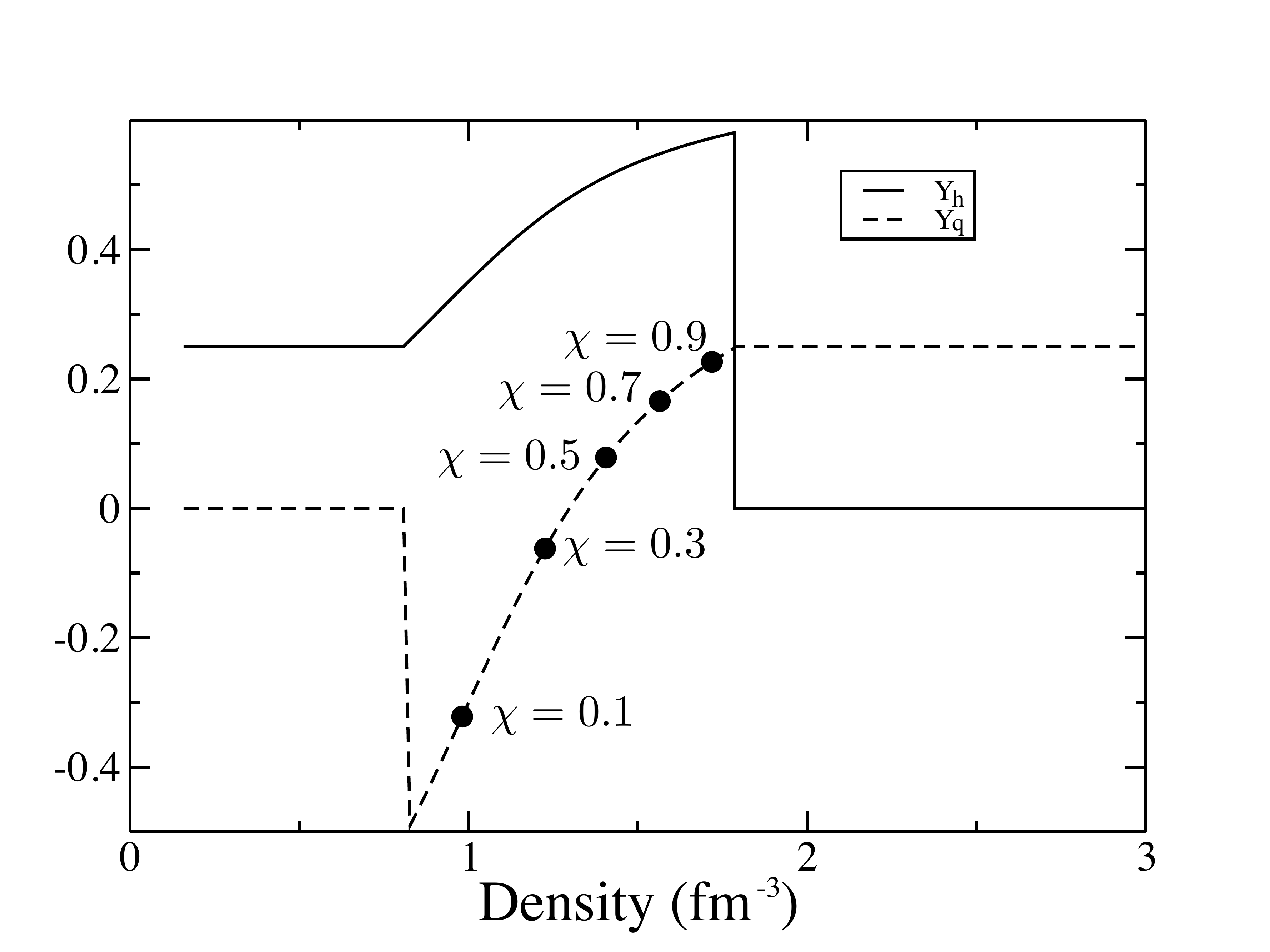}
  \caption{Charge fractions of the mixed quark and hadronic phase. Due to the redistribution of charge, the hadronic phase becomes isospin symmetric even exceeding $Y_p >$ 0.5. This has the effect of lowering the symmetry energy and thus reducing the pressure in the hadronic phase. This figure was made using the GSkI Skyrme parameterization at $T = 1$ MeV and a bag constant of $B^{1/4} = 190$ MeV.}
 \label{fig:isospin}
\end{figure}

Fig.~\ref{fig:isospin} shows the charge fractions of the mixed phase and hadronic phases for the GsKI Skyrme parameterization at $T =1$ MeV and a bag constant of $B^{1/4} = 190$ MeV. From this we see that the internal mixed phase region of a hot, proto-neutron star contains a positively charged region of nuclear matter and a negatively charged region of quark-gluon plasma until a density of \mbox{$n_0 \sim 1.3~\text{fm}^{-3}$}. The presence of the isospin restoring force causes the thermal pionic contribution to the state variables to be negligible. This is due to the dependence of the pion chemical potential on the isospin asymmetry parameter $(1-2Y_p)$.  As the hadronic phase becomes more isospin symmetric, the pion chemical potential remains small compared to its effective mass. 

Since the system contains two conserved quantities, electric charge and baryon number, the coexistence region cannot be treated as a single substance, but must be evolved as a complex multi-component fluid.  It is common in Nature to have global conservation laws and not necessarily locally conserved quantities. Hence, within the Gibbs construction the pressure is a monotonically increasing function of density.  

\begin{figure}[h]
  \centering
  \includegraphics[width=0.5\textwidth]{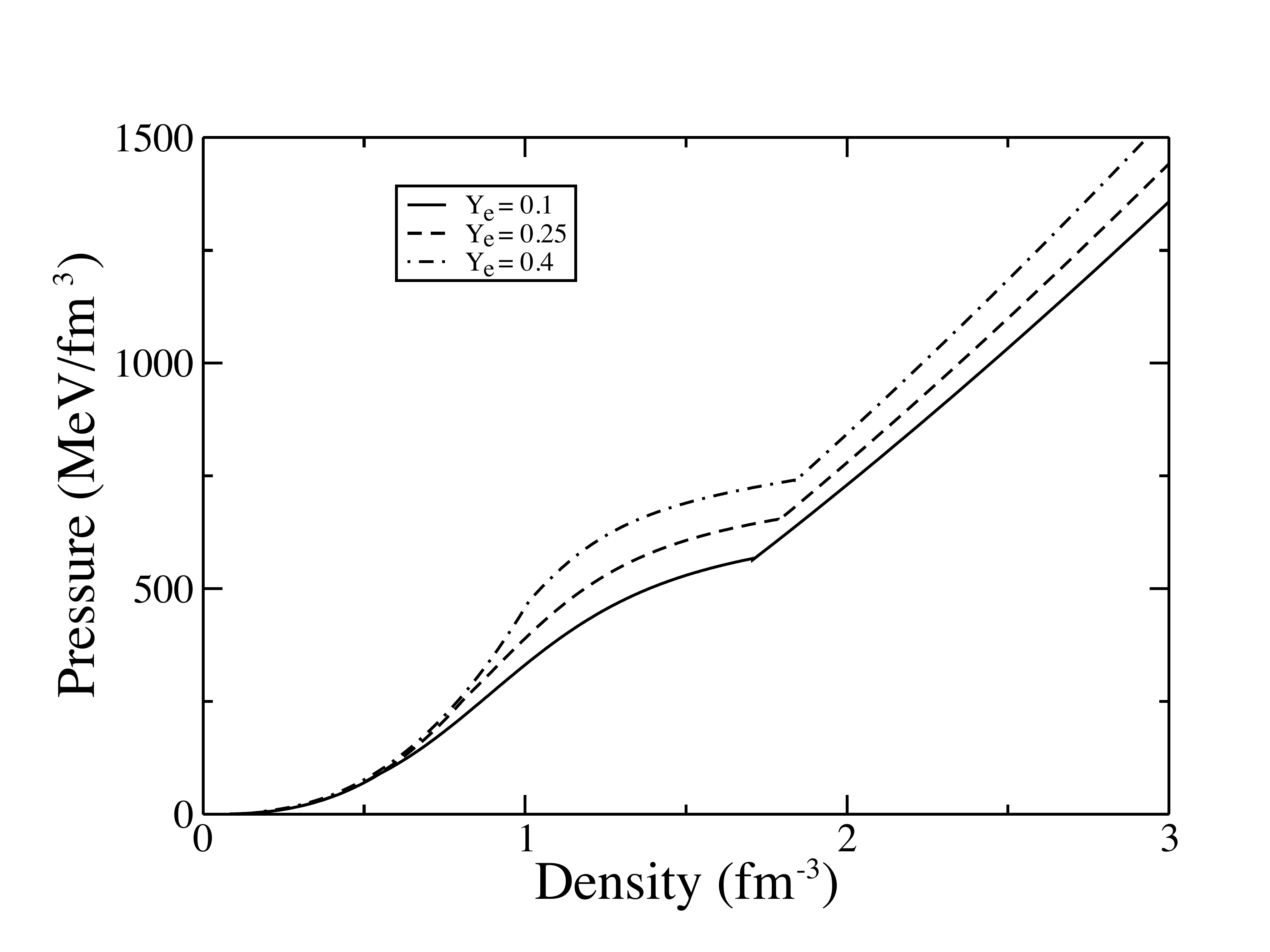}
  \caption{Pressure as a function of baryon number density through the mixed phase transition for $T = 1$~MeV and the GSkI Skyrme parameterization and a bag constant of $B^{1/4} = 190$ MeV. The EoS softens significantly upon entering the mixed phase at $n_B \sim 0.6 \text{ fm}^{-3}$ due to the larger number of available degrees of freedom.}
 \label{fig:mixed_pressure}
\end{figure}

Fig.~\ref{fig:mixed_pressure} shows pressure versus density for various values of $Y_e$ and the GsKI Skyrme parameterization, through the mixed phase region into a phase of pure QGP.  One of the features shown is that, as the density increases to about 2-3 times the saturation density the slope of the pressure versus density decreases as one enters the mixed phase.  
Note that in a simple crossover there is no mixed phase so that the onset of the QCD phase causes an immediate jump toward the  asymptotic behavior shown on Fig.~\ref{fig:mixed_pressure}.  

However, in a simple crossover transition the EoS at first becomes stiff then asymptotes to a simple $\Gamma \approx  4/3$ EoS.  The reason for this straightforward to explain.
In the low temperature limit for $N_f$ flavors of massless quarks, Eq.~(\ref{q20}) leads to \cite{Gentile} the following approximate expressions for  the  chemical potential, baryon density and pressure  
\begin{equation}
\mu = 3\biggl(\frac{3 \pi^2}{N_f}\biggr)^{1/3} \biggl[ 1 + \frac{2 \alpha_s}{3 \pi} \biggr] n^{1/3} + B
\end{equation}
\begin{equation}
n = \frac{9}{4} \biggl(\frac{3 \pi^2}{N_f}\biggr)^{1/3} \biggl[ 1 + \frac{2 \alpha_s}{3 \pi} \biggr] n^{4/3} + B
\end{equation}
\begin{equation}
P = \frac{3}{4} \biggl(\frac{3 \pi^2}{N_f}\biggr)^{1/3} \biggl[ 1 + \frac{2 \alpha_s}{3 \pi} \biggr] n^{1/3} - B
\end{equation}
The adiabatic index is then
\begin{equation}
\Gamma = \frac{n}{P} \biggl( \frac{\partial P}{\partial n} \biggr)= \frac{4}{3} \frac{P + B}{P} ~~.
\end{equation}
From this one can see that immediately after a crossover transition when the bag pressure is comparable to the baryonic pressure the adiabatic index is stiff $\sim 2$.  
It then asymptotes toward $\Gamma = 4/3$ as the pressure  increases.

\begin{figure}[h]
  \centering
  \includegraphics[width=0.5\textwidth]{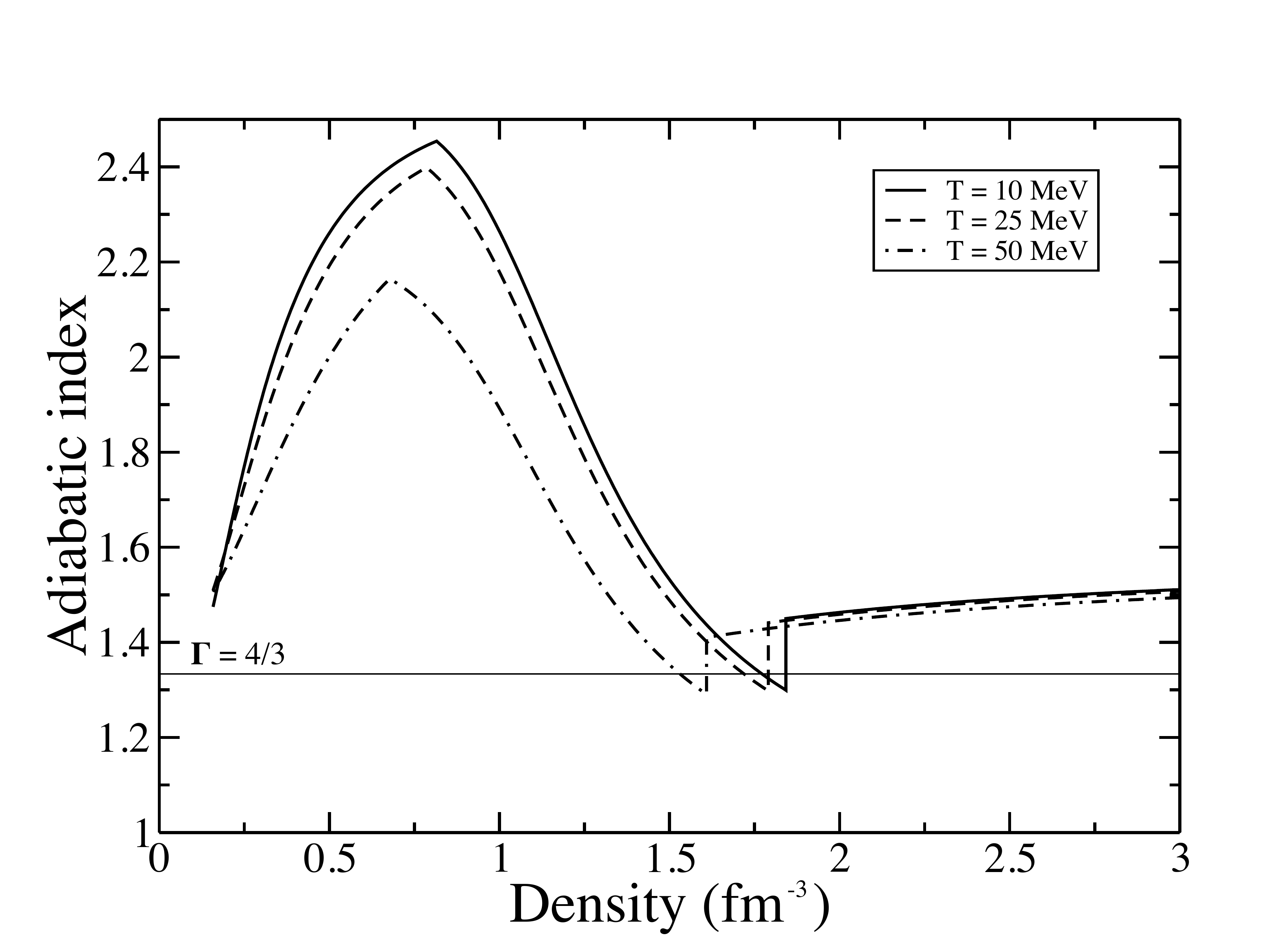}
  \caption{$\Gamma$ as a function of baryon number density showing the softening of the EoS as it enters the mixed phase regime. The EoS promptly stiffens as it exits the mixed phase into the pure quark-gluon plasma phase due to losing the extra degrees of freedom supplied by the hadrons.  The horizontal line indicates	 $\Gamma = 4/3$.  Systems with $\Gamma > 4/3$ will be stable. This figure was made using the GsKI Skyrme parameterization and a bag constant of $B^{1/4} = 190$ MeV.}
 \label{fig:gamma}
\end{figure}

The effect on the adiabatic index for a mixed phase is shown in Fig.~\ref{fig:gamma}. Here, one can see that for for a mixed phase the EoS softens abruptly upon entering the mixed phase due to the fact that increasing density leads to a larger volume fraction of QGP rather than an increase in pressure. If this occurs while forming a proto-neutron star, its evolution will be affected as $\Gamma$ falls below the stability point of $\Gamma < 4/3$ and a second collapse can ensue.  

\begin{figure}[h]
  \centering
  \includegraphics[width=0.5\textwidth]{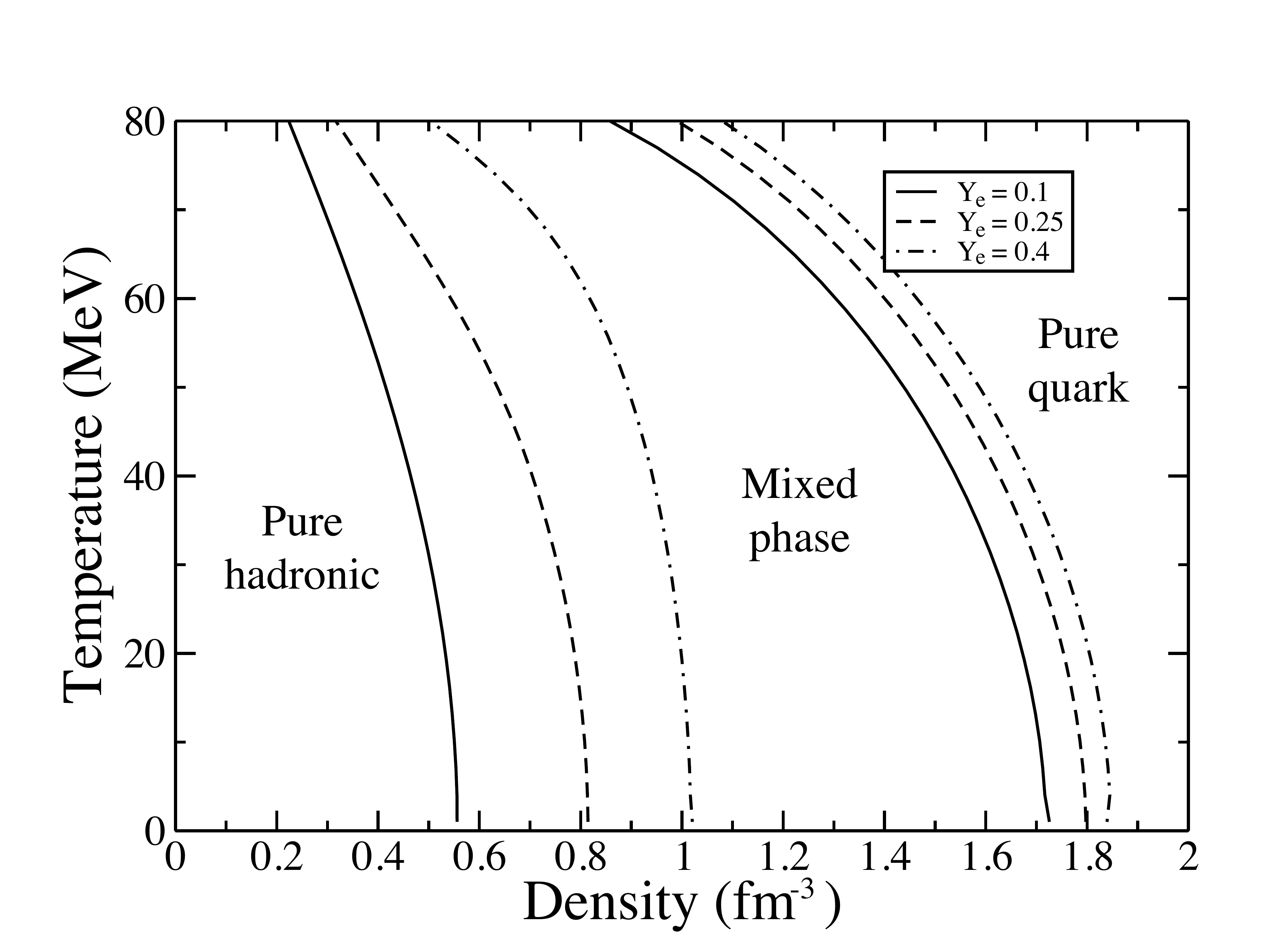}	
  \caption{Density-temperature phase diagram showing the density range of the mixed phase coexistence region for three values of $Y_e$ (solid line for $Y_{e} = 0.1$, dashed line for $Y_e=0.25$ and dash-dotted line for $Y_e=0.4$), including pions. The onset of the mixed phase is indicated by the set of curves on the left, while the curves on the right show the completion of the mixed phase. For low temperatures it is seen that the onset density is highly $Y_e$ dependent.}
 \label{fig:phase_plot}
\end{figure}



Another feature seen in Figs.~\ref{fig:mixed_pressure} and~\ref{fig:gamma}  is the $Y_e$ dependence of the onset density of the mixed phase. Fig.~\ref{fig:phase_plot} shows a phase diagram indicating the mixed phase transition temperature as a function of density for three values of $Y_e$. Pions were included in the making of this figure.  For higher temperatures the onset happens at lower densities as would be expected. However, for high electron fractions ($Y_e \sim 0.3$) such as those that can be found deep inside the cores of a proto-neutron star, the transition density remains quite high \mbox{$n_B \sim 0.7 ~\text{fm}^{-3}$}.  We note that similar phase diagrams have been made (e.g. \cite{mueller1997,DiToro06,DiToro11,cavagnoli2011}) indicating that the phase diagram is quite model dependent.

The supernova simulations of Refs.~\cite{Fischer2010, Gentile} show that, as the interior of the PNS reaches the onset of the mixed phase, the equation of state softens considerably and a secondary core collapse ensues. The matter then sharply stiffens upon entering the pure quark phase and a secondary shock wave can be  generated. As this shock catches up to the initially stalled accretion shock, a more robust explosion ensues.  However, their simulations assumed a Bag constants of $B^{1/4}\leq 165$ MeV, since they did not enforce the $M > 2 M_{\odot}$ maximum neutron star mass limit, and thus the mixed phase had a much lower onset density ($n_{\text{onset}}\sim 0.1 \text{ fm}^{-3}$) than what we have found here.  

\subsection{Neutron Stars with QGP Interiors} \label{sec:bag}

As stated previously, the observations~\cite{Demorest,Antoniadis} of a \mbox{$1.97 \pm 0.04~M_{\odot}$} and a \mbox{$2.01 \pm 0.04 ~M_\odot$} neutron star have ruled out many soft  EoSs including many hyperonic models~\cite{Lattimer2012}. We explore here how neutron stars with QGP interiors may be constrained by this consideration.  To do so, we enforce $n+p+e$ beta equilibrium in hadronic matter (and $u+d+e$ equilibrium in quark-gluon plasma).

\begin{figure}[b]
  \centering
  \includegraphics[width=0.5\textwidth]{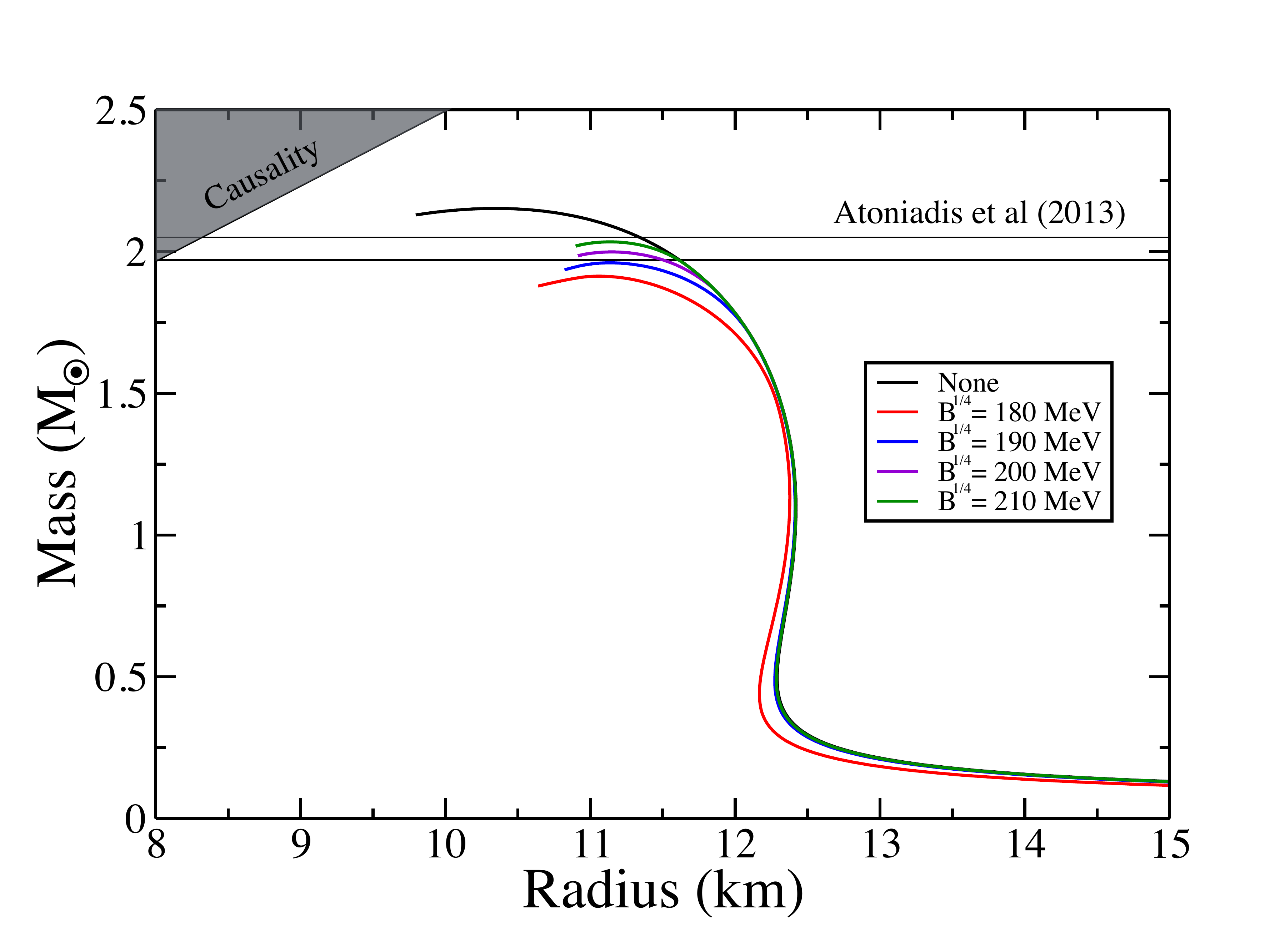}
  \caption{Neutron star mass-radius relation for various values of the bag constant B$^{1/4}$ using the GSkI parameter set. We find that a first order phase transition is consistent with the maximum mass neutron star measurement for our adopted value of B$^{1/4}$ = 190 MeV.   Horizontal lines show $2.01\pm 0.04 \text{ M}_{\odot}$ measurement from Ref.~\cite{Antoniadis}.}
 \label{fig:MaxMassBag}
\end{figure}

Fig.~\ref{fig:MaxMassBag} compares the neutron star mass radius relation for the GSKI Skyrme parameter set of Table~\ref{tab:skyrmemodels}.  
Using the range of bag constants determined by Eq.~(\ref{eqn:Tc}), we find that a first order phase transition to a QGP can be  consistent with the high maximum neutron star mass constraint~\cite{Demorest,Antoniadis} the GsKI Skyrme parameterization in the NDL EoS. From Fig.~\ref{fig:MaxMassBag}  we deduce  that a bag constant \mbox{$B^{1/4} > 190$ MeV} is required to satisfy the maximum neutron star mass constraint. This imposes a low baryon density transition temperature of $T_c > 150$~MeV, which is in the low end of the range of crossover temperatures allowed  from LGT~\cite{Kronfeld}.  Hence, all allowed values of the Bag constant inferred from LGT are consistent with the neutron star mass constraint in this formulation.  For our purpose we will adopt \mbox{$B^{1/4} = 190$ MeV} (corresponding to \mbox{$T_c \sim 150$~MeV}).

\subsection{Neutron Stars without QGP Interiors} 
\begin{widetext}

\begin{figure}[h]
  \centering
  \includegraphics[width=\textwidth]{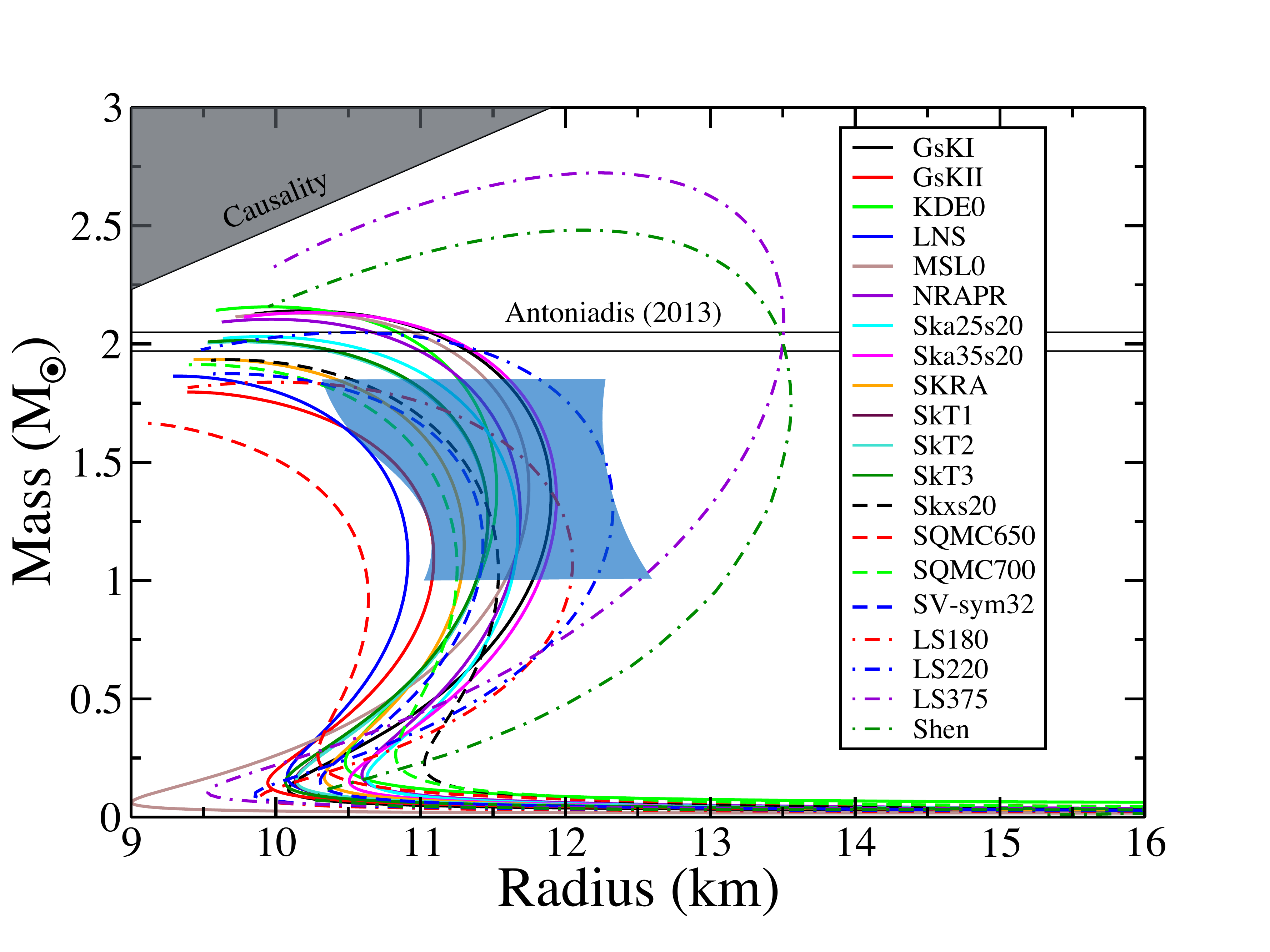}
  \caption{Mass-radius relation for the NDL~EoS with the Skyrme parameterizations from Table~\ref{tab:skyrmemodels}. The horizontal lines indicate the \mbox{$2.01 \pm 0.04~M_{\odot}$} maximum neutron star mass constraint from Ref.~\cite{Antoniadis}.  The grey shaded region shows the causality constraint and the blue shaded region shows the $2\sigma$ bounds from Ref.~\cite{steiner2010}. Note that not all of these curves satisfy the maximum mass or radius constraints.  (Color available online) }
 \label{fig:MaxMass}
\end{figure}

\end{widetext}

Fig.~\ref{fig:MaxMass} compares the neutron star mass radius relation for the Skyrme parameter sets in Table~\ref{tab:skyrmemodels} without a transition to QGP.  
Note that not all Skyrme parameter sets accommodate a maximum neutron star \mbox{mass $\ge 2.01 \pm 0.04 ~M_\odot$} \cite{Antoniadis,Demorest}. These we disregard in supernova simulations.

\section{Supernova simulations}

To demonstrate the viability of the NDL EoS for astrophysical simulations, we have run a series of core-collapse supernova simulations.  We have utilized University of Notre Dame/Lawrence Livermore National Laboratory supernova model \cite{Bowers82,WilsonMathews}, a spherically symmetric general relativistic hydrodynamic simulation with neutrino transport via multigroup flux limited diffusion.  We present here the explosion dynamics of simulations run with  the Skyrme models in Table~\ref{tab:skyrmemodels} that satisfy the  mass maximum neutron star $> 2$ M$_\odot$
constraint.  For this study we do not consider the transition to QGP.  That we leave to a subsequent paper.

\begin{figure}[h]
\centering
\includegraphics[width = 0.5\textwidth]{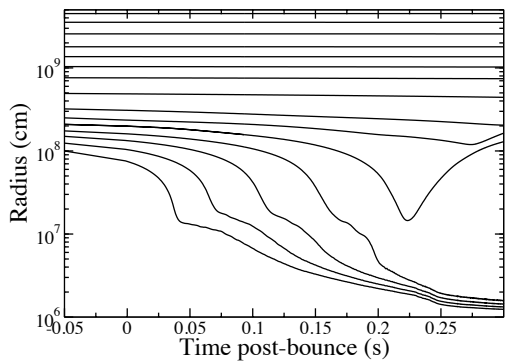}
\caption{Radius versus time post-bounce for the NDL EoS with the GsKI Skyrme parameter.}
\label{fig:radius}
\end{figure}

As an example, Fig.~\ref{fig:radius} shows the radial evolution versus time post-bounce of mass elements in a simulation using the NDL EoS with the GSkI Skyrme parameters.  For ease of comparison with other simulations in the lliterature, this simulation was run using the 20 M$_\odot$ progenitor model of Ref.~\cite{woosley_1995}.  As one can see in this figure the late time neutrino heating at $t \sim 200$ ms leads to an explosion.  Indeed, the expansion of the neutrino heated bubble is quite similar to the simulation in \cite{WilsonMathews} based upon the Bower and Wilson EoS. Hence, the stiffer EoS has not diminished the explosion.

To explore the impact of the new EoSs on the explosion, Fig.~\ref{fig:eos-ke}  shows kinetic energy versus time post-bounce for the various Skyrme parameter sets compared to the Bowers and Wilson equation of state as a point of reference.  The early evolution from core bounce ($t_{pb} = 0$ s) to shock stagnation ($t_{pb} \sim 0.2$ s) is relatively unchanged for different Skyrme models, but differs from the evolution of the Bowers and Wilson equation of state.  Even though the neutrino luminosity is greater for $t > 0.26$ s in the Bowers and Wilson EoS, this does not significantly affect the explosion.  This difference will, however, have an affect on the subsequent $r$-process nucleosynthesis as will be explored in a subsequent paper.  

Although the Bowers and Wilson EoS is softer and leads to a higher neutrino luminosity at later times the kinetic energy of the explosion with the Skyrme models is greater for $t > 0.2 $ s.  However, the kinetic energy of supernova simulations with different Skyrme parameters diverge after $t_{pb} \sim 0.2$ s.  These effects can be related to differences in the neutrino luminosities, and thus in the efficiency of neutrino reheating of the shock as illustrated in Fig.~\ref{fig:eos-lum}a-c. 

Fig.~\ref{fig:eos-lum}a shows the electron neutrino luminosity versus time post-bounce, while Figs. ~\ref{fig:eos-lum}b-c show the electron anti-neutrino and $\mu,\tau$ neutrino luminosities, respectively.  As is expected, the neutrino luminosities of the various simulations are similar until $t_{pb} \sim 0.2$ s, after which the different properties of the Skyrme parameter sets lead to differences in the electron neutrino luminosity.  The Skyrme parameter sets that result in higher electron neutrino luminosities also result in higher kinetic energies and shorter explosion timescales, as is expected. For all three neutrino flavors, however there is a bump in neutrino luminosity at $t \sim 0.23$ s.  We attribute this  to a softening of the core by the formation of thermal pions.  This leads to an enhanced neutrino luminosity and a more energetic reheated shock as shown in Fig.~\ref{fig:eos-ke}.

There is a general correlation among Skyrme parameter sets in that those which result result in the highest maximun neutron star mass (and are relatively ``stiff'')  lead to lower explosion energies.  However, the Bowers and Wilson equation of state, which is too soft to meet the modern neutron star mass constraint, has the lowest kinetic energy of all of the simulations shown.  We attribute this at least in part  to differences in the density dependence in the symmetry energy and the formation of pions.  Both of these cause the the Skyrme DFT EoSs to soften relative to the Bowers and Wilson EoS at the highest densities and temperatures.  However,  we caution that there is not one nuclear physics parameter that describes the relative ``softness'' or ``stiffness'' of the EoS at the highest temperatures and densities, as the response of the nuclear matter relies on an interplay among the symmetry energy, compressibility, etc.  Nevertheless, it is clear that differences  in nuclear properties, such as the symmetry energy, result in different behavior for the Skyrme parameter sets, particularly at supra-nuclear densities.  This leads to divergences at later times in the neutrino luminosities and evolution of the supernova simulation.

We leave a more detailed analysis of the EoS dependence of core-collapse supernova, including the possible phase transition to quark-gluon plasma, to a later publication.  Nevertheless, we have shown here that the NDL EoS can be used to successfully simulate core-collapse supernovae.  Moreover, we show that the new EoS parameter sets presented here lead to an even more robust explosion than the softer Bowers and Wilson  EoS.  This we believe is  due in part to the critical roles of the density dependence of the symmetry energy and  thermal pion formation in the core.

\begin{widetext}

\begin{figure}[h]
\centering
\includegraphics[height= 0.5\textheight]{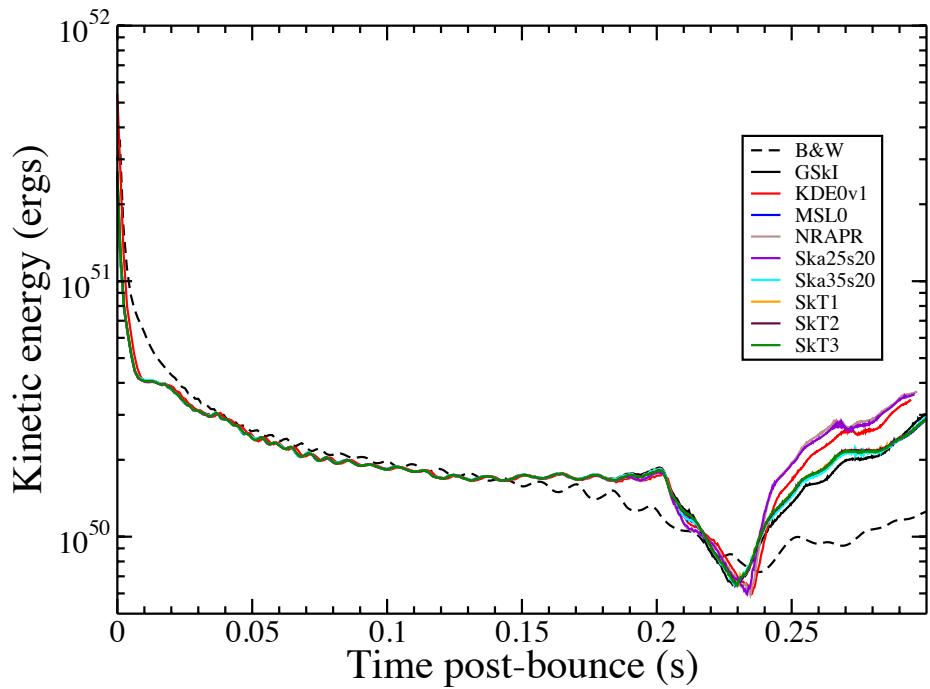}
\caption{Kinetic energy versus time post-bounce for the NDL EoS with Skyrme parameters found in Table~\ref{tab:skyrmemodels} and the Bowers and Wilson (B\&W) equation of state.  (Color available online)}
\label{fig:eos-ke}
\end{figure}

\begin{figure}[h]
\centering
\begin{subfigure}[t]{0.7\textwidth}
\includegraphics[width = \textwidth]{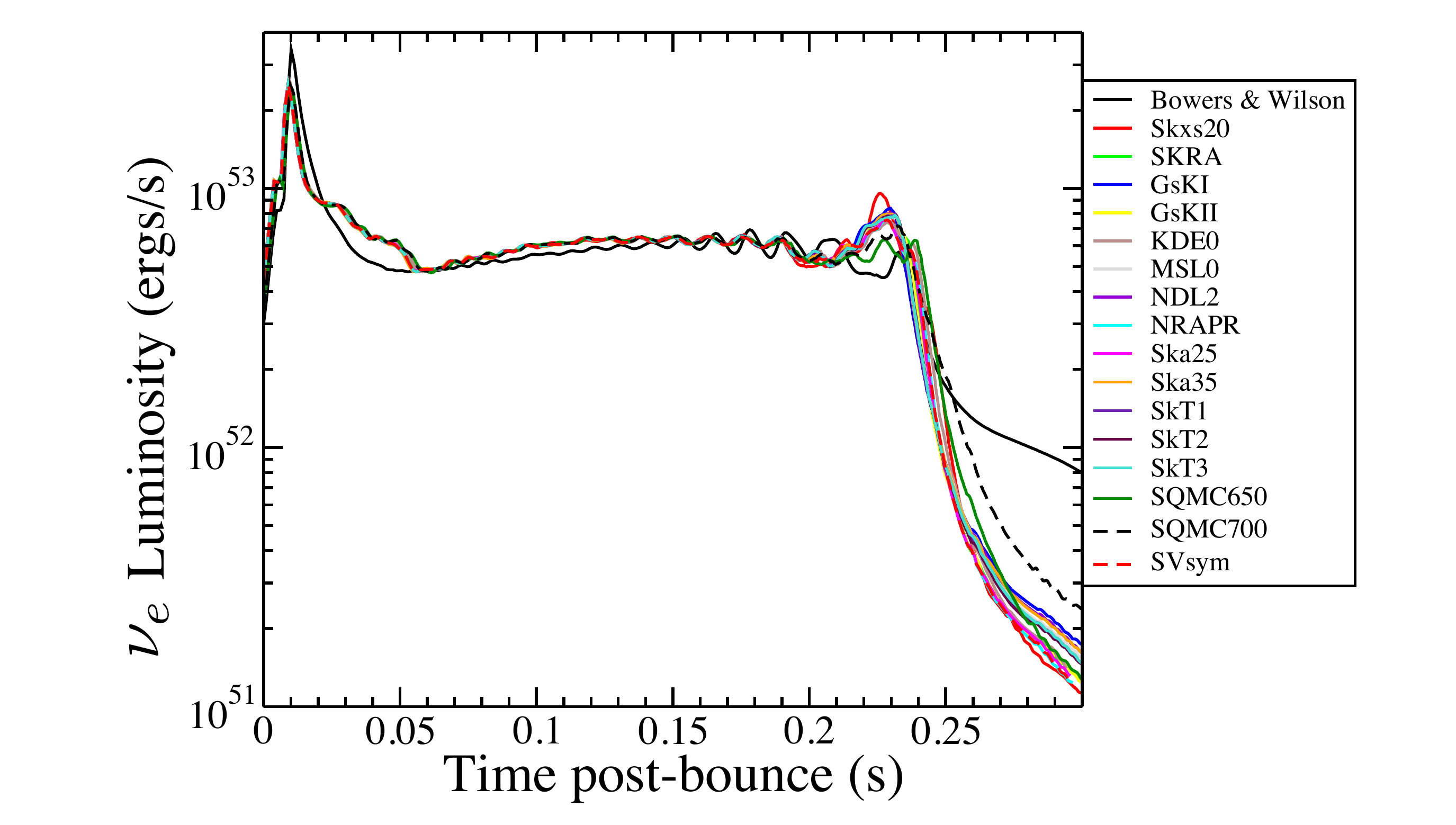}
\caption{Electron neutrino luminosity}
\end{subfigure}

\begin{subfigure}[t]{0.7\textwidth}
\includegraphics[width = \textwidth]{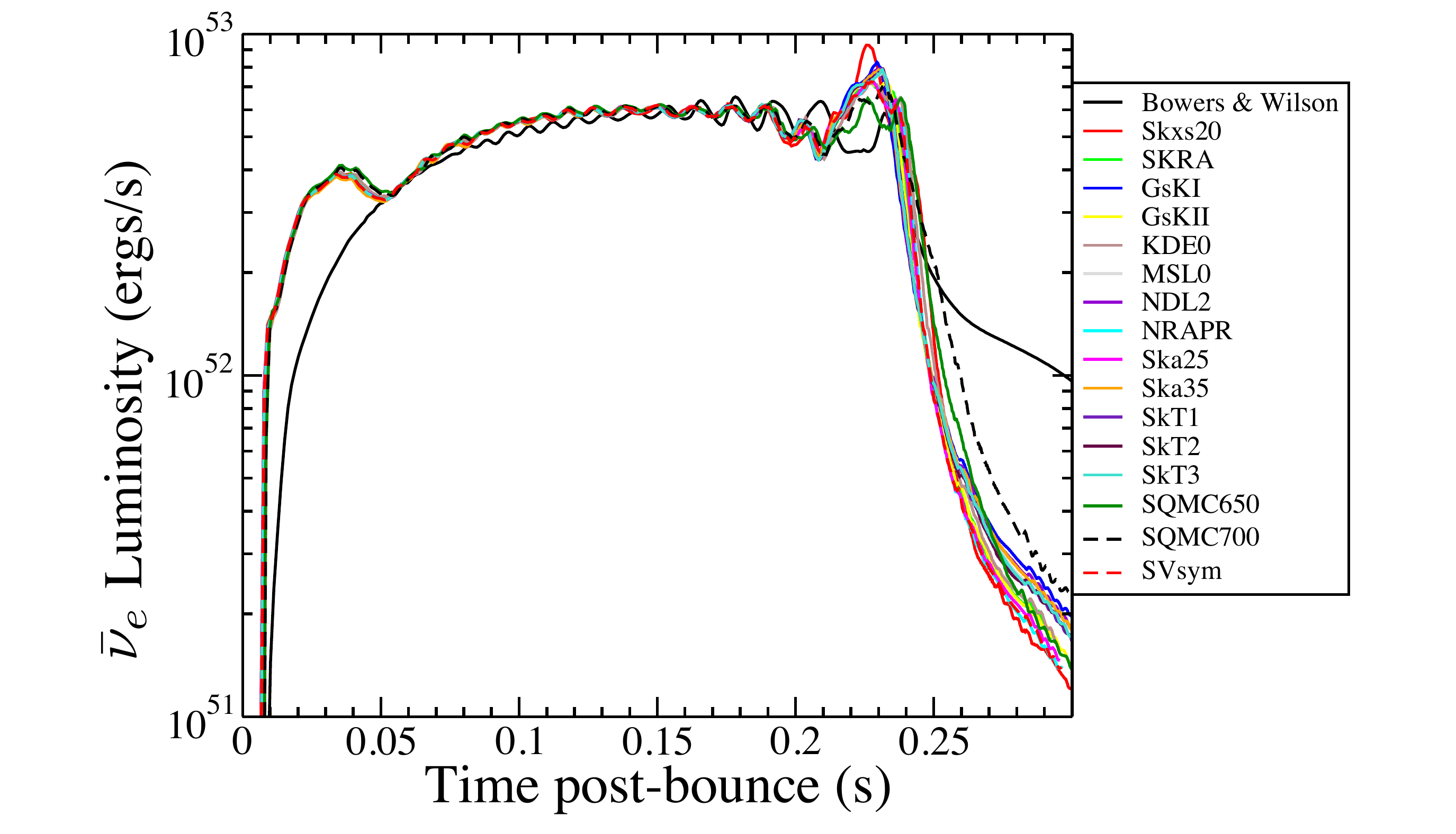}
\caption{Electron antineutrino luminosity}
\end{subfigure}

\begin{subfigure}[t]{0.7\textwidth}
\includegraphics[width = \textwidth]{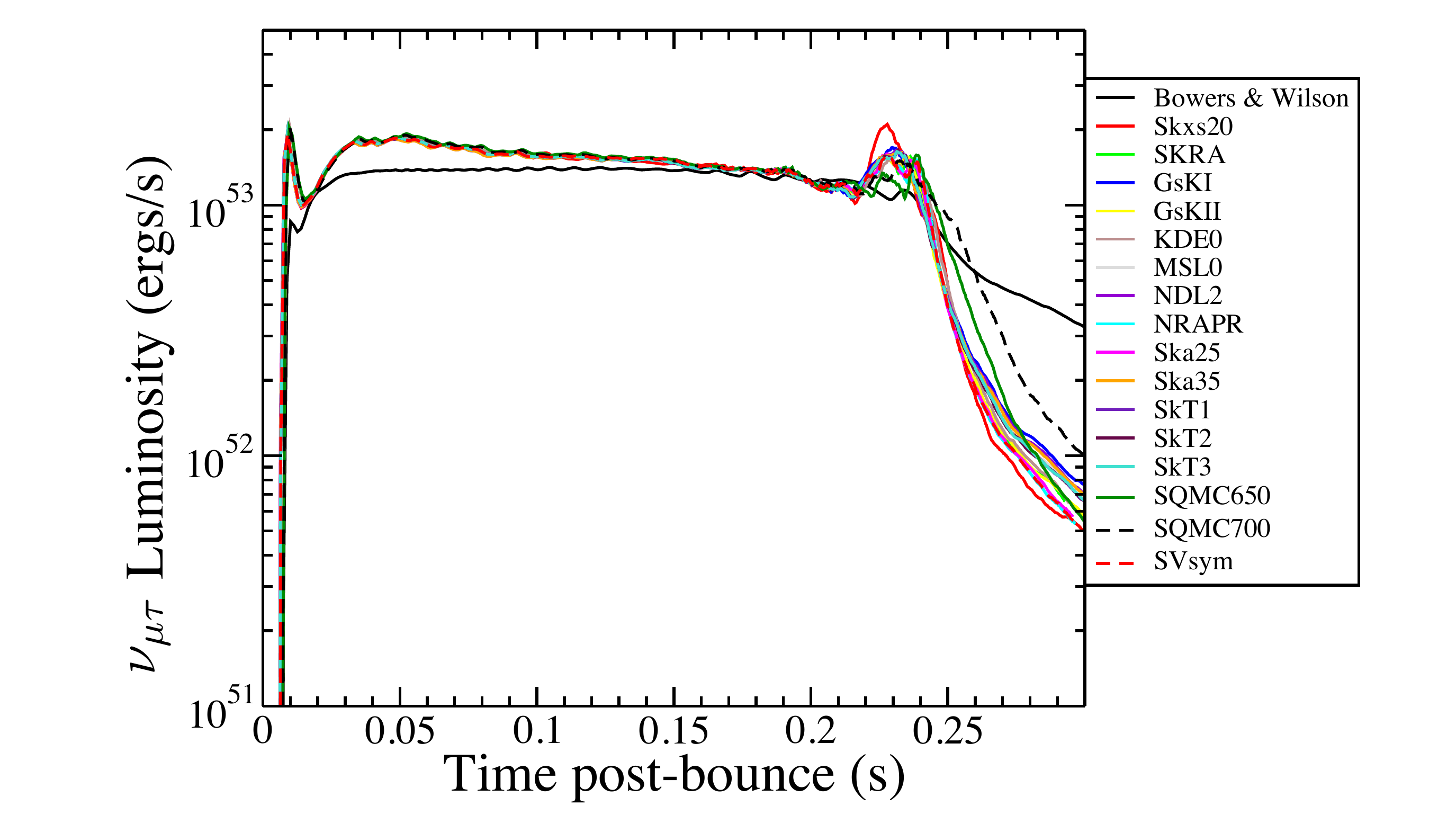}
\caption{$\mu,\tau$ neutrino luminosity}
\end{subfigure}
\caption{Neutrino luminosities versus time post-bounce for the NDL EoS with Skyrme parameters found in Table~\ref{tab:skyrmemodels} and the Bowers and Wilson (B\&W) equation of state.  (Color available online)}
\label{fig:eos-lum}
\end{figure}

\end{widetext}

\section{Conclusion}
We have discussed a new equation of state for astrophysical applications based upon a  density functional theory approach at nuclear matter density. We have shown that it is complementary to the most frequently employed nuclear equations of state for astrophysics: Shen~et~al.~\cite{Shen98a, Shen98b,Shen2011}, which is based upon the RMF, and the Lattimer-Swesty EoS~\cite{LS91}, which utilizes a liquid drop model approach.  We have adopted  a set Skyrme DFT parameterizations that are consistent with all known constraints on nuclei, nuclear matter, and observed properties of neutron stars and pulsars.   We made a first exploration of their effect  on the dynamics of core collapse supernovae and find that the Skyrme EoSs all lead to a somewhat more robust explosion.

A first order phase transition to a QGP phase was also added to the EoS in the context of a Gibbs construction. Applying the constraints from LGT for the range of low-baryon-density crossover temperatures, we were able to match the known constraints of the current maximum neutron star mass measurement for a bag constant \mbox{$B^{1/4} \ge 190$ MeV} if the finite baryon density transition if first order. On the other hand, if the high baryon-density QCD transition is a simple crossover, then the neutron star mass constraint can be easily accommodated for any value of the bag constant.  In future work, we plan to address the possible effects of a transition to QGP in CCSNe and failed supernovae and also explore possible effects of a color superconducting phase \cite{rodrigues2010,docarmo2013}.

We have also shown that the NDL EoS can lead to observable consequences in the late time neutrino luminosities.  Uncertainties in nuclear properties, such as the symmetry energy, cause the different Skyrme parameter sets to result in diverging evolution at late times ($> 250$ ms) in the supernova simulations.   This work confirms that, in the a core-collapse explosion paradigm, the equation of state may produce observable effects in the neutrino light curve, shock dynamics, and heavy element nucleosynthesis both in core-collapse supernovae and black hole formation in failed supernova events. The consequences of this new NDL~EoS for the dynamics of core collapse supernovae, including a QCD transition and the  impact on nucleosynthesis, will be further explored in forthcoming manuscripts.

Additionally, we have several future updates and improvements planned for the NDL EoS, including the inclusion of hyperons, an improved treatment of the nuclear pasta phases, and an improved treatment of the thermal component at supra-nuclear densities \cite{constantinou2015}.

With the availability  of the Notre~Dame-Livermore  (NDL)-DFT Equation of State, there is now another EoS which allows for an investigation of the dependence of the equation of state  in astrophysical simulations -- be it supernova core-collapse, neutron star mergers or black hole formation.  The density functional theory approach represents a fundamentally different theoretical approach from relativistic mean field theories and liquid drop models, and draws from numerous experimental laboratory constraints.  The beneficial nature of gaining insight from the nuclear domain is mutual, as the continued observation of heavy neutron stars and core-collapse supernovae can further constrain Skyrme parameters.  

\begin{acknowledgements}
Work at the University of Notre Dame is supported by the U.S. Department of Energy under Nuclear Theory Grant DE-FG02-95-ER40934. Two of the authors (N.Q.L. and M.W.) were supported in part by the National Science Foundation under Grant No. PHY-1430152 (JINA Center for the Evolution of the Elements).
\end{acknowledgements}

\bibliography{References}

\begin{thebibliography}{77}%
\makeatletter
\providecommand \@ifxundefined [1]{%
 \@ifx{#1\undefined}
}%
\providecommand \@ifnum [1]{%
 \ifnum #1\expandafter \@firstoftwo
 \else \expandafter \@secondoftwo
 \fi
}%
\providecommand \@ifx [1]{%
 \ifx #1\expandafter \@firstoftwo
 \else \expandafter \@secondoftwo
 \fi
}%
\providecommand \natexlab [1]{#1}%
\providecommand \enquote  [1]{``#1''}%
\providecommand \bibnamefont  [1]{#1}%
\providecommand \bibfnamefont [1]{#1}%
\providecommand \citenamefont [1]{#1}%
\providecommand \href@noop [0]{\@secondoftwo}%
\providecommand \href [0]{\begingroup \@sanitize@url \@href}%
\providecommand \@href[1]{\@@startlink{#1}\@@href}%
\providecommand \@@href[1]{\endgroup#1\@@endlink}%
\providecommand \@sanitize@url [0]{\catcode `\\12\catcode `\$12\catcode
  `\&12\catcode `\#12\catcode `\^12\catcode `\_12\catcode `\%12\relax}%
\providecommand \@@startlink[1]{}%
\providecommand \@@endlink[0]{}%
\providecommand \url  [0]{\begingroup\@sanitize@url \@url }%
\providecommand \@url [1]{\endgroup\@href {#1}{\urlprefix }}%
\providecommand \urlprefix  [0]{URL }%
\providecommand \Eprint [0]{\href }%
\providecommand \doibase [0]{http://dx.doi.org/}%
\providecommand \selectlanguage [0]{\@gobble}%
\providecommand \bibinfo  [0]{\@secondoftwo}%
\providecommand \bibfield  [0]{\@secondoftwo}%
\providecommand \translation [1]{[#1]}%
\providecommand \BibitemOpen [0]{}%
\providecommand \bibitemStop [0]{}%
\providecommand \bibitemNoStop [0]{.\EOS\space}%
\providecommand \EOS [0]{\spacefactor3000\relax}%
\providecommand \BibitemShut  [1]{\csname bibitem#1\endcsname}%
\let\auto@bib@innerbib\@empty
\bibitem [{\citenamefont {Lattimer}\ and\ \citenamefont {Swesty}(1991)}]{LS91}%
  \BibitemOpen
  \bibfield  {author} {\bibinfo {author} {\bibfnamefont {J.~M.}\ \bibnamefont
  {Lattimer}}\ and\ \bibinfo {author} {\bibfnamefont {F.~D.}\ \bibnamefont
  {Swesty}},\ }\href {\doibase 10.1016/0375-9474(91)90452-C} {\bibfield
  {journal} {\bibinfo  {journal} {Nuclear Physics A}\ }\textbf {\bibinfo
  {volume} {535}},\ \bibinfo {pages} {331 } (\bibinfo {year}
  {1991})}\BibitemShut {NoStop}%
\bibitem [{\citenamefont {Shen}\ \emph
  {et~al.}(1998{\natexlab{a}})\citenamefont {Shen}, \citenamefont {Toki},
  \citenamefont {Oyamatsu},\ and\ \citenamefont {Sumiyoshi}}]{Shen98a}%
  \BibitemOpen
  \bibfield  {author} {\bibinfo {author} {\bibfnamefont {H.}~\bibnamefont
  {Shen}}, \bibinfo {author} {\bibfnamefont {H.}~\bibnamefont {Toki}}, \bibinfo
  {author} {\bibfnamefont {K.}~\bibnamefont {Oyamatsu}}, \ and\ \bibinfo
  {author} {\bibfnamefont {K.}~\bibnamefont {Sumiyoshi}},\ }\href {\doibase
  10.1016/S0375-9474(98)00236-X} {\bibfield  {journal} {\bibinfo  {journal}
  {Nuclear Physics A}\ }\textbf {\bibinfo {volume} {637}},\ \bibinfo {pages}
  {435 } (\bibinfo {year} {1998}{\natexlab{a}})}\BibitemShut {NoStop}%
\bibitem [{\citenamefont {Shen}\ \emph
  {et~al.}(1998{\natexlab{b}})\citenamefont {Shen}, \citenamefont {Toki},
  \citenamefont {Oyamatsu},\ and\ \citenamefont {Sumiyoshi}}]{Shen98b}%
  \BibitemOpen
  \bibfield  {author} {\bibinfo {author} {\bibfnamefont {H.}~\bibnamefont
  {Shen}}, \bibinfo {author} {\bibfnamefont {H.}~\bibnamefont {Toki}}, \bibinfo
  {author} {\bibfnamefont {K.}~\bibnamefont {Oyamatsu}}, \ and\ \bibinfo
  {author} {\bibfnamefont {K.}~\bibnamefont {Sumiyoshi}},\ }\href {\doibase
  10.1143/PTP.100.1013} {\bibfield  {journal} {\bibinfo  {journal} {Progress of
  Theoretical Physics}\ }\textbf {\bibinfo {volume} {100}},\ \bibinfo {pages}
  {1013} (\bibinfo {year} {1998}{\natexlab{b}})}\BibitemShut {NoStop}%
\bibitem [{\citenamefont {{Shen}}\ \emph {et~al.}(2011)\citenamefont {{Shen}},
  \citenamefont {{Toki}}, \citenamefont {{Oyamatsu}},\ and\ \citenamefont
  {{Sumiyoshi}}}]{Shen2011}%
  \BibitemOpen
  \bibfield  {author} {\bibinfo {author} {\bibfnamefont {H.}~\bibnamefont
  {{Shen}}}, \bibinfo {author} {\bibfnamefont {H.}~\bibnamefont {{Toki}}},
  \bibinfo {author} {\bibfnamefont {K.}~\bibnamefont {{Oyamatsu}}}, \ and\
  \bibinfo {author} {\bibfnamefont {K.}~\bibnamefont {{Sumiyoshi}}},\ }\href
  {\doibase 10.1088/0067-0049/197/2/20} {\bibfield  {journal} {\bibinfo
  {journal} {Astrophysical Journal Supplement Series}\ }\textbf {\bibinfo
  {volume} {197}},\ \bibinfo {eid} {035802} (\bibinfo {year} {2011})},\ \Eprint
  {http://arxiv.org/abs/1105.1666} {arXiv:1105.1666 [astro-ph.HE]} \BibitemShut
  {NoStop}%
\bibitem [{\citenamefont {{Ishizuka}}\ \emph {et~al.}(2008)\citenamefont
  {{Ishizuka}}, \citenamefont {{Ohnishi}}, \citenamefont {{Tsubakihara}},
  \citenamefont {{Sumiyoshi}},\ and\ \citenamefont {{Yamada}}}]{Ishizuka2008}%
  \BibitemOpen
  \bibfield  {author} {\bibinfo {author} {\bibfnamefont {C.}~\bibnamefont
  {{Ishizuka}}}, \bibinfo {author} {\bibfnamefont {A.}~\bibnamefont
  {{Ohnishi}}}, \bibinfo {author} {\bibfnamefont {K.}~\bibnamefont
  {{Tsubakihara}}}, \bibinfo {author} {\bibfnamefont {K.}~\bibnamefont
  {{Sumiyoshi}}}, \ and\ \bibinfo {author} {\bibfnamefont {S.}~\bibnamefont
  {{Yamada}}},\ }\href {\doibase 10.1088/0954-3899/35/8/085201} {\bibfield
  {journal} {\bibinfo  {journal} {Journal of Physics G Nuclear Physics}\
  }\textbf {\bibinfo {volume} {35}},\ \bibinfo {pages} {085201} (\bibinfo
  {year} {2008})},\ \Eprint {http://arxiv.org/abs/0802.2318} {arXiv:0802.2318
  [nucl-th]} \BibitemShut {NoStop}%
\bibitem [{\citenamefont {{Fischer}}\ \emph {et~al.}(2011)\citenamefont
  {{Fischer}}, \citenamefont {{Sagert}}, \citenamefont {{Pagliara}},
  \citenamefont {{Hempel}}, \citenamefont {{Schaffner-Bielich}}, \citenamefont
  {{Rauscher}}, \citenamefont {{Thielemann}}, \citenamefont {{K{\"a}ppeli}},
  \citenamefont {{Mart{\'{\i}}nez-Pinedo}},\ and\ \citenamefont
  {{Liebend{\"o}rfer}}}]{Fischer2011}%
  \BibitemOpen
  \bibfield  {author} {\bibinfo {author} {\bibfnamefont {T.}~\bibnamefont
  {{Fischer}}}, \bibinfo {author} {\bibfnamefont {I.}~\bibnamefont {{Sagert}}},
  \bibinfo {author} {\bibfnamefont {G.}~\bibnamefont {{Pagliara}}}, \bibinfo
  {author} {\bibfnamefont {M.}~\bibnamefont {{Hempel}}}, \bibinfo {author}
  {\bibfnamefont {J.}~\bibnamefont {{Schaffner-Bielich}}}, \bibinfo {author}
  {\bibfnamefont {T.}~\bibnamefont {{Rauscher}}}, \bibinfo {author}
  {\bibfnamefont {F.-K.}\ \bibnamefont {{Thielemann}}}, \bibinfo {author}
  {\bibfnamefont {R.}~\bibnamefont {{K{\"a}ppeli}}}, \bibinfo {author}
  {\bibfnamefont {G.}~\bibnamefont {{Mart{\'{\i}}nez-Pinedo}}}, \ and\ \bibinfo
  {author} {\bibfnamefont {M.}~\bibnamefont {{Liebend{\"o}rfer}}},\ }\href
  {\doibase 10.1088/0067-0049/194/2/39} {\bibfield  {journal} {\bibinfo
  {journal} {Astrophysical Journal Supplement Series}\ }\textbf {\bibinfo
  {volume} {194}},\ \bibinfo {eid} {39} (\bibinfo {year} {2011})},\ \Eprint
  {http://arxiv.org/abs/1011.3409} {arXiv:1011.3409 [astro-ph.HE]} \BibitemShut
  {NoStop}%
\bibitem [{\citenamefont {{Fischer}}\ \emph
  {et~al.}(2010{\natexlab{a}})\citenamefont {{Fischer}}, \citenamefont
  {{Sagert}}, \citenamefont {{Hempel}}, \citenamefont {{Pagliara}},
  \citenamefont {{Schaffner-Bielich}},\ and\ \citenamefont
  {{Liebend{o}rfer}}}]{Sagert}%
  \BibitemOpen
  \bibfield  {author} {\bibinfo {author} {\bibfnamefont {T.}~\bibnamefont
  {{Fischer}}}, \bibinfo {author} {\bibfnamefont {I.}~\bibnamefont {{Sagert}}},
  \bibinfo {author} {\bibfnamefont {M.}~\bibnamefont {{Hempel}}}, \bibinfo
  {author} {\bibfnamefont {G.}~\bibnamefont {{Pagliara}}}, \bibinfo {author}
  {\bibfnamefont {J.}~\bibnamefont {{Schaffner-Bielich}}}, \ and\ \bibinfo
  {author} {\bibfnamefont {M.}~\bibnamefont {{Liebend{o}rfer}}},\ }\href
  {\doibase 10.1088/0264-9381/27/11/114102} {\bibfield  {journal} {\bibinfo
  {journal} {Classical and Quantum Gravity}\ }\textbf {\bibinfo {volume}
  {27}},\ \bibinfo {pages} {114102} (\bibinfo {year}
  {2010}{\natexlab{a}})}\BibitemShut {NoStop}%
\bibitem [{\citenamefont {{Hempel}}\ and\ \citenamefont
  {{Schaffner-Bielich}}(2010)}]{Hempel2010}%
  \BibitemOpen
  \bibfield  {author} {\bibinfo {author} {\bibfnamefont {M.}~\bibnamefont
  {{Hempel}}}\ and\ \bibinfo {author} {\bibfnamefont {J.}~\bibnamefont
  {{Schaffner-Bielich}}},\ }\href {\doibase 10.1016/j.nuclphysa.2010.02.010}
  {\bibfield  {journal} {\bibinfo  {journal} {Nuclear Physics A}\ }\textbf
  {\bibinfo {volume} {837}},\ \bibinfo {pages} {210} (\bibinfo {year}
  {2010})},\ \Eprint {http://arxiv.org/abs/0911.4073} {arXiv:0911.4073
  [nucl-th]} \BibitemShut {NoStop}%
\bibitem [{\citenamefont {{Steiner}}\ \emph {et~al.}(2012)\citenamefont
  {{Steiner}}, \citenamefont {{Hempel}},\ and\ \citenamefont
  {{Fischer}}}]{Steiner2012}%
  \BibitemOpen
  \bibfield  {author} {\bibinfo {author} {\bibfnamefont {A.~W.}\ \bibnamefont
  {{Steiner}}}, \bibinfo {author} {\bibfnamefont {M.}~\bibnamefont {{Hempel}}},
  \ and\ \bibinfo {author} {\bibfnamefont {T.}~\bibnamefont {{Fischer}}},\
  }\href@noop {} {\bibfield  {journal} {\bibinfo  {journal} {ArXiv e-prints}\ }
  (\bibinfo {year} {2012})},\ \Eprint {http://arxiv.org/abs/1207.2184}
  {arXiv:1207.2184 [astro-ph.SR]} \BibitemShut {NoStop}%
\bibitem [{\citenamefont {Brown}(2013)}]{brown2013}%
  \BibitemOpen
  \bibfield  {author} {\bibinfo {author} {\bibfnamefont {B.~A.}\ \bibnamefont
  {Brown}},\ }\href@noop {} {\bibfield  {journal} {\bibinfo  {journal}
  {Phys.Rev.Lett}\ }\textbf {\bibinfo {volume} {111}} (\bibinfo {year}
  {2013})}\BibitemShut {NoStop}%
\bibitem [{\citenamefont {Brown}\ and\ \citenamefont
  {Schwenk}(2014)}]{brown2014}%
  \BibitemOpen
  \bibfield  {author} {\bibinfo {author} {\bibfnamefont {B.~A.}\ \bibnamefont
  {Brown}}\ and\ \bibinfo {author} {\bibfnamefont {A.}~\bibnamefont
  {Schwenk}},\ }\href@noop {} {\bibfield  {journal} {\bibinfo  {journal}
  {Phys.Rev.C}\ }\textbf {\bibinfo {volume} {89}} (\bibinfo {year}
  {2014})}\BibitemShut {NoStop}%
\bibitem [{\citenamefont {{Janka}}(2012)}]{janka2012}%
  \BibitemOpen
  \bibfield  {author} {\bibinfo {author} {\bibfnamefont {H.-T.}\ \bibnamefont
  {{Janka}}},\ }\href {\doibase 10.1146/annurev-nucl-102711-094901} {\bibfield
  {journal} {\bibinfo  {journal} {Annual Review of Nuclear and Particle
  Science}\ }\textbf {\bibinfo {volume} {62}},\ \bibinfo {pages} {407}
  (\bibinfo {year} {2012})},\ \Eprint {http://arxiv.org/abs/1206.2503}
  {arXiv:1206.2503 [astro-ph.SR]} \BibitemShut {NoStop}%
\bibitem [{\citenamefont {{Kitaura}}\ \emph {et~al.}(2006)\citenamefont
  {{Kitaura}}, \citenamefont {{Janka}},\ and\ \citenamefont
  {{Hillebrandt}}}]{kitaura2006}%
  \BibitemOpen
  \bibfield  {author} {\bibinfo {author} {\bibfnamefont {F.~S.}\ \bibnamefont
  {{Kitaura}}}, \bibinfo {author} {\bibfnamefont {H.-T.}\ \bibnamefont
  {{Janka}}}, \ and\ \bibinfo {author} {\bibfnamefont {W.}~\bibnamefont
  {{Hillebrandt}}},\ }\href {\doibase 10.1051/0004-6361:20054703} {\bibfield
  {journal} {\bibinfo  {journal} {A\&A}\ }\textbf {\bibinfo {volume} {450}},\
  \bibinfo {pages} {345} (\bibinfo {year} {2006})},\ \Eprint
  {http://arxiv.org/abs/astro-ph/0512065} {astro-ph/0512065} \BibitemShut
  {NoStop}%
\bibitem [{\citenamefont {{Burrows}}\ \emph {et~al.}(2007)\citenamefont
  {{Burrows}}, \citenamefont {{Dessart}},\ and\ \citenamefont
  {{Livne}}}]{burrows2007}%
  \BibitemOpen
  \bibfield  {author} {\bibinfo {author} {\bibfnamefont {A.}~\bibnamefont
  {{Burrows}}}, \bibinfo {author} {\bibfnamefont {L.}~\bibnamefont
  {{Dessart}}}, \ and\ \bibinfo {author} {\bibfnamefont {E.}~\bibnamefont
  {{Livne}}},\ }in\ \href {\doibase 10.1063/1.3682931} {\emph {\bibinfo
  {booktitle} {Supernova 1987A: 20 Years After: Supernovae and Gamma-Ray
  Bursters}}},\ \bibinfo {series} {American Institute of Physics Conference
  Series}, Vol.\ \bibinfo {volume} {937},\ \bibinfo {editor} {edited by\
  \bibinfo {editor} {\bibfnamefont {S.}~\bibnamefont {{Immler}}}, \bibinfo
  {editor} {\bibfnamefont {K.}~\bibnamefont {{Weiler}}}, \ and\ \bibinfo
  {editor} {\bibfnamefont {R.}~\bibnamefont {{McCray}}}}\ (\bibinfo {year}
  {2007})\ pp.\ \bibinfo {pages} {370--380}\BibitemShut {NoStop}%
\bibitem [{\citenamefont {{Burrows}}\ \emph {et~al.}(2012)\citenamefont
  {{Burrows}}, \citenamefont {{Dolence}},\ and\ \citenamefont
  {{Murphy}}}]{burrows2012}%
  \BibitemOpen
  \bibfield  {author} {\bibinfo {author} {\bibfnamefont {A.}~\bibnamefont
  {{Burrows}}}, \bibinfo {author} {\bibfnamefont {J.~C.}\ \bibnamefont
  {{Dolence}}}, \ and\ \bibinfo {author} {\bibfnamefont {J.~W.}\ \bibnamefont
  {{Murphy}}},\ }\href {\doibase 10.1088/0004-637X/759/1/5} {\bibfield
  {journal} {\bibinfo  {journal} {Ap.J.}\ }\textbf {\bibinfo {volume} {759}},\
  \bibinfo {eid} {5} (\bibinfo {year} {2012})},\ \Eprint
  {http://arxiv.org/abs/1204.3088} {arXiv:1204.3088 [astro-ph.SR]} \BibitemShut
  {NoStop}%
\bibitem [{\citenamefont {{Blondin}}\ \emph {et~al.}(2003)\citenamefont
  {{Blondin}}, \citenamefont {{Mezzacappa}},\ and\ \citenamefont
  {{DeMarino}}}]{blondin2003}%
  \BibitemOpen
  \bibfield  {author} {\bibinfo {author} {\bibfnamefont {J.~M.}\ \bibnamefont
  {{Blondin}}}, \bibinfo {author} {\bibfnamefont {A.}~\bibnamefont
  {{Mezzacappa}}}, \ and\ \bibinfo {author} {\bibfnamefont {C.}~\bibnamefont
  {{DeMarino}}},\ }\href {\doibase 10.1086/345812} {\bibfield  {journal}
  {\bibinfo  {journal} {Ap.J.}\ }\textbf {\bibinfo {volume} {584}},\ \bibinfo
  {pages} {971} (\bibinfo {year} {2003})},\ \Eprint
  {http://arxiv.org/abs/astro-ph/0210634} {astro-ph/0210634} \BibitemShut
  {NoStop}%
\bibitem [{\citenamefont {Couch}\ and\ \citenamefont {Ott}(2015)}]{Couch15}%
  \BibitemOpen
  \bibfield  {author} {\bibinfo {author} {\bibfnamefont {S.}~\bibnamefont
  {Couch}}\ and\ \bibinfo {author} {\bibfnamefont {C.}~\bibnamefont {Ott}},\
  }\href@noop {} {\bibfield  {journal} {\bibinfo  {journal} {Ap.J.}\ }\textbf
  {\bibinfo {volume} {779}} (\bibinfo {year} {2015})}\BibitemShut {NoStop}%
\bibitem [{\citenamefont {{Wilson}}\ and\ \citenamefont
  {{Mayle}}(1988)}]{wilson1988}%
  \BibitemOpen
  \bibfield  {author} {\bibinfo {author} {\bibfnamefont {J.~R.}\ \bibnamefont
  {{Wilson}}}\ and\ \bibinfo {author} {\bibfnamefont {R.~W.}\ \bibnamefont
  {{Mayle}}},\ }\href@noop {} {\bibfield  {journal} {\bibinfo  {journal}
  {Phys.Rep.}\ }\textbf {\bibinfo {volume} {163}},\ \bibinfo {pages} {63}
  (\bibinfo {year} {1988})}\BibitemShut {NoStop}%
\bibitem [{\citenamefont {{Wilson}}\ and\ \citenamefont
  {{Mayle}}(1993)}]{wilson1993}%
  \BibitemOpen
  \bibfield  {author} {\bibinfo {author} {\bibfnamefont {J.~R.}\ \bibnamefont
  {{Wilson}}}\ and\ \bibinfo {author} {\bibfnamefont {R.~W.}\ \bibnamefont
  {{Mayle}}},\ }\href@noop {} {\bibfield  {journal} {\bibinfo  {journal}
  {Phys.Rep.}\ }\textbf {\bibinfo {volume} {227}},\ \bibinfo {pages} {97}
  (\bibinfo {year} {1993})}\BibitemShut {NoStop}%
\bibitem [{\citenamefont {{Wilson}}\ and\ \citenamefont
  {{Mathews}}(2003)}]{WilsonMathews}%
  \BibitemOpen
  \bibfield  {author} {\bibinfo {author} {\bibfnamefont {J.~R.}\ \bibnamefont
  {{Wilson}}}\ and\ \bibinfo {author} {\bibfnamefont {G.~J.}\ \bibnamefont
  {{Mathews}}},\ }\href@noop {} {\emph {\bibinfo {title} {Relativistic
  Numerical Hydrodynamics, by James R.~Wilson and Grant J.~Mathews,
  pp.~232.~ISBN 0521631556.~Cambridge, UK: Cambridge University Press, December
  2003.}}}\ (\bibinfo  {publisher} {Cambridge University Press},\ \bibinfo
  {year} {2003})\BibitemShut {NoStop}%
\bibitem [{\citenamefont {{Wilson}}\ \emph {et~al.}(2005)\citenamefont
  {{Wilson}}, \citenamefont {{Mathews}},\ and\ \citenamefont
  {{Dalhed}}}]{wilson2005}%
  \BibitemOpen
  \bibfield  {author} {\bibinfo {author} {\bibfnamefont {J.~R.}\ \bibnamefont
  {{Wilson}}}, \bibinfo {author} {\bibfnamefont {G.~J.}\ \bibnamefont
  {{Mathews}}}, \ and\ \bibinfo {author} {\bibfnamefont {H.~E.}\ \bibnamefont
  {{Dalhed}}},\ }\href {\doibase 10.1086/430297} {\bibfield  {journal}
  {\bibinfo  {journal} {Ap.J.}\ }\textbf {\bibinfo {volume} {628}},\ \bibinfo
  {pages} {335} (\bibinfo {year} {2005})},\ \Eprint
  {http://arxiv.org/abs/astro-ph/0508146} {astro-ph/0508146} \BibitemShut
  {NoStop}%
\bibitem [{\citenamefont {{Sagert}}\ \emph {et~al.}(2009)\citenamefont
  {{Sagert}}, \citenamefont {{Fischer}}, \citenamefont {{Hempel}},
  \citenamefont {{Pagliara}}, \citenamefont {{Schaffner-Bielich}},
  \citenamefont {{Mezzacappa}}, \citenamefont {{Thielemann}},\ and\
  \citenamefont {{Liebend{\"o}rfer}}}]{Sagert2009}%
  \BibitemOpen
  \bibfield  {author} {\bibinfo {author} {\bibfnamefont {I.}~\bibnamefont
  {{Sagert}}}, \bibinfo {author} {\bibfnamefont {T.}~\bibnamefont {{Fischer}}},
  \bibinfo {author} {\bibfnamefont {M.}~\bibnamefont {{Hempel}}}, \bibinfo
  {author} {\bibfnamefont {G.}~\bibnamefont {{Pagliara}}}, \bibinfo {author}
  {\bibfnamefont {J.}~\bibnamefont {{Schaffner-Bielich}}}, \bibinfo {author}
  {\bibfnamefont {A.}~\bibnamefont {{Mezzacappa}}}, \bibinfo {author}
  {\bibfnamefont {F.-K.}\ \bibnamefont {{Thielemann}}}, \ and\ \bibinfo
  {author} {\bibfnamefont {M.}~\bibnamefont {{Liebend{\"o}rfer}}},\ }\href
  {\doibase 10.1103/PhysRevLett.102.081101} {\bibfield  {journal} {\bibinfo
  {journal} {Physical Review Letters}\ }\textbf {\bibinfo {volume} {102}},\
  \bibinfo {eid} {081101} (\bibinfo {year} {2009})},\ \Eprint
  {http://arxiv.org/abs/0809.4225} {arXiv:0809.4225} \BibitemShut {NoStop}%
\bibitem [{\citenamefont {Fischer}\ \emph {et~al.}(2011)\citenamefont
  {Fischer}, \citenamefont {Sagert}, \citenamefont {Pagliara}, \citenamefont
  {Hempel}, \citenamefont {Schaffner-Bielich}, \citenamefont {Rauscher},
  \citenamefont {Thielemann}, \citenamefont {KÃ¤ppeli}, \citenamefont
  {MartÃ­nez-Pinedo},\ and\ \citenamefont
  {LiebendÃ¶rfer}}]{FischerSagert2011}%
  \BibitemOpen
  \bibfield  {author} {\bibinfo {author} {\bibfnamefont {T.}~\bibnamefont
  {Fischer}}, \bibinfo {author} {\bibfnamefont {I.}~\bibnamefont {Sagert}},
  \bibinfo {author} {\bibfnamefont {G.}~\bibnamefont {Pagliara}}, \bibinfo
  {author} {\bibfnamefont {M.}~\bibnamefont {Hempel}}, \bibinfo {author}
  {\bibfnamefont {J.}~\bibnamefont {Schaffner-Bielich}}, \bibinfo {author}
  {\bibfnamefont {T.}~\bibnamefont {Rauscher}}, \bibinfo {author}
  {\bibfnamefont {F.-K.}\ \bibnamefont {Thielemann}}, \bibinfo {author}
  {\bibfnamefont {R.}~\bibnamefont {KÃ¤ppeli}}, \bibinfo {author}
  {\bibfnamefont {G.}~\bibnamefont {MartÃ­nez-Pinedo}}, \ and\ \bibinfo
  {author} {\bibfnamefont {M.}~\bibnamefont {LiebendÃ¶rfer}},\ }\href
  {http://stacks.iop.org/0067-0049/194/i=2/a=39} {\bibfield  {journal}
  {\bibinfo  {journal} {The Astrophysical Journal Supplement Series}\ }\textbf
  {\bibinfo {volume} {194}},\ \bibinfo {pages} {39} (\bibinfo {year}
  {2011})}\BibitemShut {NoStop}%
\bibitem [{\citenamefont {Warren}\ \emph {et~al.}(2014)\citenamefont {Warren},
  \citenamefont {Meixner}, \citenamefont {Mathews}, \citenamefont {Hidaka},\
  and\ \citenamefont {Kajino}}]{Warren14}%
  \BibitemOpen
  \bibfield  {author} {\bibinfo {author} {\bibfnamefont {M.}~\bibnamefont
  {Warren}}, \bibinfo {author} {\bibfnamefont {M.}~\bibnamefont {Meixner}},
  \bibinfo {author} {\bibfnamefont {G.}~\bibnamefont {Mathews}}, \bibinfo
  {author} {\bibfnamefont {J.}~\bibnamefont {Hidaka}}, \ and\ \bibinfo {author}
  {\bibfnamefont {T.}~\bibnamefont {Kajino}},\ }\href@noop {} {\bibfield
  {journal} {\bibinfo  {journal} {Phys.Rev.D}\ }\textbf {\bibinfo {volume}
  {90}} (\bibinfo {year} {2014})}\BibitemShut {NoStop}%
\bibitem [{\citenamefont {Warren}\ \emph {et~al.}(2016)\citenamefont {Warren},
  \citenamefont {Mathews}, \citenamefont {Meixner}, \citenamefont {Hidaka},\
  and\ \citenamefont {Kajino}}]{Warren16}%
  \BibitemOpen
  \bibfield  {author} {\bibinfo {author} {\bibfnamefont {M.}~\bibnamefont
  {Warren}}, \bibinfo {author} {\bibfnamefont {G.}~\bibnamefont {Mathews}},
  \bibinfo {author} {\bibfnamefont {M.}~\bibnamefont {Meixner}}, \bibinfo
  {author} {\bibfnamefont {J.}~\bibnamefont {Hidaka}}, \ and\ \bibinfo {author}
  {\bibfnamefont {T.}~\bibnamefont {Kajino}},\ }\href@noop {} {\bibfield
  {journal} {\bibinfo  {journal} {Int.J.Mod.Phys.A}\ }\textbf {\bibinfo
  {volume} {31}} (\bibinfo {year} {2016})}\BibitemShut {NoStop}%
\bibitem [{\citenamefont {{Demorest}}\ \emph {et~al.}(2010)\citenamefont
  {{Demorest}}, \citenamefont {{Pennucci}}, \citenamefont {{Ransom}},
  \citenamefont {{Roberts}},\ and\ \citenamefont {{Hessels}}}]{Demorest}%
  \BibitemOpen
  \bibfield  {author} {\bibinfo {author} {\bibfnamefont {P.~B.}\ \bibnamefont
  {{Demorest}}}, \bibinfo {author} {\bibfnamefont {T.}~\bibnamefont
  {{Pennucci}}}, \bibinfo {author} {\bibfnamefont {S.~M.}\ \bibnamefont
  {{Ransom}}}, \bibinfo {author} {\bibfnamefont {M.~S.~E.}\ \bibnamefont
  {{Roberts}}}, \ and\ \bibinfo {author} {\bibfnamefont {J.~W.~T.}\
  \bibnamefont {{Hessels}}},\ }\href {\doibase 10.1038/nature09466} {\bibfield
  {journal} {\bibinfo  {journal} {\nat}\ }\textbf {\bibinfo {volume} {467}},\
  \bibinfo {pages} {1081} (\bibinfo {year} {2010})},\ \Eprint
  {http://arxiv.org/abs/1010.5788} {arXiv:1010.5788 [astro-ph.HE]} \BibitemShut
  {NoStop}%
\bibitem [{\citenamefont {{Antoniadis}}\ \emph {et~al.}(2013)\citenamefont
  {{Antoniadis}}, \citenamefont {{Freire}}, \citenamefont {{Wex}},
  \citenamefont {{Tauris}}, \citenamefont {{Lynch}}, \citenamefont {{van
  Kerkwijk}}, \citenamefont {{Kramer}}, \citenamefont {{Bassa}}, \citenamefont
  {{Dhillon}}, \citenamefont {{Driebe}}, \citenamefont {{Hessels}},
  \citenamefont {{Kaspi}}, \citenamefont {{Kondratiev}}, \citenamefont
  {{Langer}}, \citenamefont {{Marsh}}, \citenamefont {{McLaughlin}},
  \citenamefont {{Pennucci}}, \citenamefont {{Ransom}}, \citenamefont
  {{Stairs}}, \citenamefont {{van Leeuwen}}, \citenamefont {{Verbiest}},\ and\
  \citenamefont {{Whelan}}}]{Antoniadis}%
  \BibitemOpen
  \bibfield  {author} {\bibinfo {author} {\bibfnamefont {J.}~\bibnamefont
  {{Antoniadis}}}, \bibinfo {author} {\bibfnamefont {P.~C.~C.}\ \bibnamefont
  {{Freire}}}, \bibinfo {author} {\bibfnamefont {N.}~\bibnamefont {{Wex}}},
  \bibinfo {author} {\bibfnamefont {T.~M.}\ \bibnamefont {{Tauris}}}, \bibinfo
  {author} {\bibfnamefont {R.~S.}\ \bibnamefont {{Lynch}}}, \bibinfo {author}
  {\bibfnamefont {M.~H.}\ \bibnamefont {{van Kerkwijk}}}, \bibinfo {author}
  {\bibfnamefont {M.}~\bibnamefont {{Kramer}}}, \bibinfo {author}
  {\bibfnamefont {C.}~\bibnamefont {{Bassa}}}, \bibinfo {author} {\bibfnamefont
  {V.~S.}\ \bibnamefont {{Dhillon}}}, \bibinfo {author} {\bibfnamefont
  {T.}~\bibnamefont {{Driebe}}}, \bibinfo {author} {\bibfnamefont {J.~W.~T.}\
  \bibnamefont {{Hessels}}}, \bibinfo {author} {\bibfnamefont {V.~M.}\
  \bibnamefont {{Kaspi}}}, \bibinfo {author} {\bibfnamefont {V.~I.}\
  \bibnamefont {{Kondratiev}}}, \bibinfo {author} {\bibfnamefont
  {N.}~\bibnamefont {{Langer}}}, \bibinfo {author} {\bibfnamefont {T.~R.}\
  \bibnamefont {{Marsh}}}, \bibinfo {author} {\bibfnamefont {M.~A.}\
  \bibnamefont {{McLaughlin}}}, \bibinfo {author} {\bibfnamefont {T.~T.}\
  \bibnamefont {{Pennucci}}}, \bibinfo {author} {\bibfnamefont {S.~M.}\
  \bibnamefont {{Ransom}}}, \bibinfo {author} {\bibfnamefont {I.~H.}\
  \bibnamefont {{Stairs}}}, \bibinfo {author} {\bibfnamefont {J.}~\bibnamefont
  {{van Leeuwen}}}, \bibinfo {author} {\bibfnamefont {J.~P.~W.}\ \bibnamefont
  {{Verbiest}}}, \ and\ \bibinfo {author} {\bibfnamefont {D.~G.}\ \bibnamefont
  {{Whelan}}},\ }\href {\doibase 10.1126/science.1233232} {\bibfield  {journal}
  {\bibinfo  {journal} {Science}\ }\textbf {\bibinfo {volume} {340}},\ \bibinfo
  {pages} {448} (\bibinfo {year} {2013})},\ \Eprint
  {http://arxiv.org/abs/1304.6875} {arXiv:1304.6875 [astro-ph.HE]} \BibitemShut
  {NoStop}%
\bibitem [{\citenamefont {{Steiner}}\ \emph {et~al.}(2010)\citenamefont
  {{Steiner}}, \citenamefont {{Lattimer}},\ and\ \citenamefont
  {{Brown}}}]{steiner2010}%
  \BibitemOpen
  \bibfield  {author} {\bibinfo {author} {\bibfnamefont {A.~W.}\ \bibnamefont
  {{Steiner}}}, \bibinfo {author} {\bibfnamefont {J.~M.}\ \bibnamefont
  {{Lattimer}}}, \ and\ \bibinfo {author} {\bibfnamefont {E.~F.}\ \bibnamefont
  {{Brown}}},\ }\href {\doibase 10.1088/0004-637X/722/1/33} {\bibfield
  {journal} {\bibinfo  {journal} {Ap.J.}\ }\textbf {\bibinfo {volume} {722}},\
  \bibinfo {pages} {33} (\bibinfo {year} {2010})},\ \Eprint
  {http://arxiv.org/abs/1005.0811} {arXiv:1005.0811 [astro-ph.HE]} \BibitemShut
  {NoStop}%
\bibitem [{\citenamefont {{Bowers}}\ and\ \citenamefont
  {{Wilson}}(1982)}]{Bowers82}%
  \BibitemOpen
  \bibfield  {author} {\bibinfo {author} {\bibfnamefont {R.~L.}\ \bibnamefont
  {{Bowers}}}\ and\ \bibinfo {author} {\bibfnamefont {J.~R.}\ \bibnamefont
  {{Wilson}}},\ }\href {\doibase 10.1086/190822} {\bibfield  {journal}
  {\bibinfo  {journal} {Astrophysical Journal Supplement Series}\ }\textbf
  {\bibinfo {volume} {50}},\ \bibinfo {pages} {115} (\bibinfo {year}
  {1982})}\BibitemShut {NoStop}%
\bibitem [{\citenamefont {{Rikovska Stone}}\ \emph {et~al.}(2003)\citenamefont
  {{Rikovska Stone}}, \citenamefont {{Miller}}, \citenamefont {{Koncewicz}},
  \citenamefont {{Stevenson}},\ and\ \citenamefont {{Strayer}}}]{stone2003}%
  \BibitemOpen
  \bibfield  {author} {\bibinfo {author} {\bibfnamefont {J.}~\bibnamefont
  {{Rikovska Stone}}}, \bibinfo {author} {\bibfnamefont {J.~C.}\ \bibnamefont
  {{Miller}}}, \bibinfo {author} {\bibfnamefont {R.}~\bibnamefont
  {{Koncewicz}}}, \bibinfo {author} {\bibfnamefont {P.~D.}\ \bibnamefont
  {{Stevenson}}}, \ and\ \bibinfo {author} {\bibfnamefont {M.~R.}\ \bibnamefont
  {{Strayer}}},\ }\href {\doibase 10.1103/PhysRevC.68.034324} {\bibfield
  {journal} {\bibinfo  {journal} {Phys.Rev.C}\ }\textbf {\bibinfo {volume}
  {68}},\ \bibinfo {eid} {034324} (\bibinfo {year} {2003})}\BibitemShut
  {NoStop}%
\bibitem [{\citenamefont {Xu}\ \emph {et~al.}(2009)\citenamefont {Xu},
  \citenamefont {Chen}, \citenamefont {Li},\ and\ \citenamefont {Ma}}]{Xu09}%
  \BibitemOpen
  \bibfield  {author} {\bibinfo {author} {\bibfnamefont {J.}~\bibnamefont
  {Xu}}, \bibinfo {author} {\bibfnamefont {L.-W.}\ \bibnamefont {Chen}},
  \bibinfo {author} {\bibfnamefont {B.-A.}\ \bibnamefont {Li}}, \ and\ \bibinfo
  {author} {\bibfnamefont {H.-R.}\ \bibnamefont {Ma}},\ }\href
  {http://stacks.iop.org/0004-637X/697/i=2/a=1549} {\bibfield  {journal}
  {\bibinfo  {journal} {The Astrophysical Journal}\ }\textbf {\bibinfo {volume}
  {697}},\ \bibinfo {pages} {1549} (\bibinfo {year} {2009})}\BibitemShut
  {NoStop}%
\bibitem [{\citenamefont {{Dutra}}\ \emph {et~al.}(2012)\citenamefont
  {{Dutra}}, \citenamefont {{Louren{\c c}o}}, \citenamefont {{S{\'a} Martins}},
  \citenamefont {{Delfino}}, \citenamefont {{Stone}},\ and\ \citenamefont
  {{Stevenson}}}]{Dutra2012}%
  \BibitemOpen
  \bibfield  {author} {\bibinfo {author} {\bibfnamefont {M.}~\bibnamefont
  {{Dutra}}}, \bibinfo {author} {\bibfnamefont {O.}~\bibnamefont {{Louren{\c
  c}o}}}, \bibinfo {author} {\bibfnamefont {J.~S.}\ \bibnamefont {{S{\'a}
  Martins}}}, \bibinfo {author} {\bibfnamefont {A.}~\bibnamefont {{Delfino}}},
  \bibinfo {author} {\bibfnamefont {J.~R.}\ \bibnamefont {{Stone}}}, \ and\
  \bibinfo {author} {\bibfnamefont {P.~D.}\ \bibnamefont {{Stevenson}}},\
  }\href {\doibase 10.1103/PhysRevC.85.035201} {\bibfield  {journal} {\bibinfo
  {journal} {\prc}\ }\textbf {\bibinfo {volume} {85}},\ \bibinfo {eid} {035201}
  (\bibinfo {year} {2012})},\ \Eprint {http://arxiv.org/abs/1202.3902}
  {arXiv:1202.3902 [nucl-th]} \BibitemShut {NoStop}%
\bibitem [{\citenamefont {{J. Beringer et al.}}(2012)}]{Beringer2012}%
  \BibitemOpen
  \bibfield  {author} {\bibinfo {author} {\bibnamefont {{J. Beringer et
  al.}}},\ }\href {\doibase 10.1103/PhysRevD.86.010001} {\bibfield  {journal}
  {\bibinfo  {journal} {Phys. Rev. D}\ }\textbf {\bibinfo {volume} {86}},\
  \bibinfo {pages} {010001} (\bibinfo {year} {2012})}\BibitemShut {NoStop}%
\bibitem [{\citenamefont {{Mart{\'{\i}}nez-Pinedo}}\ \emph
  {et~al.}(2012)\citenamefont {{Mart{\'{\i}}nez-Pinedo}}, \citenamefont
  {{Fischer}}, \citenamefont {{Lohs}},\ and\ \citenamefont
  {{Huther}}}]{fischer2012}%
  \BibitemOpen
  \bibfield  {author} {\bibinfo {author} {\bibfnamefont {G.}~\bibnamefont
  {{Mart{\'{\i}}nez-Pinedo}}}, \bibinfo {author} {\bibfnamefont
  {T.}~\bibnamefont {{Fischer}}}, \bibinfo {author} {\bibfnamefont
  {A.}~\bibnamefont {{Lohs}}}, \ and\ \bibinfo {author} {\bibfnamefont
  {L.}~\bibnamefont {{Huther}}},\ }\href {\doibase
  10.1103/PhysRevLett.109.251104} {\bibfield  {journal} {\bibinfo  {journal}
  {Physical Review Letters}\ }\textbf {\bibinfo {volume} {109}},\ \bibinfo
  {eid} {251104} (\bibinfo {year} {2012})},\ \Eprint
  {http://arxiv.org/abs/1205.2793} {arXiv:1205.2793 [astro-ph.HE]} \BibitemShut
  {NoStop}%
\bibitem [{\citenamefont {Bethe}\ \emph {et~al.}(1979)\citenamefont {Bethe},
  \citenamefont {Brown}, \citenamefont {Applegate},\ and\ \citenamefont
  {Lattimer}}]{bethe1979}%
  \BibitemOpen
  \bibfield  {author} {\bibinfo {author} {\bibfnamefont {H.}~\bibnamefont
  {Bethe}}, \bibinfo {author} {\bibfnamefont {G.}~\bibnamefont {Brown}},
  \bibinfo {author} {\bibfnamefont {J.}~\bibnamefont {Applegate}}, \ and\
  \bibinfo {author} {\bibfnamefont {J.}~\bibnamefont {Lattimer}},\ }\href@noop
  {} {\bibfield  {journal} {\bibinfo  {journal} {Nuclear Physics A}\ }\textbf
  {\bibinfo {volume} {324}},\ \bibinfo {pages} {487} (\bibinfo {year}
  {1979})}\BibitemShut {NoStop}%
\bibitem [{\citenamefont {{Ravenhall}}\ \emph {et~al.}(1983)\citenamefont
  {{Ravenhall}}, \citenamefont {{Pethick}},\ and\ \citenamefont
  {{Wilson}}}]{RavenhallPethickWilson1983}%
  \BibitemOpen
  \bibfield  {author} {\bibinfo {author} {\bibfnamefont {D.~G.}\ \bibnamefont
  {{Ravenhall}}}, \bibinfo {author} {\bibfnamefont {C.~J.}\ \bibnamefont
  {{Pethick}}}, \ and\ \bibinfo {author} {\bibfnamefont {J.~R.}\ \bibnamefont
  {{Wilson}}},\ }\href {\doibase 10.1103/PhysRevLett.50.2066} {\bibfield
  {journal} {\bibinfo  {journal} {Physical Review Letters}\ }\textbf {\bibinfo
  {volume} {50}},\ \bibinfo {pages} {2066} (\bibinfo {year}
  {1983})}\BibitemShut {NoStop}%
\bibitem [{\citenamefont {{Horowitz}}\ \emph {et~al.}(2015)\citenamefont
  {{Horowitz}}, \citenamefont {{Berry}}, \citenamefont {{Briggs}},
  \citenamefont {{Caplan}}, \citenamefont {{Cumming}},\ and\ \citenamefont
  {{Schneider}}}]{horowitz2015}%
  \BibitemOpen
  \bibfield  {author} {\bibinfo {author} {\bibfnamefont {C.~J.}\ \bibnamefont
  {{Horowitz}}}, \bibinfo {author} {\bibfnamefont {D.~K.}\ \bibnamefont
  {{Berry}}}, \bibinfo {author} {\bibfnamefont {C.~M.}\ \bibnamefont
  {{Briggs}}}, \bibinfo {author} {\bibfnamefont {M.~E.}\ \bibnamefont
  {{Caplan}}}, \bibinfo {author} {\bibfnamefont {A.}~\bibnamefont {{Cumming}}},
  \ and\ \bibinfo {author} {\bibfnamefont {A.~S.}\ \bibnamefont
  {{Schneider}}},\ }\href {\doibase 10.1103/PhysRevLett.114.031102} {\bibfield
  {journal} {\bibinfo  {journal} {Physical Review Letters}\ }\textbf {\bibinfo
  {volume} {114}},\ \bibinfo {eid} {031102} (\bibinfo {year} {2015})},\ \Eprint
  {http://arxiv.org/abs/1410.2197} {arXiv:1410.2197 [astro-ph.HE]} \BibitemShut
  {NoStop}%
\bibitem [{\citenamefont {{Schneider}}\ \emph {et~al.}(2016)\citenamefont
  {{Schneider}}, \citenamefont {{Berry}}, \citenamefont {{Caplan}},
  \citenamefont {{Horowitz}},\ and\ \citenamefont {{Lin}}}]{schneider2016}%
  \BibitemOpen
  \bibfield  {author} {\bibinfo {author} {\bibfnamefont {A.~S.}\ \bibnamefont
  {{Schneider}}}, \bibinfo {author} {\bibfnamefont {D.~K.}\ \bibnamefont
  {{Berry}}}, \bibinfo {author} {\bibfnamefont {M.~E.}\ \bibnamefont
  {{Caplan}}}, \bibinfo {author} {\bibfnamefont {C.~J.}\ \bibnamefont
  {{Horowitz}}}, \ and\ \bibinfo {author} {\bibfnamefont {Z.}~\bibnamefont
  {{Lin}}},\ }\href@noop {} {\bibfield  {journal} {\bibinfo  {journal} {ArXiv
  e-prints}\ } (\bibinfo {year} {2016})},\ \Eprint
  {http://arxiv.org/abs/1602.03215} {arXiv:1602.03215 [nucl-th]} \BibitemShut
  {NoStop}%
\bibitem [{\citenamefont {{Lamb}}\ \emph {et~al.}(1978)\citenamefont {{Lamb}},
  \citenamefont {{Lattimer}}, \citenamefont {{Pethick}},\ and\ \citenamefont
  {{Ravenhall}}}]{PastaRef1}%
  \BibitemOpen
  \bibfield  {author} {\bibinfo {author} {\bibfnamefont {D.~Q.}\ \bibnamefont
  {{Lamb}}}, \bibinfo {author} {\bibfnamefont {J.~M.}\ \bibnamefont
  {{Lattimer}}}, \bibinfo {author} {\bibfnamefont {C.~J.}\ \bibnamefont
  {{Pethick}}}, \ and\ \bibinfo {author} {\bibfnamefont {D.~G.}\ \bibnamefont
  {{Ravenhall}}},\ }\href {\doibase 10.1103/PhysRevLett.41.1623} {\bibfield
  {journal} {\bibinfo  {journal} {Physical Review Letters}\ }\textbf {\bibinfo
  {volume} {41}},\ \bibinfo {pages} {1623} (\bibinfo {year}
  {1978})}\BibitemShut {NoStop}%
\bibitem [{\citenamefont {{Brink}}\ and\ \citenamefont
  {{Boeker}}(1967)}]{Brink}%
  \BibitemOpen
  \bibfield  {author} {\bibinfo {author} {\bibfnamefont {D.~M.}\ \bibnamefont
  {{Brink}}}\ and\ \bibinfo {author} {\bibfnamefont {E.}~\bibnamefont
  {{Boeker}}},\ }\href {\doibase 10.1016/0375-9474(67)90446-0} {\bibfield
  {journal} {\bibinfo  {journal} {Nuclear Physics A}\ }\textbf {\bibinfo
  {volume} {91}},\ \bibinfo {pages} {1} (\bibinfo {year} {1967})}\BibitemShut
  {NoStop}%
\bibitem [{\citenamefont {{Moszkowski}}(1970)}]{Moszkowski}%
  \BibitemOpen
  \bibfield  {author} {\bibinfo {author} {\bibfnamefont {S.~A.}\ \bibnamefont
  {{Moszkowski}}},\ }\href {\doibase 10.1103/PhysRevC.2.402} {\bibfield
  {journal} {\bibinfo  {journal} {\prc}\ }\textbf {\bibinfo {volume} {2}},\
  \bibinfo {pages} {402} (\bibinfo {year} {1970})}\BibitemShut {NoStop}%
\bibitem [{\citenamefont {{Skyrme}}(1956)}]{Skyrme1956}%
  \BibitemOpen
  \bibfield  {author} {\bibinfo {author} {\bibfnamefont {T.~H.~R.}\
  \bibnamefont {{Skyrme}}},\ }\href {\doibase 10.1080/14786435608238186}
  {\bibfield  {journal} {\bibinfo  {journal} {Philosophical Magazine}\ }\textbf
  {\bibinfo {volume} {1}},\ \bibinfo {pages} {1043} (\bibinfo {year}
  {1956})}\BibitemShut {NoStop}%
\bibitem [{\citenamefont {{Vautherin}}\ and\ \citenamefont
  {{Brink}}(1972)}]{Vautherin72}%
  \BibitemOpen
  \bibfield  {author} {\bibinfo {author} {\bibfnamefont {D.}~\bibnamefont
  {{Vautherin}}}\ and\ \bibinfo {author} {\bibfnamefont {D.~M.}\ \bibnamefont
  {{Brink}}},\ }\href {\doibase 10.1103/PhysRevC.5.626} {\bibfield  {journal}
  {\bibinfo  {journal} {\prc}\ }\textbf {\bibinfo {volume} {5}},\ \bibinfo
  {pages} {626} (\bibinfo {year} {1972})}\BibitemShut {NoStop}%
\bibitem [{\citenamefont {Ring}\ and\ \citenamefont {Schuck}(2000)}]{Ring}%
  \BibitemOpen
  \bibfield  {author} {\bibinfo {author} {\bibfnamefont {P.}~\bibnamefont
  {Ring}}\ and\ \bibinfo {author} {\bibfnamefont {P.}~\bibnamefont {Schuck}},\
  }\href@noop {} {\emph {\bibinfo {title} {The nuclear many-body problem}}}\
  (\bibinfo  {publisher} {Springer},\ \bibinfo {year} {2000})\BibitemShut
  {NoStop}%
\bibitem [{\citenamefont {{Mansour}}(1990)}]{Mansour}%
  \BibitemOpen
  \bibfield  {author} {\bibinfo {author} {\bibfnamefont {H.~M.~M.}\
  \bibnamefont {{Mansour}}},\ }\href {\doibase 10.1016/0375-9474(89)90002-X}
  {\bibfield  {journal} {\bibinfo  {journal} {Acta Physica Polonica}\ }\textbf
  {\bibinfo {volume} {B21}},\ \bibinfo {pages} {741} (\bibinfo {year}
  {1990})}\BibitemShut {NoStop}%
\bibitem [{\citenamefont {{K{\"o}hler}}(1965)}]{Kohler}%
  \BibitemOpen
  \bibfield  {author} {\bibinfo {author} {\bibfnamefont {H.~S.}\ \bibnamefont
  {{K{\"o}hler}}},\ }\href {\doibase 10.1103/PhysRev.138.B831} {\bibfield
  {journal} {\bibinfo  {journal} {Physical Review}\ }\textbf {\bibinfo {volume}
  {138}},\ \bibinfo {pages} {831} (\bibinfo {year} {1965})}\BibitemShut
  {NoStop}%
\bibitem [{\citenamefont {{Krivine}}\ \emph {et~al.}(1980)\citenamefont
  {{Krivine}}, \citenamefont {{Treiner}},\ and\ \citenamefont
  {{Bohigas}}}]{Krivine}%
  \BibitemOpen
  \bibfield  {author} {\bibinfo {author} {\bibfnamefont {H.}~\bibnamefont
  {{Krivine}}}, \bibinfo {author} {\bibfnamefont {J.}~\bibnamefont
  {{Treiner}}}, \ and\ \bibinfo {author} {\bibfnamefont {O.}~\bibnamefont
  {{Bohigas}}},\ }\href {\doibase 10.1016/0375-9474(80)90618-1} {\bibfield
  {journal} {\bibinfo  {journal} {Nuclear Physics A}\ }\textbf {\bibinfo
  {volume} {336}},\ \bibinfo {pages} {155} (\bibinfo {year}
  {1980})}\BibitemShut {NoStop}%
\bibitem [{\citenamefont {{Newton}}\ \emph {et~al.}(2013)\citenamefont
  {{Newton}}, \citenamefont {{Gearheart}},\ and\ \citenamefont
  {{Li}}}]{Newton2013}%
  \BibitemOpen
  \bibfield  {author} {\bibinfo {author} {\bibfnamefont {W.~G.}\ \bibnamefont
  {{Newton}}}, \bibinfo {author} {\bibfnamefont {M.}~\bibnamefont
  {{Gearheart}}}, \ and\ \bibinfo {author} {\bibfnamefont {B.-A.}\ \bibnamefont
  {{Li}}},\ }\href {\doibase 10.1088/0067-0049/204/1/9} {\bibfield  {journal}
  {\bibinfo  {journal} {The Astrophysical Journal Supplement}\ }\textbf
  {\bibinfo {volume} {204}},\ \bibinfo {eid} {9} (\bibinfo {year} {2013})},\
  \Eprint {http://arxiv.org/abs/1110.4043} {arXiv:1110.4043 [astro-ph.SR]}
  \BibitemShut {NoStop}%
\bibitem [{\citenamefont {{Mayle}}\ \emph {et~al.}(1993)\citenamefont
  {{Mayle}}, \citenamefont {{Tavani}},\ and\ \citenamefont {{Wilson}}}]{Mayle}%
  \BibitemOpen
  \bibfield  {author} {\bibinfo {author} {\bibfnamefont {R.~W.}\ \bibnamefont
  {{Mayle}}}, \bibinfo {author} {\bibfnamefont {M.}~\bibnamefont {{Tavani}}}, \
  and\ \bibinfo {author} {\bibfnamefont {J.~R.}\ \bibnamefont {{Wilson}}},\
  }\href {\doibase 10.1086/173400} {\bibfield  {journal} {\bibinfo  {journal}
  {\apj}\ }\textbf {\bibinfo {volume} {418}},\ \bibinfo {pages} {398} (\bibinfo
  {year} {1993})}\BibitemShut {NoStop}%
\bibitem [{\citenamefont {{Migdal}}(1978)}]{Migdal}%
  \BibitemOpen
  \bibfield  {author} {\bibinfo {author} {\bibfnamefont {A.~B.}\ \bibnamefont
  {{Migdal}}},\ }\href {\doibase 10.1103/RevModPhys.50.107} {\bibfield
  {journal} {\bibinfo  {journal} {Reviews of Modern Physics}\ }\textbf
  {\bibinfo {volume} {50}},\ \bibinfo {pages} {107} (\bibinfo {year}
  {1978})}\BibitemShut {NoStop}%
\bibitem [{\citenamefont {{Lattimer}}(2012)}]{Lattimer2012}%
  \BibitemOpen
  \bibfield  {author} {\bibinfo {author} {\bibfnamefont {J.~M.}\ \bibnamefont
  {{Lattimer}}},\ }\href {\doibase 10.1146/annurev-nucl-102711-095018}
  {\bibfield  {journal} {\bibinfo  {journal} {Annual Review of Nuclear and
  Particle Science}\ }\textbf {\bibinfo {volume} {62}},\ \bibinfo {pages} {485}
  (\bibinfo {year} {2012})}\BibitemShut {NoStop}%
\bibitem [{\citenamefont {{Stone}}\ and\ \citenamefont
  {{Reinhard}}(2007)}]{stone2007}%
  \BibitemOpen
  \bibfield  {author} {\bibinfo {author} {\bibfnamefont {J.~R.}\ \bibnamefont
  {{Stone}}}\ and\ \bibinfo {author} {\bibfnamefont {P.-G.}\ \bibnamefont
  {{Reinhard}}},\ }\href {\doibase 10.1016/j.ppnp.2006.07.001} {\bibfield
  {journal} {\bibinfo  {journal} {Progress in Particle and Nuclear Physics}\
  }\textbf {\bibinfo {volume} {58}},\ \bibinfo {pages} {587} (\bibinfo {year}
  {2007})},\ \Eprint {http://arxiv.org/abs/nucl-th/0607002} {nucl-th/0607002}
  \BibitemShut {NoStop}%
\bibitem [{\citenamefont {{Danielewicz}}\ and\ \citenamefont
  {{Lee}}(2014)}]{danielewicz2013}%
  \BibitemOpen
  \bibfield  {author} {\bibinfo {author} {\bibfnamefont {P.}~\bibnamefont
  {{Danielewicz}}}\ and\ \bibinfo {author} {\bibfnamefont {J.}~\bibnamefont
  {{Lee}}},\ }\href {\doibase 10.1016/j.nuclphysa.2013.11.005} {\bibfield
  {journal} {\bibinfo  {journal} {Nuclear Physics A}\ }\textbf {\bibinfo
  {volume} {922}},\ \bibinfo {pages} {1} (\bibinfo {year} {2014})},\ \Eprint
  {http://arxiv.org/abs/1307.4130} {arXiv:1307.4130 [nucl-th]} \BibitemShut
  {NoStop}%
\bibitem [{\citenamefont {{McAbee}}\ and\ \citenamefont
  {{Wilson}}(1994)}]{McAbee}%
  \BibitemOpen
  \bibfield  {author} {\bibinfo {author} {\bibfnamefont {T.~L.}\ \bibnamefont
  {{McAbee}}}\ and\ \bibinfo {author} {\bibfnamefont {J.~R.}\ \bibnamefont
  {{Wilson}}},\ }\href {\doibase 10.1016/0375-9474(94)90747-1} {\bibfield
  {journal} {\bibinfo  {journal} {Nuclear Physics A}\ }\textbf {\bibinfo
  {volume} {576}},\ \bibinfo {pages} {626} (\bibinfo {year}
  {1994})}\BibitemShut {NoStop}%
\bibitem [{\citenamefont {{Friedman}}\ \emph {et~al.}(1981)\citenamefont
  {{Friedman}}, \citenamefont {{Pandharipande}},\ and\ \citenamefont
  {{Usmani}}}]{Friedman}%
  \BibitemOpen
  \bibfield  {author} {\bibinfo {author} {\bibfnamefont {B.}~\bibnamefont
  {{Friedman}}}, \bibinfo {author} {\bibfnamefont {V.~R.}\ \bibnamefont
  {{Pandharipande}}}, \ and\ \bibinfo {author} {\bibfnamefont {Q.~N.}\
  \bibnamefont {{Usmani}}},\ }\href {\doibase 10.1016/0375-9474(81)90048-8}
  {\bibfield  {journal} {\bibinfo  {journal} {Nuclear Physics A}\ }\textbf
  {\bibinfo {volume} {372}},\ \bibinfo {pages} {483} (\bibinfo {year}
  {1981})}\BibitemShut {NoStop}%
\bibitem [{\citenamefont {{Ericson}}\ and\ \citenamefont
  {{Delorme}}(1978)}]{Ericson}%
  \BibitemOpen
  \bibfield  {author} {\bibinfo {author} {\bibfnamefont {M.}~\bibnamefont
  {{Ericson}}}\ and\ \bibinfo {author} {\bibfnamefont {J.}~\bibnamefont
  {{Delorme}}},\ }\href {\doibase 10.1016/0370-2693(78)90270-8} {\bibfield
  {journal} {\bibinfo  {journal} {Physics Letters B}\ }\textbf {\bibinfo
  {volume} {76}},\ \bibinfo {pages} {182} (\bibinfo {year} {1978})}\BibitemShut
  {NoStop}%
\bibitem [{\citenamefont {{Harris}}\ \emph {et~al.}(1987)\citenamefont
  {{Harris}}, \citenamefont {{Odyniec}}, \citenamefont {{Pugh}}, \citenamefont
  {{Schroeder}}, \citenamefont {{Tincknell}}, \citenamefont {{Rauch}},
  \citenamefont {{Stock}}, \citenamefont {{Bock}}, \citenamefont {{Brockmann}},
  \citenamefont {{Sandoval}}, \citenamefont {{Str{\"o}bele}}, \citenamefont
  {{Renfordt}}, \citenamefont {{Schall}}, \citenamefont {{Bangert}},
  \citenamefont {{Sullivan}},\ and\ \citenamefont {{et al.}}}]{Harris}%
  \BibitemOpen
  \bibfield  {author} {\bibinfo {author} {\bibfnamefont {J.~W.}\ \bibnamefont
  {{Harris}}}, \bibinfo {author} {\bibfnamefont {G.}~\bibnamefont {{Odyniec}}},
  \bibinfo {author} {\bibfnamefont {H.~G.}\ \bibnamefont {{Pugh}}}, \bibinfo
  {author} {\bibfnamefont {L.~S.}\ \bibnamefont {{Schroeder}}}, \bibinfo
  {author} {\bibfnamefont {M.~L.}\ \bibnamefont {{Tincknell}}}, \bibinfo
  {author} {\bibfnamefont {W.}~\bibnamefont {{Rauch}}}, \bibinfo {author}
  {\bibfnamefont {R.}~\bibnamefont {{Stock}}}, \bibinfo {author} {\bibfnamefont
  {R.}~\bibnamefont {{Bock}}}, \bibinfo {author} {\bibfnamefont
  {R.}~\bibnamefont {{Brockmann}}}, \bibinfo {author} {\bibfnamefont
  {A.}~\bibnamefont {{Sandoval}}}, \bibinfo {author} {\bibfnamefont
  {H.}~\bibnamefont {{Str{\"o}bele}}}, \bibinfo {author} {\bibfnamefont
  {R.~E.}\ \bibnamefont {{Renfordt}}}, \bibinfo {author} {\bibfnamefont
  {D.}~\bibnamefont {{Schall}}}, \bibinfo {author} {\bibfnamefont
  {D.}~\bibnamefont {{Bangert}}}, \bibinfo {author} {\bibfnamefont {J.~P.}\
  \bibnamefont {{Sullivan}}}, \ and\ \bibinfo {author} {\bibnamefont {{et
  al.}}},\ }\href {\doibase 10.1103/PhysRevLett.58.463} {\bibfield  {journal}
  {\bibinfo  {journal} {Physical Review Letters}\ }\textbf {\bibinfo {volume}
  {58}},\ \bibinfo {pages} {463} (\bibinfo {year} {1987})}\BibitemShut
  {NoStop}%
\bibitem [{\citenamefont {{M{\"u}ller}}(1997)}]{mueller1997}%
  \BibitemOpen
  \bibfield  {author} {\bibinfo {author} {\bibfnamefont {H.}~\bibnamefont
  {{M{\"u}ller}}},\ }\href {\doibase 10.1016/S0375-9474(97)00018-3} {\bibfield
  {journal} {\bibinfo  {journal} {Nuclear Physics A}\ }\textbf {\bibinfo
  {volume} {618}},\ \bibinfo {pages} {349} (\bibinfo {year} {1997})},\ \Eprint
  {http://arxiv.org/abs/nucl-th/9701035} {nucl-th/9701035} \BibitemShut
  {NoStop}%
\bibitem [{\citenamefont {{Cavagnoli}}\ \emph {et~al.}(2011)\citenamefont
  {{Cavagnoli}}, \citenamefont {{Provid{\^e}ncia}},\ and\ \citenamefont
  {{Menezes}}}]{cavagnoli2011}%
  \BibitemOpen
  \bibfield  {author} {\bibinfo {author} {\bibfnamefont {R.}~\bibnamefont
  {{Cavagnoli}}}, \bibinfo {author} {\bibfnamefont {C.}~\bibnamefont
  {{Provid{\^e}ncia}}}, \ and\ \bibinfo {author} {\bibfnamefont {D.~P.}\
  \bibnamefont {{Menezes}}},\ }\href {\doibase 10.1103/PhysRevC.83.045201}
  {\bibfield  {journal} {\bibinfo  {journal} {\prc}\ }\textbf {\bibinfo
  {volume} {83}},\ \bibinfo {eid} {045201} (\bibinfo {year} {2011})},\ \Eprint
  {http://arxiv.org/abs/1009.3596} {arXiv:1009.3596 [nucl-th]} \BibitemShut
  {NoStop}%
\bibitem [{\citenamefont {{McLerran}}(1986)}]{McLerran}%
  \BibitemOpen
  \bibfield  {author} {\bibinfo {author} {\bibfnamefont {L.}~\bibnamefont
  {{McLerran}}},\ }\href {\doibase 10.1103/RevModPhys.58.1021} {\bibfield
  {journal} {\bibinfo  {journal} {Reviews of Modern Physics}\ }\textbf
  {\bibinfo {volume} {58}},\ \bibinfo {pages} {1021} (\bibinfo {year}
  {1986})}\BibitemShut {NoStop}%
\bibitem [{\citenamefont {{Kronfeld}}(2012)}]{Kronfeld}%
  \BibitemOpen
  \bibfield  {author} {\bibinfo {author} {\bibfnamefont {A.~S.}\ \bibnamefont
  {{Kronfeld}}},\ }\href {\doibase 10.1146/annurev-nucl-102711-094942}
  {\bibfield  {journal} {\bibinfo  {journal} {Annual Review of Nuclear and
  Particle Science}\ }\textbf {\bibinfo {volume} {62}},\ \bibinfo {pages} {265}
  (\bibinfo {year} {2012})},\ \Eprint {http://arxiv.org/abs/1203.1204}
  {arXiv:1203.1204 [hep-lat]} \BibitemShut {NoStop}%
\bibitem [{\citenamefont {{Bors{\'a}nyi}}\ \emph {et~al.}(2012)\citenamefont
  {{Bors{\'a}nyi}}, \citenamefont {{D{\"u}rr}}, \citenamefont {{Fodor}},
  \citenamefont {{Hoelbling}}, \citenamefont {{Katz}}, \citenamefont {{Krieg}},
  \citenamefont {{N{\'o}gr{\'a}di}}, \citenamefont {{Szab{\'o}}}, \citenamefont
  {{T{\'o}th}},\ and\ \citenamefont {{Trombit{\'a}s}}}]{Borsanyi}%
  \BibitemOpen
  \bibfield  {author} {\bibinfo {author} {\bibfnamefont {S.}~\bibnamefont
  {{Bors{\'a}nyi}}}, \bibinfo {author} {\bibfnamefont {S.}~\bibnamefont
  {{D{\"u}rr}}}, \bibinfo {author} {\bibfnamefont {Z.}~\bibnamefont {{Fodor}}},
  \bibinfo {author} {\bibfnamefont {C.}~\bibnamefont {{Hoelbling}}}, \bibinfo
  {author} {\bibfnamefont {S.~D.}\ \bibnamefont {{Katz}}}, \bibinfo {author}
  {\bibfnamefont {S.}~\bibnamefont {{Krieg}}}, \bibinfo {author} {\bibfnamefont
  {D.}~\bibnamefont {{N{\'o}gr{\'a}di}}}, \bibinfo {author} {\bibfnamefont
  {K.~K.}\ \bibnamefont {{Szab{\'o}}}}, \bibinfo {author} {\bibfnamefont
  {B.~C.}\ \bibnamefont {{T{\'o}th}}}, \ and\ \bibinfo {author} {\bibfnamefont
  {N.}~\bibnamefont {{Trombit{\'a}s}}},\ }\href {\doibase
  10.1007/JHEP08(2012)126} {\bibfield  {journal} {\bibinfo  {journal} {Journal
  of High Energy Physics}\ }\textbf {\bibinfo {volume} {8}},\ \bibinfo {pages}
  {126} (\bibinfo {year} {2012})},\ \Eprint {http://arxiv.org/abs/1205.0440}
  {arXiv:1205.0440 [hep-lat]} \BibitemShut {NoStop}%
\bibitem [{\citenamefont {{Bazavov}}\ \emph {et~al.}(2012)\citenamefont
  {{Bazavov}}, \citenamefont {{Bhattacharya}}, \citenamefont {{Buchoff}},
  \citenamefont {{Cheng}}, \citenamefont {{Christ}}, \citenamefont {{Ding}},
  \citenamefont {{Gupta}}, \citenamefont {{Hegde}}, \citenamefont {{Jung}},
  \citenamefont {{Karsch}}, \citenamefont {{Lin}}, \citenamefont {{Mawhinney}},
  \citenamefont {{Mukherjee}}, \citenamefont {{Petreczky}}, \citenamefont
  {{Soltz}}, \citenamefont {{Vranas}},\ and\ \citenamefont {{Yin}}}]{Bazavov1}%
  \BibitemOpen
  \bibfield  {author} {\bibinfo {author} {\bibfnamefont {A.}~\bibnamefont
  {{Bazavov}}}, \bibinfo {author} {\bibfnamefont {T.}~\bibnamefont
  {{Bhattacharya}}}, \bibinfo {author} {\bibfnamefont {M.~I.}\ \bibnamefont
  {{Buchoff}}}, \bibinfo {author} {\bibfnamefont {M.}~\bibnamefont {{Cheng}}},
  \bibinfo {author} {\bibfnamefont {N.~H.}\ \bibnamefont {{Christ}}}, \bibinfo
  {author} {\bibfnamefont {H.-T.}\ \bibnamefont {{Ding}}}, \bibinfo {author}
  {\bibfnamefont {R.}~\bibnamefont {{Gupta}}}, \bibinfo {author} {\bibfnamefont
  {P.}~\bibnamefont {{Hegde}}}, \bibinfo {author} {\bibfnamefont
  {C.}~\bibnamefont {{Jung}}}, \bibinfo {author} {\bibfnamefont
  {F.}~\bibnamefont {{Karsch}}}, \bibinfo {author} {\bibfnamefont
  {Z.}~\bibnamefont {{Lin}}}, \bibinfo {author} {\bibfnamefont {R.~D.}\
  \bibnamefont {{Mawhinney}}}, \bibinfo {author} {\bibfnamefont
  {S.}~\bibnamefont {{Mukherjee}}}, \bibinfo {author} {\bibfnamefont
  {P.}~\bibnamefont {{Petreczky}}}, \bibinfo {author} {\bibfnamefont {R.~A.}\
  \bibnamefont {{Soltz}}}, \bibinfo {author} {\bibfnamefont {P.~M.}\
  \bibnamefont {{Vranas}}}, \ and\ \bibinfo {author} {\bibfnamefont
  {H.}~\bibnamefont {{Yin}}},\ }\href {\doibase 10.1103/PhysRevD.86.094503}
  {\bibfield  {journal} {\bibinfo  {journal} {\prd}\ }\textbf {\bibinfo
  {volume} {86}},\ \bibinfo {eid} {094503} (\bibinfo {year} {2012})},\ \Eprint
  {http://arxiv.org/abs/1205.3535} {arXiv:1205.3535 [hep-lat]} \BibitemShut
  {NoStop}%
\bibitem [{\citenamefont {{Aoki}}\ \emph {et~al.}(2006)\citenamefont {{Aoki}},
  \citenamefont {{Endr{\H o}di}}, \citenamefont {{Fodor}}, \citenamefont
  {{Katz}},\ and\ \citenamefont {{Szab{\'o}}}}]{Aoki}%
  \BibitemOpen
  \bibfield  {author} {\bibinfo {author} {\bibfnamefont {Y.}~\bibnamefont
  {{Aoki}}}, \bibinfo {author} {\bibfnamefont {G.}~\bibnamefont {{Endr{\H
  o}di}}}, \bibinfo {author} {\bibfnamefont {Z.}~\bibnamefont {{Fodor}}},
  \bibinfo {author} {\bibfnamefont {S.~D.}\ \bibnamefont {{Katz}}}, \ and\
  \bibinfo {author} {\bibfnamefont {K.~K.}\ \bibnamefont {{Szab{\'o}}}},\
  }\href {\doibase 10.1038/nature05120} {\bibfield  {journal} {\bibinfo
  {journal} {\nat}\ }\textbf {\bibinfo {volume} {443}},\ \bibinfo {pages} {675}
  (\bibinfo {year} {2006})},\ \Eprint
  {http://arxiv.org/abs/arXiv:hep-lat/0611014} {arXiv:hep-lat/0611014}
  \BibitemShut {NoStop}%
\bibitem [{\citenamefont {{Bazavov}}\ \emph {et~al.}(2009)\citenamefont
  {{Bazavov}}, \citenamefont {{Bhattacharya}}, \citenamefont {{Cheng}},
  \citenamefont {{Christ}}, \citenamefont {{DeTar}}, \citenamefont {{Ejiri}},
  \citenamefont {{Gottlieb}}, \citenamefont {{Gupta}}, \citenamefont
  {{Heller}}, \citenamefont {{Huebner}}, \citenamefont {{Jung}}, \citenamefont
  {{Karsch}}, \citenamefont {{Laermann}}, \citenamefont {{Levkova}},
  \citenamefont {{Miao}}, \citenamefont {{Mawhinney}}, \citenamefont
  {{Petreczky}}, \citenamefont {{Schmidt}}, \citenamefont {{Soltz}},
  \citenamefont {{Soeldner}}, \citenamefont {{Sugar}}, \citenamefont
  {{Toussaint}},\ and\ \citenamefont {{Vranas}}}]{Bazavov2}%
  \BibitemOpen
  \bibfield  {author} {\bibinfo {author} {\bibfnamefont {A.}~\bibnamefont
  {{Bazavov}}}, \bibinfo {author} {\bibfnamefont {T.}~\bibnamefont
  {{Bhattacharya}}}, \bibinfo {author} {\bibfnamefont {M.}~\bibnamefont
  {{Cheng}}}, \bibinfo {author} {\bibfnamefont {N.~H.}\ \bibnamefont
  {{Christ}}}, \bibinfo {author} {\bibfnamefont {C.}~\bibnamefont {{DeTar}}},
  \bibinfo {author} {\bibfnamefont {S.}~\bibnamefont {{Ejiri}}}, \bibinfo
  {author} {\bibfnamefont {S.}~\bibnamefont {{Gottlieb}}}, \bibinfo {author}
  {\bibfnamefont {R.}~\bibnamefont {{Gupta}}}, \bibinfo {author} {\bibfnamefont
  {U.~M.}\ \bibnamefont {{Heller}}}, \bibinfo {author} {\bibfnamefont
  {K.}~\bibnamefont {{Huebner}}}, \bibinfo {author} {\bibfnamefont
  {C.}~\bibnamefont {{Jung}}}, \bibinfo {author} {\bibfnamefont
  {F.}~\bibnamefont {{Karsch}}}, \bibinfo {author} {\bibfnamefont
  {E.}~\bibnamefont {{Laermann}}}, \bibinfo {author} {\bibfnamefont
  {L.}~\bibnamefont {{Levkova}}}, \bibinfo {author} {\bibfnamefont
  {C.}~\bibnamefont {{Miao}}}, \bibinfo {author} {\bibfnamefont {R.~D.}\
  \bibnamefont {{Mawhinney}}}, \bibinfo {author} {\bibfnamefont
  {P.}~\bibnamefont {{Petreczky}}}, \bibinfo {author} {\bibfnamefont
  {C.}~\bibnamefont {{Schmidt}}}, \bibinfo {author} {\bibfnamefont {R.~A.}\
  \bibnamefont {{Soltz}}}, \bibinfo {author} {\bibfnamefont {W.}~\bibnamefont
  {{Soeldner}}}, \bibinfo {author} {\bibfnamefont {R.}~\bibnamefont {{Sugar}}},
  \bibinfo {author} {\bibfnamefont {D.}~\bibnamefont {{Toussaint}}}, \ and\
  \bibinfo {author} {\bibfnamefont {P.}~\bibnamefont {{Vranas}}},\ }\href
  {\doibase 10.1103/PhysRevD.80.014504} {\bibfield  {journal} {\bibinfo
  {journal} {\prd}\ }\textbf {\bibinfo {volume} {80}},\ \bibinfo {eid} {014504}
  (\bibinfo {year} {2009})},\ \Eprint {http://arxiv.org/abs/0903.4379}
  {arXiv:0903.4379 [hep-lat]} \BibitemShut {NoStop}%
\bibitem [{\citenamefont {{Fuller}}\ \emph {et~al.}(1988)\citenamefont
  {{Fuller}}, \citenamefont {{Mathews}},\ and\ \citenamefont
  {{Alcock}}}]{Fuller}%
  \BibitemOpen
  \bibfield  {author} {\bibinfo {author} {\bibfnamefont {G.~M.}\ \bibnamefont
  {{Fuller}}}, \bibinfo {author} {\bibfnamefont {G.~J.}\ \bibnamefont
  {{Mathews}}}, \ and\ \bibinfo {author} {\bibfnamefont {C.~R.}\ \bibnamefont
  {{Alcock}}},\ }\href {\doibase 10.1103/PhysRevD.37.1380} {\bibfield
  {journal} {\bibinfo  {journal} {\prd}\ }\textbf {\bibinfo {volume} {37}},\
  \bibinfo {pages} {1380} (\bibinfo {year} {1988})}\BibitemShut {NoStop}%
\bibitem [{\citenamefont {{Gentile}}\ \emph {et~al.}(1993)\citenamefont
  {{Gentile}}, \citenamefont {{Aufderheide}}, \citenamefont {{Mathews}},
  \citenamefont {{Swesty}},\ and\ \citenamefont {{Fuller}}}]{Gentile}%
  \BibitemOpen
  \bibfield  {author} {\bibinfo {author} {\bibfnamefont {N.~A.}\ \bibnamefont
  {{Gentile}}}, \bibinfo {author} {\bibfnamefont {M.~B.}\ \bibnamefont
  {{Aufderheide}}}, \bibinfo {author} {\bibfnamefont {G.~J.}\ \bibnamefont
  {{Mathews}}}, \bibinfo {author} {\bibfnamefont {F.~D.}\ \bibnamefont
  {{Swesty}}}, \ and\ \bibinfo {author} {\bibfnamefont {G.~M.}\ \bibnamefont
  {{Fuller}}},\ }\href {\doibase 10.1086/173116} {\bibfield  {journal}
  {\bibinfo  {journal} {\apj}\ }\textbf {\bibinfo {volume} {414}},\ \bibinfo
  {pages} {701} (\bibinfo {year} {1993})}\BibitemShut {NoStop}%
\bibitem [{\citenamefont {{Fischer}}\ \emph
  {et~al.}(2010{\natexlab{b}})\citenamefont {{Fischer}}, \citenamefont
  {{Sagert}}, \citenamefont {{Hempel}}, \citenamefont {{Pagliara}},
  \citenamefont {{Schaffner-Bielich}},\ and\ \citenamefont
  {{Liebend{\"o}rfer}}}]{Fischer2010}%
  \BibitemOpen
  \bibfield  {author} {\bibinfo {author} {\bibfnamefont {T.}~\bibnamefont
  {{Fischer}}}, \bibinfo {author} {\bibfnamefont {I.}~\bibnamefont {{Sagert}}},
  \bibinfo {author} {\bibfnamefont {M.}~\bibnamefont {{Hempel}}}, \bibinfo
  {author} {\bibfnamefont {G.}~\bibnamefont {{Pagliara}}}, \bibinfo {author}
  {\bibfnamefont {J.}~\bibnamefont {{Schaffner-Bielich}}}, \ and\ \bibinfo
  {author} {\bibfnamefont {M.}~\bibnamefont {{Liebend{\"o}rfer}}},\ }\href
  {\doibase 10.1088/0264-9381/27/11/114102} {\bibfield  {journal} {\bibinfo
  {journal} {Classical and Quantum Gravity}\ }\textbf {\bibinfo {volume}
  {27}},\ \bibinfo {pages} {114102} (\bibinfo {year}
  {2010}{\natexlab{b}})}\BibitemShut {NoStop}%
\bibitem [{\citenamefont {{Nakazato}}\ \emph {et~al.}(2008)\citenamefont
  {{Nakazato}}, \citenamefont {{Sumiyoshi}},\ and\ \citenamefont
  {{Yamada}}}]{Nakazato}%
  \BibitemOpen
  \bibfield  {author} {\bibinfo {author} {\bibfnamefont {K.}~\bibnamefont
  {{Nakazato}}}, \bibinfo {author} {\bibfnamefont {K.}~\bibnamefont
  {{Sumiyoshi}}}, \ and\ \bibinfo {author} {\bibfnamefont {S.}~\bibnamefont
  {{Yamada}}},\ }\href {\doibase 10.1103/PhysRevD.77.103006} {\bibfield
  {journal} {\bibinfo  {journal} {\prd}\ }\textbf {\bibinfo {volume} {77}},\
  \bibinfo {eid} {103006} (\bibinfo {year} {2008})},\ \Eprint
  {http://arxiv.org/abs/0804.0661} {arXiv:0804.0661} \BibitemShut {NoStop}%
\bibitem [{\citenamefont {Landau}\ and\ \citenamefont
  {Lifshitz}(1969)}]{Landau69}%
  \BibitemOpen
  \bibfield  {author} {\bibinfo {author} {\bibfnamefont {L.}~\bibnamefont
  {Landau}}\ and\ \bibinfo {author} {\bibfnamefont {E.}~\bibnamefont
  {Lifshitz}},\ }\href@noop {} {\emph {\bibinfo {title} {Statistical
  Physics}}}\ (\bibinfo  {publisher} {Pergamon},\ \bibinfo {year}
  {1969})\BibitemShut {NoStop}%
\bibitem [{Kap()}]{Kapusta}%
  \BibitemOpen
  \href@noop {} {Ph.D. thesis}\BibitemShut {NoStop}%
\bibitem [{\citenamefont {Toro}\ \emph {et~al.}(2006)\citenamefont {Toro},
  \citenamefont {Drago}, \citenamefont {Gaitanos}, \citenamefont {Greco},\ and\
  \citenamefont {Lavagno}}]{DiToro06}%
  \BibitemOpen
  \bibfield  {author} {\bibinfo {author} {\bibfnamefont {M.~D.}\ \bibnamefont
  {Toro}}, \bibinfo {author} {\bibfnamefont {A.}~\bibnamefont {Drago}},
  \bibinfo {author} {\bibfnamefont {T.}~\bibnamefont {Gaitanos}}, \bibinfo
  {author} {\bibfnamefont {V.}~\bibnamefont {Greco}}, \ and\ \bibinfo {author}
  {\bibfnamefont {A.}~\bibnamefont {Lavagno}},\ }\href@noop {} {\bibfield
  {journal} {\bibinfo  {journal} {NPA}\ }\textbf {\bibinfo {volume} {775}}
  (\bibinfo {year} {2006})}\BibitemShut {NoStop}%
\bibitem [{\citenamefont {Toro}\ \emph {et~al.}(2011)\citenamefont {Toro},
  \citenamefont {Liu}, \citenamefont {Greco}, \citenamefont {Baran},
  \citenamefont {Colonna},\ and\ \citenamefont {Plumari}}]{DiToro11}%
  \BibitemOpen
  \bibfield  {author} {\bibinfo {author} {\bibfnamefont {M.~D.}\ \bibnamefont
  {Toro}}, \bibinfo {author} {\bibfnamefont {B.}~\bibnamefont {Liu}}, \bibinfo
  {author} {\bibfnamefont {V.}~\bibnamefont {Greco}}, \bibinfo {author}
  {\bibfnamefont {V.}~\bibnamefont {Baran}}, \bibinfo {author} {\bibfnamefont
  {M.}~\bibnamefont {Colonna}}, \ and\ \bibinfo {author} {\bibfnamefont
  {S.}~\bibnamefont {Plumari}},\ }\href@noop {} {\bibfield  {journal} {\bibinfo
   {journal} {Phys.Rev.C}\ }\textbf {\bibinfo {volume} {83}},\ \bibinfo {pages}
  {014911} (\bibinfo {year} {2011})}\BibitemShut {NoStop}%
\bibitem [{\citenamefont {Woosley}\ and\ \citenamefont
  {Weaver}(1995)}]{woosley_1995}%
  \BibitemOpen
  \bibfield  {author} {\bibinfo {author} {\bibfnamefont {S.}~\bibnamefont
  {Woosley}}\ and\ \bibinfo {author} {\bibfnamefont {T.}~\bibnamefont
  {Weaver}},\ }\href@noop {} {\bibfield  {journal} {\bibinfo  {journal}
  {Ap.J.Supp.}\ }\textbf {\bibinfo {volume} {101}} (\bibinfo {year}
  {1995})}\BibitemShut {NoStop}%
\bibitem [{\citenamefont {Rodrigues}\ \emph {et~al.}(2010)\citenamefont
  {Rodrigues}, \citenamefont {Duarte},\ and\ \citenamefont
  {Oliviera}}]{rodrigues2010}%
  \BibitemOpen
  \bibfield  {author} {\bibinfo {author} {\bibfnamefont {H.}~\bibnamefont
  {Rodrigues}}, \bibinfo {author} {\bibfnamefont {S.}~\bibnamefont {Duarte}}, \
  and\ \bibinfo {author} {\bibfnamefont {J.}~\bibnamefont {Oliviera}},\
  }\href@noop {} {\bibfield  {journal} {\bibinfo  {journal} {Nuc.Phys.B}\
  }\textbf {\bibinfo {volume} {199}} (\bibinfo {year} {2010})}\BibitemShut
  {NoStop}%
\bibitem [{\citenamefont {do~Carmo}\ and\ \citenamefont
  {Lugones}(2013)}]{docarmo2013}%
  \BibitemOpen
  \bibfield  {author} {\bibinfo {author} {\bibfnamefont {T.}~\bibnamefont
  {do~Carmo}}\ and\ \bibinfo {author} {\bibfnamefont {G.}~\bibnamefont
  {Lugones}},\ }\href@noop {} {\bibfield  {journal} {\bibinfo  {journal}
  {Physica A}\ }\textbf {\bibinfo {volume} {392}} (\bibinfo {year}
  {2013})}\BibitemShut {NoStop}%
\bibitem [{\citenamefont {Constantinou}\ \emph {et~al.}(2015)\citenamefont
  {Constantinou}, \citenamefont {Muccioli}, \citenamefont {Prakash},\ and\
  \citenamefont {Lattimer}}]{constantinou2015}%
  \BibitemOpen
  \bibfield  {author} {\bibinfo {author} {\bibfnamefont {C.}~\bibnamefont
  {Constantinou}}, \bibinfo {author} {\bibfnamefont {B.}~\bibnamefont
  {Muccioli}}, \bibinfo {author} {\bibfnamefont {M.}~\bibnamefont {Prakash}}, \
  and\ \bibinfo {author} {\bibfnamefont {J.}~\bibnamefont {Lattimer}},\
  }\href@noop {} {\bibfield  {journal} {\bibinfo  {journal} {Phys.Rev.C}\
  }\textbf {\bibinfo {volume} {92}},\ \bibinfo {pages} {025801} (\bibinfo
  {year} {2015})}\BibitemShut {NoStop}%
\end{thebibliography}%

\end{document}